\documentclass{article}
\usepackage[utf8]{inputenc}
\usepackage[T1]{fontenc}
\usepackage{amsmath, amssymb}
\usepackage{hyperref}

\providecommand{\keywords}[1]
{
  \small	
  \textbf{\textit{Keywords---}} #1
}

\title{Was Einstein a Lone Genius?}
\author{Galina Weinstein
  \thanks{Reichman University, The Efi Arazi School of Computer Science and The Department of Philosophy, University of Haifa.} 
  }

\begin{document}

\maketitle

\begin{abstract}
Albert Einstein's journey to formulate the theory of general relativity involved significant shifts in his approach to gravitational field equations. Starting in 1912, he collaborated with Marcel Grossmann and initially explored broadly covariant field equations. However, in 1913, Einstein discarded these equations in favor of the non-covariant \emph{Entwurf} field equations. This shift was a crucial step in his path to the final theory of general relativity, which he eventually achieved in 1915. This analysis explores Einstein's evolving ideas and decisions regarding the mathematical framework of his theory of gravity during the critical period 1912-1916.
My findings in this paper highlight that Einstein's brilliance did not exist in isolation but thrived within a vibrant scientific discourse. His work was significantly enriched through contributions and discussions with friends and colleagues, notably Michele Besso and Marcel Grossmann, illustrating the collaborative essence of scientific advancement. Nevertheless, it is essential to recognize that while these individuals made substantial contributions, they did not share an equal role in shaping the fundamental framework of general relativity, which remained predominantly Einstein's intellectual creation. This nuanced perspective enhances our grasp of the pivotal role played by collaborative dynamics in the development of groundbreaking theories.
\end{abstract}


\tableofcontents

\section{Introduction}

In the winter of 1912-1913, Einstein began working on his new gravitational theory, aiming to develop field equations that describe gravity in a way consistent with his theory of relativity.
He used a small notebook, the \emph{Zurich Notebook}, to jot down gravitational equations during this period \cite{CPAE4}, Doc 10.
Collaborating with his friend Marcel Grossmann, Einstein gradually gained access to new mathematical tools that would help him in his quest.

In his early collaboration with Grossmann, Einstein's gravitational field equations were closely related to the ones he would eventually present in his November 1915 theory of general relativity \cite{Einstein1}.
These early equations exhibited broad covariance, indicating their general applicability. 

In the spring of 1913, Einstein discarded what he later considered correct gravitational field equations.
He abandoned the calculations in the final pages of the \emph{Zurich Notebook}, representing equations with broad covariance.
Instead, he adopted the non-covariant \emph{Entwurf} field equations in 1913, collaborating with Grossmann on this formulation \cite{Stachel}.

I examine Einstein's evolving approach to gravitational field equations, tracing his journey from the early stages of collaboration with Grossmann and the exploration of general covariance in the \emph{Zurich Notebook} to his subsequent abandonment of it in favor of adopting the \emph{Entwurf} field equations, which he worked on with Grossmann and solved with Michele Besso. Additionally, I delve into his challenges in disentangling himself from the constraints of the \emph{Entwurf} theory and his eventual return to the principle of general covariance in 1915.

\section{From the \emph{Zurich Notebook} to the \emph{Entwurf} theory} \label{1}

During the early stages of developing his theory of gravitation, Albert Einstein conducted research focused on static gravitational fields. This research ultimately led to the formulation of a non-linear equation in 1912, as documented in reference \cite{Einstein3}:

\begin{equation} \label{Eq 101}
\boxed{c\Delta c - \frac{1}{2} (\nabla c)^2 = \kappa c^2 \sigma \quad.} 
\end{equation}

\noindent On the left-hand side, $c$ represents the speed of light; $\Delta$ is the Laplacian operator, representing the second spatial derivative $(\nabla^2)$. $\Delta c$ represents the Laplacian of the gravitational potential $c$.  
$\nabla c$ represents the function's gradient; $\kappa$ is a constant related to the gravitational constant. $\frac{1}{2\kappa} (\nabla c)^2$ is a term that Einstein referred to as the energy density of the gravitational field. 
$\sigma$ represents the energy density associated with the matter in the gravitational field; $c\sigma$ is the product of the speed of light and the mass density. 
Equation (\ref{Eq 101}) shows that the Laplacian of the gravitational potential $\Delta c$ is related to both the energy density associated with matter ($c \sigma$) and the energy density associated with the gravitational field itself [$\frac{1}{2\kappa} (\nabla c)^2$]. 
\vspace{2mm} 

Starting in the summer of 1912, Einstein embarked on a journey to explore the correct form of the field equations for his new theory of gravitation. His quest led him to delve into the realm of tensors and the intricacies of gravitational fields. During the winter months of 1912-1913, he diligently recorded his gravitational equations in a small notebook that has since become known as the \emph{Zurich notebook}.
Throughout this pivotal period, Einstein received invaluable assistance from his old schoolmate turned mathematics professor, Marcel Grossmann. This collaboration marked the beginning of their partnership, with a second significant collaboration occurring a year later when they jointly authored the \emph{Entwurf} paper \cite{Einstein16}.

During his time as a student at the Zurich Polytechnic, Einstein encountered challenges when studying subjects that did not pique his interest. He predominantly devoted his hours to solitary pursuits, embarking on a self-guided journey through physics. This involved conducting experiments and immersing himself in the works of the pioneering figures in the field. Einstein was, in essence, an autodidact. 
Einstein's approach to his education at the Polytechnic diverged from the conventional path. He often bypassed lectures, refrained from diligently taking notes, and "regrettably" missed numerous mandatory classes, particularly in mathematics. For instance, he never attended Professor Adolf Hurwitz's mathematical seminars, even though the opportunity for a solid mathematical foundation under Hurwitz's tutelage was within reach. Einstein's lack of enthusiasm extended to other mathematical courses at the Polytechnic, which he habitually skipped, much like he did with Hurwitz's classes.
Fortunately, before facing the crucial examinations mandated during his four-year course, Einstein turned to the notes of his dedicated friend and classmate, Grossmann. When Einstein found himself in dire need of employment, Grossmann's father intervened and secured him a position at the patent office in 1902. 

In the summer of 1912, Einstein again sought Grossmann's assistance, the man who had come to his rescue before.
Einstein confided in Grossmann about his predicament and asked him to investigate whether there was a suitable geometry to address the questions. The following day, Grossmann returned with the revelation that Riemannian geometry was the appropriate choice. 
Thus, Einstein, the autodidact, embarked on yet another self-guided educational journey, this time delving into the mathematical tools developed by luminaries such as Riemann, Ricci, Christoffel, and Levi-Civita.

In the \emph{Zurich notebook}, we witness Einstein's tireless search for field equations that bore a striking resemblance to the ones he would eventually unveil on November 4, 1915, as part of his general theory of relativity \cite{Einstein1}. These early gravitational field equations possessed a broad covariance, indicating their encompassing nature.
However, in the spring of 1913, Einstein made a fateful decision. He discarded what he later claimed to be the fundamentally correct gravitational field equation, which he had meticulously worked on in the final pages of the \emph{Zurich notebook}. Instead, he adopted the non-covariant \emph{Entwurf} field equations of 1913 in collaboration with Grossmann \cite{Einstein16}. This decision had profound consequences and set the stage for what would be referred to as Einstein's fall into the \emph{Entwurf} trap, discussed in section \ref{4}.
But first, I will discuss the \emph{Zurich notebook}.

Throughout the \emph{Zurich notebook}, Einstein presented equations without offering any explanations or derivations. Consequently, comprehending the methodology behind his equations in the \emph{Zurich notebook} requires extrapolation and educated guesses \cite{Einstein7}.

Einstein initiated his work on gravitational field equations with equation (\ref{30}). On page $19L$ of the notebook, he formulated field equations based on the Ricci tensor in a first-order approximation \cite{Einstein7}: 

\begin{equation} \label{30}
\Box g_{ik} = \kappa \rho_0 \frac{{dx_i}}{{d\tau}} \frac{{dx_k}}{{d\tau}} = \kappa T_{ik}.    
\end{equation}
\vspace{1mm} 

\noindent The left-hand side contains the d'Alembertian operator $\Box$, a wave operator that measures the second derivatives of the metric tensor components, a generalization of the Laplacian operator in flat spacetime. 
On the right-hand side, we have $\kappa$ (the constant related to the gravitational constant) times the contravariant stress-energy tensor $T_{ik}$, which describes the distribution of matter and energy in spacetime. 
Equation (\ref{30}) is a linearized equation that captures the gravitational effects of matter and energy to first-order approximation. 

On page $19R$, Einstein imposed three coordinate conditions. He aimed to obtain field equations consistent with energy-momentum conservation. 
He also sought to recover the Newtonian limit, corresponding to classical gravity in weak-field conditions \cite{Einstein7}, \cite{JR}. The coordinate conditions are:

1) \emph{The linearized harmonic coordinate condition}.

\begin{equation} \label{Eq 206}
\boxed{\sum_{\kappa} \gamma_{\kappa\kappa} \left(2 \frac{\partial g_{i\kappa}}{\partial x_\kappa} - \frac{\partial g_{\kappa\kappa}}{\partial x_i}\right) = 0.}    
\end{equation}

\noindent In equation (\ref{Eq 206}), $\gamma_{\kappa\kappa}$ represents the components of the inverse metric tensor (the contravariant metric tensor). $g_{i\kappa}$ and $g_{\kappa\kappa}$ are components of the metric tensor (the covariant metric tensor). $\frac{\partial g_{i\kappa}}{\partial x_\kappa}$ and $\frac{\partial g_{\kappa\kappa}}{\partial x_i}$ are the derivatives of these metric components with respect to spacetime coordinates.

Utilizing the linearized harmonic coordinate condition, represented by equation (\ref{Eq 206}), and working with the Ricci tensor, referred to as "Grossman's tensor" in Einstein's notebook (as seen in equation (\ref{Eq 270})), Einstein derived the field equations for gravitation, denoted as equation (\ref{30}).
Furthermore, Einstein was motivated to ensure that his theory would converge to Newtonian gravity under the appropriate conditions. To move closer to this objective, he enforced the harmonic coordinate condition.

2) \emph{The linearized "Hertz" condition}.

\begin{equation} \label{Eq 201}
\boxed{\sum_{\kappa}\gamma_{\kappa\kappa} \left(\frac{\partial g_{i\kappa}}{\partial x_\kappa}\right) = 0.}    
\end{equation}

This condition imposes a further constraint on the metric tensor and its derivatives. It effectively restricts how the metric tensor components can vary for spacetime coordinates. This condition was called by scholars the linearized "Hertz" condition because Einstein later mentioned it in a letter to Paul Hertz \cite{JR}. 

The linearized "Hertz" condition, as expressed in equation (\ref{Eq 201}), plays a role in simplifying the linearized field equations.
Like the linearized harmonic coordinate condition, the linearized "Hertz" condition contributes to the consistency of the field equations with energy-momentum conservation. 

3) \emph{A condition on the trace of the metric}.

\begin{equation} \label{Eq 242}
\boxed{\sum_{\kappa} \gamma_{\kappa\kappa} g_{\kappa\kappa} = \text{const}.}    
\end{equation}

\noindent In this equation, $g_{\kappa\kappa}$ are the diagonal components of the covariant metric tensor.

Einstein imposed a condition on the weak-field metric tensor, specifically its diagonal components. By setting the trace of the metric equal to a constant, Einstein ensured that the total energy and momentum associated with the metric tensor remained constant. 
In other words, by controlling the trace of the metric, Einstein aimed to prevent violations of energy-momentum conservation. 

By setting the sources $T_{ik}$ to zero in equation (\ref{30}), Einstein studied the behavior of $\frac{\partial g_{i\kappa}}{\partial x_{\kappa}}$ in the absence of external sources. He imposed the linearized Hertz condition on equations (\ref{30}) without sources. I will proceed with the derivation, illustrating the steps that Einstein likely followed. He applied the "Hertz" condition (\ref{Eq 201}) to the metric component $g_{ik}$, 
and took the second derivative of this equation with respect to $x_i$ using the box operator $\Box$:

\begin{equation}
\Box\left(\sum_{\kappa}\gamma_{\kappa\kappa} \left(\frac{\partial g_{i\kappa}}{\partial x_\kappa}\right)\right) = \Box(0).
\end{equation}

\noindent The right-hand side is zero because $\Box (0) = 0$:

\begin{equation} \label{Eq 272}
\Box\left(\sum_{\kappa}\gamma_{\kappa\kappa} \left(\frac{\partial g_{i\kappa}}{\partial x_\kappa}\right)\right) = 0.
\end{equation}

\noindent The d'Alembertian operator distributes across the sum:

\begin{equation}
\Box\left(\sum_{\kappa}\gamma_{\kappa\kappa} \left(\frac{\partial g_{i\kappa}}{\partial x_\kappa}\right)\right) = \sum_{\kappa}\Box\left(\gamma_{\kappa\kappa} \frac{\partial g_{i\kappa}}{\partial x_\kappa}\right) = \sum_{\kappa}\left(\Box\gamma_{\kappa\kappa}\right) \frac{\partial g_{i\kappa}}{\partial x_\kappa} = 0.
\end{equation}

\noindent We end up with the following result:

\begin{equation}
\sum_{\kappa}\left(\Box\gamma_{\kappa\kappa}\right) \frac{\partial g_{i\kappa}}{\partial x_\kappa} = \Box \gamma_{\kappa\kappa} \left(\frac{\partial g_{i\kappa}}{\partial x_\kappa}\right) = 0.
\end{equation}

\noindent Now we assume that $\gamma_{\kappa\kappa} = 1$, and the equation simplifies to:

\begin{equation} \label{Eq 203}
\Box\left(\frac{\partial g_{i\kappa}}{\partial x_{\kappa}}\right) = 0.    
\end{equation}

\noindent This assumption reflects a flat metric. In the weak-field limit, the spatial metric of a weak gravitational field was assumed to be approximately flat. This assumption was crucial for Einstein to obtain results consistent with Newtonian gravity in the appropriate limit. In special relativity, the metric tensor reduces to the Minkowski metric $\eta_{\mu \nu}$, which has diagonal elements of $(+1, -1, -1, -1)$, where the speed of light $c$ is treated as a constant and plays a crucial role. This metric accurately describes the behavior of spacetime in the absence of gravitational fields. In this context, $\gamma_{\kappa\kappa} = 1$ because spacetime is considered flat. In the weak-field limit, Einstein aimed to recover Newtonian gravity in a weak gravitational field. 
The $\gamma_{\kappa\kappa} = 1$ reflects the apparent coincidence in using the same metric form in both the special relativity and the weak-field limit. The similarity in form between the Minkowski metric and the metric in the weak-field limit arises because both situations describe spacetime when it is nearly flat. However, there is a fundamental difference: In special relativity, spacetime is flat due to the absence of gravitational fields, while in the weak-field limit of general relativity, spacetime is effectively flat because gravitational effects are small.  
In this context, when $\gamma_{\kappa\kappa} = 1$ for the spatial components of the metric, it reflects the assumption that the spatial part of spacetime is nearly flat.     

Einstein modified equation (\ref{30}) to include an additional term on the right-hand side involving the trace of the stress-energy tensor \cite{Einstein7}. This modification was an attempt to ensure energy-momentum conservation within the framework of these equations. 

Suppose we have a cloud of pressureless dust. In that case, the energy-momentum tensor for pressureless dust is given by:

\begin{equation} \label{Eq 356}
T_{ik} =\rho_0 u_k u^k,
\end{equation}

\noindent where $\rho_0$ is the mass density, and $u_i$ is the four-velocity of the dust.

\noindent In this case, the trace of the energy-momentum tensor is:

\begin{equation}
T = T_{k}^{k} =\rho_0 u_k u^k,
\end{equation}

\noindent So, equation (\ref{30}) becomes:

\begin{equation} \label{Eq 250}
\Box g_{ik} = \kappa\rho_0 u_k u^k =\text{const.}   
\end{equation} 

Now, it is important to notice that it is not that $g_{ik}$ is constant; rather, the equation describes the linearized gravitational response to a constant mass density of pressureless dust. The solution for $g_{ik}$ will depend on the specific distribution of dust and the boundary conditions of the problem \cite{Einstein7}. 

Einstein obtained the result equation (\ref{Eq 250}) in his calculations, and it suggested to him the presence of an additional term on the right-hand side of the equation (\ref{30}) to account for this constant. 
On page $20L$, Einstein writes the following equation \cite{Einstein7}:

\begin{equation} \label{Eq 36}
\Box(g_{ik}) = \rho_0 \frac{dx_i}{d\tau} \frac{dx_\kappa}{d\tau} - \frac{1}{4}\rho_0 \frac{dx_\kappa}{d\tau} \frac{dx_\kappa}{d\tau}.    
\end{equation}

\noindent The term involving the trace of the stress-energy tensor (\(T\)) is given by \(T = T_k^k\), which represents the sum of the diagonal components of the tensor. It can be considered a measure of the total energy density associated with matter and fields in a given region of spacetime. 
The right-hand side of equation (\ref{Eq 36}) involves terms related to the stress-energy tensor and its trace. The first term on the right (\(\kappa\rho_0\frac{dx_i}{d\tau}\frac{dx_\kappa}{d\tau}\)) accounts for the energy-momentum associated with matter moving through spacetime. The additional term on the right \(-\frac{1}{4}\kappa\rho_0\frac{dx_\kappa}{d\tau}\frac{dx_\kappa}{d\tau}\) is introduced specifically to account for the contribution of the trace of the stress-energy tensor to the gravitational field.

The additional term involving the trace of $T$ guarantees that the energy and momentum associated with the gravitational field are properly included in the theory's overall conservation of energy and momentum.
By adding this term, Einstein wanted the equations describing how matter and energy interact with the curvature of spacetime to account for the contribution of the trace of the stress-energy tensor. This was crucial for maintaining energy-momentum conservation in the presence of gravity. 

Researchers have noted that Einstein discarded equations (\ref{Eq 36}) due to their incompatibility with the condition presented in equation (\ref{Eq 242}) \cite{JR}. This setback prompted Einstein to revise equation (\ref{30}) once more, introducing an additional term involving the trace of the metric on the left-hand side of the equation. This demonstrates a shift in Einstein's approach, as he had previously introduced a term involving the trace of the stress-energy tensor on the right-hand side in equation (\ref{Eq 36}) \cite{Weinstein2}. The new equations introduced by Einstein take the following forms, with the first equation presented as \cite{Einstein7}:

\begin{equation} \label{Eq 280}
\Delta(g_{11}-\frac{1}{2}U) = T_{11}.
\end{equation}

\noindent This linearized field equation represents a relationship involving the Laplacian operator $\Delta$, the metric component $g_{11}$, the scalar field - the gravitational potential $U$, and the component of the energy-momentum tensor $T_{11}$. 

\noindent Here are the second, third, and fourth equations \cite{Einstein7}:

\begin{equation}
\Delta g_{12} = T_{12}, \Delta g_{13} = T_{13}, \Delta g_{14} = T_{14}.
\end{equation}

\noindent These equations are similar to the first one but involve the metric components $g_{12}$, $g_{13}$, and $g_{14}$, and the components of the energy-momentum tensor $T_{12}$, $T_{13}$, and $T_{14}$. They represent other relationships derived under the same conditions.

Using the Kronecker delta $\delta$, equation (\ref{Eq 280}) can be written in compact form: 

\begin{equation} \label{Eq 299}
\boxed{\Box \left(g_{ik} - \frac{1}{2}\delta_{ij}U\right)= \kappa T_{ik}.} 
\end{equation}

Scholars have noted a remarkable similarity between equation (\ref{Eq 299}) and Einstein's tensor \cite{JR}:

\begin{equation}
\boxed{R - \frac{1}{2}g_{\mu \nu}R.}
\end{equation}

\noindent $R$ is the Ricci scalar and $g_{\mu \nu}$ is the metric tensor.

However, equation (\ref{Eq 299}) is linearized.
The linearized version of the 1915 Einstein field equations (\ref{Eq 20}) takes the form:

\begin{equation} \label{Eq 277}
\boxed{\Box h_{\mu \nu} - \frac{1}{2} \eta_{\mu \nu} \Box h = \kappa T{\mu \nu},}   
\end{equation}

\noindent where, $\eta_{\mu \nu}$ is the Minkowski metric, and $h$ is the trace of the perturbation $h_{\mu \nu}$. 

The linearized form of the Einstein field equations [equation (\ref{Eq 277})] is used to describe small perturbations around a flat spacetime. In his \emph{Entwurf} theory, Einstein considered an equation of this nature; see equation (\ref{Eq 95}). To ensure comparability between both equations (\ref{Eq 299}) and (\ref{Eq 277}), it is necessary to state that in equation (\ref{Eq 299}), the perturbation is represented by $U$. Then, the equation describes the linearized behavior of gravitational perturbations in the presence of a source represented by $T_{ik}$. The term $g_{ik} - \frac{1}{2}\delta_{ij}U$ is then the perturbed metric tensor, where $U$ represents the gravitational potential. This perturbation to the metric accounts for deviations from the background (i.e., flat) metric. The equation describes how these perturbations interact with the source term $T_{ik}$ in a linearized approximation. 

Einstein then simplified the left-hand side of the first equation (\ref{Eq 280}) by noticing that the Laplacian of $U$ is $-2\Delta U$. 
Finally, he recognized that \cite{Einstein7}:

\begin{equation} \label{Eq 281}
-2\Delta U = \sum T_{\kappa\kappa} = \sum T.
\end{equation}

\noindent This equation represents the Laplace operator applied to the scalar field $U$, equal to the sum of the diagonal components of the tensor $T_{\kappa\kappa}$ which is also equal to the sum $T$, the trace of $T_{\kappa\kappa}$ over the index $\kappa$. 

\noindent He plugged equation (\ref{Eq 281}) into equation (\ref{Eq 280}) and obtained for the first equation:

\begin{equation} \label{Eq 275}
\Delta g_{11} = T_{11} - \frac{1}{2} \sum T_{\kappa\kappa}.
\end{equation}

\noindent This equation represents the Laplace operator applied to the component $g_{11}$, which is equal to $T_{11}$ minus half of the sum of the diagonal components of the tensor $T_{\kappa \kappa}$ (the trace of tensor $T_{\kappa\kappa}$).

\noindent Using the Kronecker delta $\delta$, equation (\ref{Eq 36}) can be written in compact form: 

\begin{equation} \label{Eq 276}
\boxed{\Box(g_{ij}) = T_{ij} - \frac{1}{2} \delta_{ij} \left(\sum T\right).}
\end{equation}

\noindent This equation describes the behavior of the metric tensor components $g_{ij}$ in a gravitational field. It relates the Laplacian of the metric components to the components of the stress-energy tensor $T_{ij}$ and the trace of the stress-energy tensor.  
Equation (\ref{Eq 276}) and the standard form of Einstein's field equations in the covariant form, equation (\ref{Eq 20}), share a similar form. In hindsight, the linearized equation (\ref{Eq 276}) represents a simplified version of the Einstein field equations \cite{Weinstein2}. 

Einstein ultimately abandoned the equations (\ref{Eq 275}) for the same reason he had previously discarded the other linearized field equations (see \cite{Weinstein2}).

Next, Einstein imposes a fourth condition \cite{Einstein7}: 

4) \emph{The $\theta_{ik\lambda}$ condition}.

\begin{equation} \label{Eq 208}
\boxed{\theta_{ik\lambda} = \frac{1}{2} \left(\frac{\partial g_{ik}}{\partial x_\lambda} + \frac{\partial g_{k\lambda}}{\partial x_i} + \frac{\partial g_{\lambda i}}{\partial x_k}\right).}    
\end{equation}

Scholars called equation (\ref{Eq 208}) the "theta condition" \cite{JR}. This expression significantly influenced Einstein's derivations and manipulations, particularly in eliminating unwanted terms from equations (\ref{Eq 37}). 

The expression (\ref{Eq 208}) involves partial derivatives of the metric tensor components \(g_{ik}\) concerning the coordinates \(x _\lambda\). 
The \(\theta_{ik\lambda}\) condition is defined similarly to the Christoffel symbols (\ref{Eq 278}). It involves partial derivatives of the metric components $g_{ik}$ with respect to the spacetime coordinates $x_\lambda, x_i, \text{and}, x_\kappa$, and it has a similar form to the expression for $\Gamma_{\mu \nu}^{\alpha}$. So, while it is not exactly the Christoffel symbol, it shares a close resemblance and serves a similar purpose in describing the connection coefficients associated with the metric tensor $g_{ik}$. 

In Einstein's calculations dealing with equation (\ref{Eq 37}), unwanted terms involving second-order derivatives of the metric could arise. The "theta expression" \(\theta_{ik\lambda}\) is used to manipulate and simplify these terms, by canceling out unwanted contributions. The expression is constructed to transform as a tensor under coordinate transformations (i.e., covariant).

Einstein titled page $22R$ "Grossmann". He derived a tensor, representing the gravitational field in terms of the Christoffel symbols and their derivatives. Initially, Einstein considered specifying a covariant differential tensor of the second rank and second-order for the Ricci tensor on page $22R$ \cite{Einstein7}:

\begin{equation} \label{Eq 38}
\boxed{T_{il} = \sum_{\kappa l} \frac{\partial}{\partial x_l} \left(\Gamma_{i\kappa}^\kappa\right) - \frac{\partial}{\partial x_\kappa} \left(\Gamma_{il}^\kappa\right) - \left(\Gamma_{i\kappa}^\lambda\right)\left(\Gamma_{\lambda l}^\kappa\right) - \left(\Gamma_{il}^\lambda\right)\left(\Gamma_{\lambda\kappa}^\kappa\right).}
\end{equation}

\noindent $\Gamma_{il}^\kappa$ represent the Christoffel symbols:

\begin{equation} \label{Eq 278}
\boxed{\Gamma_{il}^\kappa = \frac{1}{2} g^{\kappa\lambda} \left( \frac{\partial g_{\lambda l}}{\partial x^i} + \frac{\partial g_{i\lambda}}{\partial x^l} - \frac{\partial g_{il}}{\partial x^\lambda} \right).}
\end{equation}

\noindent In these expressions:
$i$, $l$, and $\kappa$ are spacetime indices.
$g^{\kappa\lambda}$ represents the components of the inverse metric tensor.
$g_{\lambda l}$, $g_{i\lambda}$, and $g_{il}$ are components of the metric tensor.
$\frac{\partial}{\partial x^i}\left(\frac{\partial g_{\lambda l}}{\partial x^i}\right)$, $\frac{\partial}{\partial x^l}\left(\frac{\partial g_{i\lambda}}{\partial x^l}\right)$, and $\frac{\partial}{\partial x^\lambda}\left(\frac{\partial g_{il}}{\partial x^\lambda}\right)$ are the partial derivatives of the metric tensor with respect to spacetime coordinates.
We can represent $\Gamma_{i\kappa}^\lambda$ and $\Gamma_{\lambda l}^\kappa$ using similar notation. $\Gamma_{i\kappa}^\lambda$ represents the Christoffel symbol with the indices $i$, $\kappa$, and $\lambda$. $\Gamma_{\lambda l}^\kappa$ represents the Christoffel symbol with the indices $\lambda$, $l$, and $\kappa$.

Let us derive Einstein's gravitational tensor from the Ricci tensor. We start with the original Ricci tensor as given in equation (\ref{Eq 38}). Now, let us use the relation from page $22$R \cite{Einstein7}:

\begin{equation} \label{Eq 295}
\frac{\partial \log \sqrt{-g}}{\partial x_i} = \Gamma_{\kappa}^{i \kappa},    
\end{equation}

\noindent to rewrite some of the Christoffel symbols. 

The presence of the square root of the determinant of the metric tensor $\sqrt{-g}$ in the expression (\ref{Eq 295}) is related to unimodular transformations. Unimodular transformations refer to coordinate transformations that preserve the determinant $g$ of the metric tensor. In other words, if we perform a coordinate transformation, the new metric tensor $g'_{\mu \nu}$ in the new coordinate system should have the same determinant as the original metric tensor $g_{\mu \nu}$, i.e., $g' = g$. Equation (\ref{Eq 295}) relates the partial derivative of the logarithm of the square root of the determinant of the metric tensor with the Christoffel symbols ($\Gamma_{\kappa}^{i \kappa}$), which describe the connection between covariant and contravariant derivatives in curved spacetime.

Now, using equation (\ref{Eq 295}), we can express the first and second terms in equation  (\ref{Eq 38}), $T_{il}$, differently:

\begin{equation} \label{Eq 297}
\frac{\partial}{\partial x_l} \left(\Gamma_{i\kappa}^\kappa\right) - \frac{\partial}{\partial x_\kappa} \left(\Gamma_{il}^\kappa\right) = \boxed{\frac{\partial}{\partial x_l} \left(\frac{\partial \log \sqrt{-g}}{\partial x_i}\right) - \frac{\partial}{\partial x_\kappa} \left(\Gamma_{il}^\kappa\right).}
\end{equation}

\noindent Using equation (\ref{Eq 295}), we can express the third and fourth terms in equation  (\ref{Eq 38}), $T_{il}$, differently:

\begin{equation} \label{Eq 298}
\left(\Gamma_{i\kappa}^\lambda\right)\left(\Gamma_{\lambda l}^\kappa\right) - \left(\Gamma_{il}^\lambda\right)\left(\Gamma_{\lambda\kappa}^\kappa\right) = \boxed{\left(\Gamma_{i\kappa}^\lambda\right)\left(\Gamma_{\lambda l}^\kappa\right) - \left(\Gamma_{il}^\lambda\right)\left(\frac{\partial \log \sqrt{-g}}{\partial x_i}\right).}
\end{equation}

\noindent Next, we will substitute equations (\ref{Eq 297}) and (\ref{Eq 298}) into the expression for $T_{il}$ and simplify:

\begin{equation} \label{Eq 296}
T_{il} = \sum_{\kappa l} \frac{\partial}{\partial x_l} \left(\frac{\partial \log \sqrt{-g}}{\partial x_i}\right) - \boxed{\frac{\partial}{\partial x_\kappa} \left(\Gamma_{il}^\kappa\right) - \left(\Gamma_{i\kappa}^\lambda\right)\left(\Gamma_{\lambda l}^\kappa\right)} - \left(\Gamma_{il}^\lambda\right)\left(\frac{\partial \log \sqrt{-g}}{\partial x_i}\right).
\end{equation}

Recall that Einstein employed unimodular transformations, where $\sqrt{g} = 1$. Consequently, when $\sqrt{g} = 1$, the logarithm of $\sqrt{-g}$ approaches zero.
\noindent This leaves the Ricci tensor with the second and third terms in equation (\ref{Eq 296}). These two terms constitute Einstein's gravitational tensor:

\begin{equation} \label{Eq 37}
\boxed{T_{il}^x = {\frac{\partial}{\partial x_\kappa} \left(\Gamma_{il}^\kappa\right) - \left(\Gamma_{i\kappa}^\lambda\right)\left(\Gamma_{\lambda l}^\kappa\right)}.}
\end{equation}

Scholars coined the term the "November tensor" to describe equation (\ref{Eq 37}). The reason it is referred to in this manner is that when we set the November tensor equal to the stress-energy tensor, multiplied by $\kappa$ (the constant related to the gravitational constant), we obtain the field equations (\ref{Eq 289}) of Einstein's first paper on general relativity of November 4, 1915, see \cite{Weinstein2}.

Einstein later abandoned the November tensor approach \cite{Einstein7}. However, he thoroughly examined the November tensor in the \emph{Zurich notebook} before doing so. Let us take a moment to analyze this briefly,

Einstein used the "theta condition" $\theta_{ik\lambda}$ (\ref{Eq 208}) to relate certain terms in the November tensor to derivatives of the metric tensor: 

\begin{equation} \label{Eq 207}
T_{il}^x = \sum_{\kappa l} \frac{\partial}{\partial x_\kappa} \gamma_{k\alpha} \left(\theta_{il\alpha} - \frac{\partial g_{il}}{\partial x_\alpha}\right) - \gamma_{\lambda\alpha} \gamma_{k\beta} \left(\theta_{ik\alpha} - \frac{\partial g_{ik}}{\partial x_\alpha}\right) \left(\theta_{i\lambda\beta} - \frac{\partial g_{l\lambda}}{\partial x_\beta}\right).
\end{equation}

In other words, Einstein manipulated the Christoffel symbols (\ref{Eq 278}) $\Gamma_{il}^\kappa$ through the "theta condition" $\theta_{ik\lambda}$ to arrive at the final expression for $T_{il}^x$, equation (\ref{Eq 207}).

The restrictions and conditions applied, including the Hertz condition, impose restrictions on the derivatives of the metric components, eliminate unwanted terms involving second-order derivatives of the metric, and simplify the November tensor. Arriving at a reduced form of the November tensor involving the metric's derivatives and potentially other terms relevant to the gravitational field equations.

By setting $\theta_{il\alpha}=0$ in equation (\ref{Eq 207}), Einstein obtained a truncated form of the November tensor \cite{Einstein7}:

\begin{equation} \label{Eq 35}
\sum_{\kappa\alpha} \gamma_{\kappa\alpha} \frac{\partial^2 g_{il}}{\partial x_\kappa \partial x_\alpha} + \sum_{\rho\alpha} \gamma_{\rho\alpha} \gamma_{k\beta} \frac{\partial g_{ik}}{\partial x_\alpha} \frac{\partial g_{l\rho}}{\partial x_\beta}.
\end{equation}

Einstein, therefore, imposed equation (\ref{Eq 208}). Recall that this condition is related to the connection coefficients, i.e., the Christoffel symbols. These symbols are introduced to simplify the covariant derivative and describe the connection between curves on a curved manifold. When we set $\theta_{il k \lambda}=0$ to the expression given in equation (\ref{Eq 208}), we are effectively making a specific choice for how the connection coefficients depend on the metric. This choice simplifies the equations and can be used when seeking the Newtonian limit. In the limit of weak gravitational fields, this choice for $\theta_{il k \lambda}=0$ leads to equations resembling Newtonian gravity's Poisson equation. The equation (\ref{Eq 35}) shows some of the terms contributing to the Newtonian limit of the gravitational field equations in general relativity.

Equation (\ref{Eq 207}) involves terms related to the second derivatives of the metric tensor and its components. This equation is derived as part of Einstein's strategy to simplify the November tensor and understand the behavior of the field equations in the weak-field limit. The goal of the weak-field limit is to recover the Newton-Poisson equation as a first approximation. With its terms involving second derivatives of the metric, equation (\ref{Eq 35}) can be simplified and compared to the Newton-Poisson equation, which describes the gravitational potential in Newtonian gravity. The Newton-Poisson equation relates the Laplacian of the gravitational potential (related to the second derivatives of the potential) to the mass density. Comparing equation (\ref{Eq 35}) to the Newton-Poisson equation and making appropriate approximations, one can show that Einstein's November tensor reduces to Newtonian gravity in the weak-field limit. By manipulating Equation (\ref{Eq 35}) and showing its resemblance to the Newton-Poisson equation, one can demonstrate that, for relatively weak gravitational fields, the November tensor reproduces the same gravitational effects as those predicted by Newton's law of gravitation.

Despite his efforts and the introduction of conditions 1), 2), 3), and 4), Einstein faced difficulties in achieving both goals (consistency with energy-momentum conservation and recovering the Newtonian limit) with the equations (\ref{30}), (\ref{Eq 36}), equation (\ref{Eq 275}), (\ref{Eq 37}), and (\ref{Eq 35}).
It appears that he could not demonstrate energy-momentum conservation satisfactorily within this framework.
Additionally, the recovery of the Newtonian limit posed challenges.
Due to these difficulties, Einstein eventually abandoned the November tensor (\ref{Eq 37}) and the associated field equations.
He moved on to the \emph{Entwurf} field equations, which were non-covariant and different from the November tensor \cite{Weinstein1}, \cite{Weinstein2}.

Grossmann collaborated with Einstein again, this time on the \emph{Entwurf} theory.
Einstein's collaboration with Grossmann resulted in a joint paper titled "Draft of a Generalized Theory of Relativity and a Theory of Gravitation" \cite{Einstein16}. This collaboration involved Einstein contributing to the physical aspects of the theory while Grossmann worked on the mathematical aspects. 

Grossmann reasoned that while it was possible to define the Ricci tensor (\ref{Eq 38}), it could not be reduced to the Poisson equation of Newton's theory of gravitation in the case of an infinitely weak static gravitational field. 
This realization raised questions about the connection between the Ricci tensor equation (\ref{Eq 38}) and the gravitational equations, particularly in arbitrary transformations. 
The key issue was whether the field equations allowed for arbitrary transformations or only a certain group of transformations (linear transformations).
Grossmann was uncertain about this distinction, which led to a lack of clarity regarding the role of the Ricci tensor equation (\ref{Eq 38}) in the gravitational field equations \cite{Einstein16}.
Notably, Grossmann seemed to adopt some of Einstein's misconceptions.

In their joint paper of 1913, Einstein formulated the covariant form of the \emph{Entwurf} vacuum field equations (\ref{Eq 1}), which describe the gravitational field in the absence of matter and energy, as follows \cite{Einstein16}:

\begin{equation} \label{Eq 1}
\boxed{D_{\mu\nu}(g) = 0.}
\end{equation}

\noindent He also expressed the covariant form of the \emph{Entwurf} field equation with sources as follows \cite{Einstein16}:

\begin{equation} \label{Eq 54}
\boxed{D_{\mu\nu}(g) = \kappa(T_{\mu\nu} + t_{\mu\nu}).}
\end{equation}

The \emph{Entwurf} field equations are about the relationship between the matter content in spacetime - described by the stress-energy tensor $T_{\mu \nu}$, and the energy-momentum carried by the gravitational field itself - described by the stress-energy pseudo-tensor of the gravitational field $t_{\mu \nu}$. $\kappa$ is a constant that relates the metric curvature to the stress-energy tensor. It is related to Newton's gravitational constant and the speed of light: $\kappa = \frac{8\pi G}{c^4}$. 
The above equations describe how spacetime curvature (as captured by the operator $D_{\mu \nu}(g)$ acting on the metric $g$) relates to the distribution of matter and energy.
$D_{\mu\nu} (g)$ represents an expression related to the curvature of spacetime. It involves the metric coefficients $\gamma_{\alpha\beta}$: 

\begin{equation} \label{Eq 143}
D_{\mu\nu} (g) = \sum_{\alpha\beta} \frac{1}{\sqrt{-g}} \frac{\partial}{\partial x_{\alpha}} \left( \gamma_{\alpha\beta} \sqrt{-g} \frac{\partial g_{\mu\nu}}{\partial x_{\beta}} \right) - \sum_{\alpha\beta\tau\rho} \gamma_{\alpha\beta} \gamma_{\tau\rho} \frac{\partial g_{\mu\tau}}{\partial x_{\alpha}} \frac{\partial g_{\nu\rho}}{\partial x_{\beta}}.
\end{equation}

\noindent $D_{\mu \nu}(g)$ is a differential operator acting on the metric tensor $g_{\mu \nu}$. $g$ is the determinant of $g_{\mu \nu}$. $D_{\mu \nu} (g)$ includes derivatives of the metric tensor and terms involving the connection $\gamma_{\alpha \beta}$, which is not the Christoffel symbol found in the final formulation of general relativity. 

The \emph{Entwurf} field equations (\ref{Eq 54}) are the covariant form of the \emph{Entwurf} field equations (\ref{Eq 279}). 
In a $4$-dimensional spacetime, the symmetric rank-$2$ tensors used in the \emph{Entwurf} field equations have $10$ independent components due to their symmetry. This symmetry property means that the components are equal when their indices are exchanged [$D_{\mu \nu}(g)=D_{\nu \mu}(g)$]. Therefore, the number of independent equations is reduced to $10$. 

Einstein then wrote an equation representing $t_{\mu\nu}$, the covariant form of the stress-energy pseudo-tensor $t_{\mu\nu}$ for the gravitational field:

\begin{equation} \label{Eq 146}
\boxed{-2\kappa t_{\mu\nu} = \sum_{\alpha\beta\tau\rho}\left(\frac{\partial g_{\tau\rho}}{\partial x_\mu}\frac{\partial \gamma_{\tau\rho}}{\partial x_\nu} - \frac{1}{2}g_{\mu\nu}\gamma_{\alpha\beta}\frac{\partial g_{\tau\rho}}{\partial x_\alpha}\frac{\partial \gamma_{\tau\rho}}{\partial x_\beta}\right).}
\end{equation}

\noindent Equation (\ref{Eq 146}) consists of two terms: The first term involves products of derivatives of the metric tensor $g_{\tau\rho}$ and the metric coefficients $\gamma_{\tau\rho}$. The second term in equation (\ref{Eq 146}) involves products of the metric tensor $g_{\mu\nu}$ and the metric coefficients $\gamma_{\alpha\beta}$ and their derivatives, with a factor of $\frac{1}{2}$. 

We plug equation (\ref{Eq 143}) into equation (\ref{Eq 54}) to replace $D_{\mu \nu}(g)$ on the left-hand side:

\begin{equation} \label{Eq 144}
\boxed{\sum_{\alpha\beta}\frac{1}{\sqrt{-g}}\frac{\partial}{\partial x_\alpha} \left(\gamma_{\alpha\beta}\sqrt{-g}\frac{\partial g_{\mu\nu}}{\partial x_\beta}\right) - \sum_{\alpha\beta\tau\rho}\gamma_{\alpha\beta}\gamma_{\tau\rho}\frac{\partial g_{\mu\tau}}{\partial x_\alpha}\frac{\partial g_{\nu\rho}}{\partial x_\beta} = -\kappa(T_{\mu\nu} + t_{\mu\nu}).}
\end{equation}

Einstein wrote the \emph{Entwurf} field equations in their contravariant form:

\begin{equation} \label{Eq 279}
\Delta_{\mu \nu}(\gamma) = \kappa(\Theta_{\mu \nu} + \theta_{\mu \nu}).
\end{equation}

\noindent $\Delta_{\mu \nu}(\gamma)$ is related to the curvature of spacetime, given by the metric coefficients $\gamma_{\alpha \beta}$. $\Theta_{\mu \nu}$ represents the energy-momentum tensor, and $\theta_{\mu \nu}$ is the pseudo-tensor related to the gravitational field itself.

The contravariant form of the $\Delta_{\mu \nu}(\gamma)$ which relates to the curvature of spacetime $\gamma_{\alpha \beta}$ is given by:

\begin{equation}
\Delta_{\mu\nu}(\gamma) = \sum_{\alpha\beta}\frac{1}{\sqrt{-g}}\frac{\partial}{\partial x_\alpha} \left(\gamma_{\alpha\beta}\sqrt{-g}\frac{\partial \gamma_{\mu\nu}}{\partial x_\beta}\right) - \sum_{\alpha\beta\tau\rho}\gamma_{\alpha\beta}\gamma_{\tau\rho}\frac{\partial \gamma_{\mu\tau}}{\partial x_\alpha}\frac{\partial \gamma_{\nu\rho}}{\partial x_\beta}.
\end{equation}

\noindent The first term involves derivatives of the metric coefficients $\gamma_{\alpha \beta}$ with respect to spacetime coordinates $x_\alpha$. The second term involves products of metric coefficients and their derivatives.

Einstein wrote the contravariant form of the pseudo-tensor $\theta_{\mu \nu}$; the relationship between the pseudo-tensor of the gravitational field and the metric coefficients \cite{Einstein16}:

\begin{equation} \label{Eq 145}
\boxed{-2\kappa\theta_{\mu\nu} = \sum_{\alpha\beta\tau\rho}\left(\gamma_{\alpha\mu}\gamma_{\beta\nu}\frac{\partial g_{\tau\rho}}{\partial x_\alpha}\frac{\partial\gamma_{\tau\rho}}{\partial x_\beta} - \frac{1}{2}\gamma_{\mu\nu}\gamma_{\alpha\beta}\frac{\partial g_{\tau\rho}}{\partial x_\alpha}\frac{\partial\gamma_{\tau\rho}}{\partial x_\beta}\right).}
\end{equation}

The pseudo-tensor $\theta_{\mu \nu}$ of the gravitational field is related to the metric coefficients $\gamma_{\alpha \beta}$ and derivatives of the metric tensor $g_{\tau \rho}$. The equation consists of two terms: The first term involves products of metric coefficients $\gamma_{\alpha\mu}\gamma_{\beta\nu}$ and their derivatives.
The second term involves products of the metric coefficients $\gamma_{\mu\nu}\gamma_{\alpha\beta}$ and their derivatives, with a factor of $\frac{1}{2}$.

The \emph{Entwurf} field equations are not generally covariant, i.e., they do not retain their form under arbitrary coordinate transformations. The absence of general covariance in the \emph{Entwurf} theory can be seen when trying to transform the equations under a general coordinate transformation. The $D_{\nu \mu}(g)$ operator, which is part of the gravitational field equations, as defined in the \emph{Entwurf} theory, will not maintain its form under such transformations. 

When Einstein and Grossmann worked together on developing the general theory of relativity, Einstein was the driving force and significantly influenced the project. Einstein was primarily responsible for physical concepts and developing the theory's foundations. He provided the initial ideas and insights that guided their collaboration. Conversely, Grossmann was crucial in providing mathematical expertise and tools to formulate the theory mathematically. He translated Einstein's physical insights into the language of tensor calculus and differential geometry, which was essential for developing the theory's equations. While both Einstein and Grossmann contributed to the collaboration, Einstein's influence and ideas were more prominent in shaping the project's direction. Einstein's authority and vision in physics were significant factors in their work together.

\section{1913: The Perihelion of Mercury} \label{2}

In June 1913, Einstein’s closest friend, Besso, a proficient engineer with an understanding of physics and mathematics, visited Einstein in Zurich. Together, they attempted to solve the \emph{Entwurf} vacuum field equations to determine Mercury's perihelion advance in the Sun's static field.

The derivation starts with assuming a weak gravitational field far from the Sun, which the Minkowski flat metric of special relativity can approximate. The metric components are represented by $g_{\mu \nu}$, where $\mu$ and $\nu$ are indices. 
Einstein kicks things off with the calculations in the \emph{Einstein-Besso manuscript}. He uses his field equation from static gravitational fields theory (a form of Poisson's equation) \cite{CPAE4}, Doc 14:

\begin{equation} \label{Eq 51}
\boxed{\nabla^2 g_{44} = c_0^2 \kappa \rho_0.}
\end{equation}

\noindent Equation (\ref{Eq 51}) relates the Laplacian of \(g_{44}\), which is the component of the metric tensor corresponding to the time-time component (\(\mu = \nu = 4\)), to the static dust cloud's mass density \(\rho_0\).
$c_0^2$ is the speed of light in vacuum.  

Einstein reduces equation (\ref{Eq 101}) to equation (\ref{Eq 51}). In equation (\ref{Eq 101}), the speed of light $c$ is interpreted as the speed of light $c_0$, representing the same constant. The term $\sigma$ in equation (\ref{Eq 101}) is considered as the mass density $\rho_0$ in equation (\ref{Eq 51}), and so equation (\ref{Eq 101}) is reduced to equation (\ref{Eq 51}) \cite{Weinstein1}. 

Einstein uses equation (\ref{Eq 51}) to find the metric field of the Sun($g_{}$) to first-order approximation when considering a spherically symmetric, static gravitational field produced by the Sun \cite{CPAE4}, Doc 14:

\begin{equation} \label{Eq 52}
g_{44} = c_0^2 \left(1 - \frac{A}{r}\right).
\end{equation}

\noindent $A$ represents a correction factor in the above metric component $g_{44}$ that considers the gravitational effects in a spherically symmetric gravitational field of the Sun: 

\begin{equation} \label{Eq 310}
A = \left(1 + \frac{\alpha}{r} \right).    
\end{equation}

\noindent The term $\frac{\alpha}{r}$ in the metric represents the deviation from flatness due to the Sun's gravity. $\alpha$ in this term is represented by:

\begin{equation} \label{Eq 73}
\alpha = \frac{\kappa M}{4\pi} = \frac{2GM}{c^2}.
\end{equation}

\noindent Here, $\kappa$ is Einstein's gravitational constant, $G$ is the Newtonian gravitational constant, $M$ is the mass of the Sun, and $c$ is the speed of light. 

To account for more accurate results, Einstein calculates the second-order metric for the gravitational field of the Sun through successive approximations, leading to a more complex metric field:

\begin{equation} \label{Eq 53}
g_{44} = c_0^2 \left(1 - \frac{A}{r} + \frac{5}{8} \frac{A^2}{r^2}\right).
\end{equation}

The metric field of the static Sun, as obtained through the \emph{Entwurf} theory, including second-order corrections, determines the geometry of spacetime around the Sun. This metric field describes how the Sun's mass and gravity affect the curvature of spacetime in its vicinity.

So, Einstein substitutes the first-order metric [equation (\ref{Eq 52})] into the \emph{Entwurf} vacuum field equations (\ref{Eq 54}). By substituting the first-order metric into these field equations and performing calculations and approximations, Einstein derives the second-order metric [equation (\ref{Eq 53})]. This second-order metric more accurately describes the Sun's gravitational field in the \emph{Entwurf} theory.

As shown above, the first-order metric [equation (\ref{Eq 52}] is derived using the static field equations [equation (\ref{Eq 51})]. These static field equations describe a spherically symmetric gravitational field and are used to find an approximate solution for the metric component, $g_{44}$. This metric describes the gravitational field near the Sun. The second-order metric is then derived by applying the \emph{Entwurf} vacuum field equations, which provide a more refined description of gravity in the \emph{Entwurf} theory. 

Now, Besso takes over and handles the calculations. The motion of a planet (in this case, Mercury) within this gravitational field is described using equations of motion for a mass point, which include the following four equations \cite{CPAE4}, Doc 14:

\begin{equation} \label{Eq 55}
2\dot{f} = \dot{\varphi} r^2 = B \frac{ds}{dt}.
\end{equation}

\noindent Equation (\ref{Eq 55}) equates two different expressions $2\dot{f}$ and $\dot{\varphi} r^2$ that describe angular momentum. The constant $B$ is introduced on the right-hand side, specifically in the expression $\frac{ds}{dt}$ as a proportionality constant to relate the angular velocity $\dot \varphi$ and the squared radial coordinate $r^2$ on the left-hand side to the rate of change of $\frac{ds}{dt}$, i.e., the velocity component in the fourth coordinate direction (representing time).
\noindent This equation describes the relationship between the rate of change of the area swept out per unit time (\(\dot{f}\)) and \(\frac{ds}{dt}\). It represents the conservation of angular momentum in polar coordinates for a planet orbiting the Sun. The angular momentum is conserved as the planet moves in its orbit. \(\dot{f}\) represents the rate of change of the area swept out by the radius vector from the Sun to the planet per unit time. 

\begin{equation} \label{Eq 56}
E = g_{44} \frac{dt}{ds} \quad \text{and} \quad J_x = \frac{\dot{x}}{\frac{ds}{dt}}.
\end{equation}

\begin{equation}
xJ_x = \frac{x \dot{x}}{\frac{ds}{dt}}.
\end{equation}

\noindent These equations represent the expressions for energy ($E$), angular momentum in the $x$-direction ($J_x$), and the product of $x$ and $J_x$ in Einstein and Grossmann's \emph{Entwurf} theory of 1913 \cite{Einstein16}. 

Besso derived an equation representing angular momentum conservation, which he called the "area law." However, it does not directly represent the planet's motion under gravity's influence. This law differs from Kepler's second law because the rate of change of the area swept out by the radius vector from the Sun to Mercury ($\frac{ds}{dt}$) is not constant in the \emph{Entwurf} theory \cite{CPAE4}, Doc 14:

\begin{equation} \label{Eq 80}
\boxed{\frac{d}{dt} \left(\frac{x\dot{y} - y\dot{x}}{\frac{ds}{dt}}\right) = 0.}
\end{equation}

\noindent Besso defined an area constant ($B$) and the area speed ($\dot{f}$) \cite{CPAE4}, Doc 14:

\begin{equation} \label{Eq 65}
\dot{f} = x\dot{y} - y\dot{x} = B\frac{ds}{dt}.
\end{equation}

\noindent This equation represents the relationship between the rate of change of the area swept out per unit time ($\dot{f}$), the coordinates $x$, $y$, and their time derivatives $\dot{x}$, and $\dot{y}$. 

\noindent Now, we substitute $\left(\frac{x\dot{y} - y\dot{x}}{\frac{ds}{dt}}\right)$ with $B$ from equation \eqref{Eq 65}:

\begin{equation}
\frac{dB}{dt} = 0, \quad \text{or using equation \eqref{Eq 55}:} 
\end{equation}

\begin{equation} \label{Eq 300}
\boxed{\frac{d}{dt}\left(\frac{\dot{\varphi} r^2}{{\frac{ds}{dt}}}\right)=0.}  
\end{equation}

Besso's "area law" equation \eqref{Eq 80} is a key component of Besso's calculations of the precession of Mercury's perihelion. 

\noindent Besso then combined the first equation (\ref{Eq 56}), $E$, with equation (\ref{Eq 55}) and derived the resulting equation:

\begin{equation} \label{Eq 301}
E \dot f = Bg_{44}.    
\end{equation}

\noindent He then inserted equation (\ref{Eq 53}) into equation (\ref{Eq 301}). 
The equation obtained is:

\begin{equation} \label{Eq 60}
E(\dot \varphi {r}^2) = Bg_{44} = Bc^2_0 \left(1 - \frac{A}{r} + \frac{5}{8}\frac{A^2}{r^2}\right).    
\end{equation}

\noindent The left-hand side of the equation, \(E(\dot \varphi {r}^2)\), represents the energy of an orbiting planet (i.e., Mercury) in polar coordinates. It is the product of the angular velocity $\dot \varphi$, the square of the radial distance (\(r^2\)), and Mercury's energy (\(E\)).
On the right-hand side of the equation, \(Bc^2_0 \left(1 - \frac{A}{r} + \frac{5}{8}\frac{A^2}{r^2}\right)\), we have the constants, i.e., \(B\) and \(c^2_0\), as well as the parameter \(A\), which is defined by equation (\ref{Eq 310}). Recall that these constants and corrections account for the effects of gravity on the object's energy.
Equation (\ref{Eq 60}) relates (\(E\)) [equation (\ref{Eq 56})], (\(\varphi\)) [equation (\ref{Eq 55})], (\(r\)) and \(B\) [defined in equation (\ref{Eq 55})], \(g_{44}\), \(c^2_0\) and \(A\) [equation (\ref{Eq 310})].

Besso derived equation (\ref{Eq 60}), which represents the conservation of energy for Mercury moving in the Sun's gravitational field, where the geometry of spacetime is described by the metric field \(g_{44}\) according to the \emph{Entwurf} theory. 
Einstein and Besso wrestled with the last set of equations across several pages in the \emph{Einstein-Besso manuscript} (\cite{CPAE4}, Doc. 14), ultimately reaching a rather disheartening outcome.
When equation (\ref{Eq 60}) is integrated over the entire orbit (from perihelion to aphelion), it yields the accumulated change in the angle $\varphi$, which corresponds to the precession of Mercury's perihelion:

\begin{equation} \label{Eq 61}
\int_{r_1}^{r_2} d\varphi = \pi\left(1 + \frac{5}{8}\frac{A}{a(1-e^2)}\right).
\end{equation}

\noindent This equation describes how the angle $\varphi$ changes as the planet Mercury moves along its elliptical orbit. Specifically, $\int_{r_1}^{r_2} d\varphi$ represents the integral of the angle $\varphi$ as Mercury moves along its elliptical orbit from perihelion ($r_1$) and aphelion($r_2$). In other words, the angle swept out by the radius vector from the Sun to Mercury as the planet orbits the Sun between its perihelion and aphelion.

We want to calculate how much the angle swept out by the radius vector during one complete orbit of Mercury changes from one orbit to the next. This change in the angle:

\begin{equation} \label{eq 61}
\Delta \varphi = \int_{r_1}^{r_2} d\varphi,    
\end{equation}

\noindent gives us the total change in the angle over one orbit. To find the change per orbit in terms of the precession, we divide it by the number of orbits per revolution. 

\noindent Substituting the expression for $\Delta \varphi$ [Equation (\ref{Eq 61})] back into the equation (\ref{Eq 60}), we get:

\begin{equation} \label{Eq 62}
\pi\frac{5}{4}\frac{A}{a(1-e^2)}.
\end{equation}

\noindent This equation represents the precession of Mercury's perihelion, giving us the amount by which the perihelion of Mercury advances in its orbit per revolution.

Einstein and his close friend Besso, in 1913, grappled with the \emph{Entwurf} vacuum field equations to discern the perihelion advance of Mercury. This feat encountered miscalculations and subsequent corrections, resulting far from the observed discrepancy. According to the \emph{Entwurf} theory, the field of a static Sun produced an advance of Mercury's perihelion per revolution, equation (\ref{Eq 62}), corresponding to $18^{''}$.

\section{Rotation and Mach's principle} \label{3}

On pages 41-42 of the \emph{Einstein-Besso manuscript}, Einstein probed if his \emph{Entwurf} theory conflicted with the rotation concept \cite{CPAE4}, Doc. 14.
He performed a series of calculations to bridge his understanding of general relativity (as per the \emph{Entwurf} theory at this stage) with Mach's principle. He was trying to see if the effects of a rotating universe, as embodied by a rotating metric, could be derived from his \emph{Entwurf} field equations. The analysis concludes that, with some approximation methods, the rotating metric is a valid solution to the \emph{Entwurf} equations, giving some credence to the notions inspired by Mach's principle.
Intriguingly, the rotation challenges and the anomaly in Mercury's perihelion advance were interrelated. For diminutive angular velocities, a rotating system's metric field largely mirrors the Sun's metric field — the metric examined in the 1913 \emph{Einstein-Besso manuscript} to deduce Mercury's perihelion advance \cite{Weinstein1}, \cite{Weinstein2}.

Consider the line element given by \cite{CPAE4}, Doc. 14:
\begin{equation}
ds^2 = dx^2 + dy^2 + 2\omega y dx dt - 2\omega x dy dt - (1 - \omega^2 r^2) dt^2,
\end{equation}

\noindent where, $r^2 = x^2 + y^2$.
\vspace{1mm} 

\noindent This is the spacetime for a rotating system in terms of Minkowski spacetime. Here, $\omega$ represents the angular velocity of the rotation, and $r$ is a radial coordinate.

Einstein then writes down the covariant and contravariant components of the metric tensor for this rotating spacetime. Einstein wishes to determine if the rotating metric is a solution to the \emph{Entwurf} field equations. He uses an approximation method similar to the one he used for the perihelion advance of Mercury. By plugging in the first-order approximation of $g_{\mu \nu}$ into the \emph{Entwurf} field equations (\ref{Eq 1}), 
Einstein aims to see if this metric describes a vacuum spacetime. Recall from section \ref{1} that the operator $D_{\mu \nu}(g)$ is used in the \emph{Entwurf} field equations (\ref{Eq 1}). 
Einstein derived the following equation \cite{CPAE4}, Doc. 14:  

\begin{equation} \label{Eq 309}
D_{44}(g) = -\Delta g_{44} - 8 \omega^2.  
\end{equation}

\noindent The above equation describes the effect of the rotating metric on $D_{\mu \nu}(g)$. The term $\Delta_{44}$ is related to the Laplacian of the $g_{44}$ component, which describes how much it deviates from a flat spacetime value. 

Einstein made a calculation error regarding the terms proportional to $\omega^2$ and a sign error. 
specifically, Einstein considered $t_{44}$, the $44$-component of the stress-energy pseudo-tensor of the gravitational field. He believed the result for $\kappa t_{44}$ was \cite{CPAE4}, Doc. 14: 

\begin{equation} \label{Eq 2}
\kappa t_{44} = -4\omega^2.    
\end{equation}   

\noindent Equation (\ref{Eq 146}) [section \ref{1}] represents $t_{\mu \nu}$. For $\mu = \nu = 4$, this equation takes the form:

\begin{equation} \label{Eq 308}
\kappa t_{44} = \frac{1}{2}\sum_{\alpha\beta\tau\rho} \left( \frac{\partial g_{\tau\rho}}{\partial x_4} \frac{\partial \gamma_{\tau\rho}}{\partial x_4} - \frac{1}{2} g_{44} \gamma_{\alpha\beta} \frac{\partial g_{\tau\rho}}{\partial x_{\alpha}} \frac{\partial \gamma_{\tau\rho}}{\partial x_{\beta}} \right).
\end{equation}

\noindent The first term in equation (\ref{Eq 308}) does not contribute to second-order terms proportional to $\omega^2$ because it involves derivatives with respect to a spatial coordinate $x_4$. The second term yields:    

\begin{equation}
-\frac{1}{2}\left(\frac{1}{2} g_{44} \gamma_{22} \frac{\partial g_{14}}{\partial x_2} \frac{\partial \gamma_{14}}{\partial x_2}\right) \rightarrow -\frac{1}{4} (4\omega)^2.
\end{equation}

\noindent Hence, $t_{44}$, in the rotating system is proportional to $\omega^2$, where recall that $\omega$ is the angular velocity of the system. 

Plugging equations (\ref{Eq 309}) and (\ref{Eq 2}) into the \emph{Entwurf} field equation: 

\begin{equation} \label{Eq 5}
D_{44}(g) + \kappa t_{44}  = 0.    
\end{equation}

\noindent Einstein obtained \cite{CPAE4}, Doc. 14: 

\begin{equation}
D_{44}(g) + \kappa t_{44} = -\Delta g_{44} - 8\omega^2 + 4\omega^2 = 0.
\end{equation}

\noindent In other words:
\begin{equation} \label{Eq 311}
\Delta g_{44} = -4\omega^2.
\end{equation}

\noindent This equation indicates a sign error on Einstein's part. Einstein mistakenly omitted the minus sign in equation (\ref{Eq 2}) and wrote $+4\omega^2$ instead of $-4\omega^2$.
$\Delta g_{44}$ is a term that arises when the $D_{44}(g)$ operator is applied to the metric. So, from Einstein's perspective, the $-8\omega^2$ term comes from applying the $D_{44}(g)$ operator to the metric. 
$\kappa t_{44}$ contributes an additional $+4\omega^2$.
When we sum up the $\omega^2$ contributions, we get a net contribution of $-4\omega^2$. 

Einstein, still unaware of the mathematical error he had committed, then proposed a solution for $g_{44}$ in the form of a term proportional to the squared distance from the axis of rotation:

\begin{equation} \label{Eq 8}
C\omega^2(x^2 + y^2),    
\end{equation}

\noindent where $C$ is a constant. In two-dimensional polar coordinates $(r, \theta), r^2$ is given by $r^2 = (x^2 + y^2)$. Hence:

\begin{equation} \label{Eq 314}
\Delta g_{44} = C\omega^2 \Delta r^2.    
\end{equation}

\noindent $\Delta r^2$ in polar coordinates is given by: 

\begin{equation}
\Delta r^2 = \frac{1}{r} \frac{\partial}{\partial r} \left(r \frac{\partial (r^2)}{\partial r}\right) + \frac{1}{r^2} \frac{\partial^2 (r^2)}{\partial \theta^2}.
\end{equation}

\noindent Now, we calculate the derivatives:

\begin{equation}
\frac{\partial (r^2)}{\partial r} = 2r,
\end{equation}

\begin{equation}
\frac{\partial^2 (r^2)}{\partial \theta^2} = 0.
\end{equation}

\noindent We substitute these derivatives back into the expression:

\begin{equation} \label{Eq 312}
\Delta r^2 = \frac{1}{r} \frac{\partial}{\partial r} (r \cdot 2r) + \frac{1}{r^2} \cdot 0 = \frac{1}{r} \frac{\partial}{\partial r} (2r^2) = \frac{1}{r} \cdot 4r = 4.
\end{equation}

\noindent So, $\Delta r^2=4$, and equation (\ref{Eq 314}) becomes: 

\begin{equation} \label{Eq 313}
\Delta g_{44} = 4C\omega^2.    
\end{equation}

\noindent Equating equation (\ref{Eq 313}) to equation (\ref{Eq 311}), results in: $C=-1$. 
This ultimately led Einstein directly to the desired outcome he had been seeking \cite{CPAE4}, Doc. 14:

\begin{equation} \label{Eq 9}
g_{44} = 1 - \omega^2 (x^2 + y^2).    
\end{equation}

\noindent This equation represents the time-time component of the metric for the rotating system. Notably, this result matches what one would obtain through a direct transformation from a non-rotating to a rotating frame. 
Given the squared terms in $x$ and $y$, it shows the influence of rotation, particularly as one moves further out radially from the center of rotation \cite{Weinstein1}. 

\section{The \emph{Entwurf} Theory and the Newtonian limit} \label{5}

In 1913, Einstein provided a sketch illustrating how Poisson's equation and the Newtonian equation of motion can be derived in the weak-field approximation in the \emph{Entwurf} theory \cite{Einstein2}.

Einstein starts with the \emph{Entwurf} field equations, equations (\ref{Eq 54}). He considers a linear approximation of these field equations. In this approximation, the metric tensor \(g_{\mu\nu}\) deviates slightly from the flat spacetime metric \(\eta_{\mu\nu}\), and this deviation is represented by \(g_{\mu\nu}^*\):

\begin{equation} \label{Eq 95}
\boxed{g_{\mu\nu} = \eta_{\mu\nu} + g_{\mu\nu}^*.}    
\end{equation}

Einstein notes that in this weak-field approximation, the spatial part of the metric tensor ($\eta_{\mu\nu}$) remains approximately flat, representing special relativity.

Einstein makes several assumptions:
\vspace{1mm} 

1) He assumes that the gravitational field is static, i.e., it does not change with time. This assumption simplifies the time derivatives in the equations.

2) Einstein assumes that the perturbation in the metric (\(g_{\mu\nu}^*\)) goes to zero at infinity, i.e., the spacetime approaches a flat Minkowski spacetime. 

3) Einstein considers a pressureless and static cloud of dust characterized by the mass density \(\rho_0\). This dust distribution serves as the source of gravity. He identifies the time-time component of the stress-energy tensor \(T_{44}\) with \(\rho_0\). This identification is crucial for relating the gravitational potential to the mass distribution. 

4) In the weak gravitational field, Einstein neglects off-diagonal terms in the metric tensor (\(\mu \neq \nu\)) because they are assumed to be small. In the weak field limit, the diagonal components of the metric tensor are the dominant ones. These diagonal components are associated with the gravitational potential and have a direct physical interpretation. On the other hand, off-diagonal components describe smaller effects in a weak gravitational field. Neglecting off-diagonal terms and working with a diagonal metric is consistent with Newtonian physics in the weak field limit. 

5) Finally, Einstein assumes that all components of the metric tensor \(g_{\mu\nu}^*\) other than the time-time component (\(\mu = \nu = 4\)) are negligible (\(g_{\mu\nu}^* = 0\)). In other words, he is simplifying the metric to a diagonal form where only the time-time component is nonzero. By simplifying the metric, we greatly reduce the number of terms and equations in the field equations, and the dominant term is the gravitational potential $g_{\mu\nu}^*$, which is directly related to the Newtonian gravitational potential.
\vspace{1mm} 

These assumptions collectively simplify the problem and lead to the derivation of the Poisson equation for \(g_{44}^*\), which describes how the gravitational potential is related to the mass density $\rho_0$ in a static, weak gravitational field.

The \emph{Entwurf} field equations (\ref{Eq 54}) are simplified in the weak-field approximation and reduce to \cite{Einstein2}:

\begin{equation} \label{Eq 92}
\Box g_{\mu\nu}^* = \frac{\partial^2 g_{\mu\nu}^*}{\partial x^2} + \frac{\partial^2 g_{\mu\nu}^*}{\partial y^2} + \frac{\partial^2 g_{\mu\nu}^*}{\partial z^2} - \frac{1}{c^2} \frac{\partial^2 g_{\mu\nu}^*}{\partial t^2} = \kappa T_{\mu\nu}.
\end{equation}
\vspace{1mm} 

\noindent With Einstein's assumptions, the equation (\ref{Eq 92} becomes \cite{Einstein2}:

\begin{equation} \label{Eq 91}
\Box g_{\mu\nu}^* = - \frac{1}{c^2} \frac{\partial^2 g_{44}^*}{\partial t^2} = \kappa T_{44}.
\end{equation}

The term $\frac{1}{c^2} \frac{\partial^2} {g_{44}^*}$ in (\ref{Eq 91}) corresponds to the time-time component of the wave equation. The right-hand side of (\ref{Eq 91}) is $\kappa T_{44}$. This term is the $t-t$ component of the stress-energy tensor multiplied by the gravitational constant, $\kappa T_{\mu\nu}$.
\vspace{1mm} 

\noindent According to the aforementioned assumptions, Einstein identifies $T_{44}$ with the Poisson equation's mass density $\rho_0$, and assumes that the spacetime is asymptotically flat (i.e., that at spatial infinity, the metric perturbations $g_{\mu\nu}^*$ approach a Minkowski metric $\eta_{\mu\nu}$). In this limit, the wave operator $\Box$ reduces to a spatial Laplacian operator as the time derivatives become negligible compared to spatial derivatives. Hence, with these assumptions, $\Box g_{\mu\nu}^*$ becomes $\Delta g_{44}^*$ and the equation (\ref{Eq 91} simplifies to the Poisson equation for $g_{44}^*$ [see equation (\ref{Eq 51})] \cite{Einstein2}:

\begin{equation} \label{Eq 90}
\boxed{\Delta g_{44}^* = \kappa c^2 \rho_0.}    
\end{equation}

\noindent Equation (\ref{Eq 90}) represents Poisson's equation \(g_{44}^*\) in this weak-field limit. 
For components other than \(g_{44}^*\) (i.e., \(\mu, \nu\) not equal to $4$), the equation states \(\Delta g_{\mu\nu}^* = T_{\mu\nu} = 0\). This condition means that for metric tensor components other than the time-time component, the Laplacian of \(g_{\mu\nu}^*\) is zero, implying no gravitational potential associated with these components. Similarly, \(T_{\mu\nu} = 0\) means that the stress-energy tensor components (\(T_{\mu\nu}\)) associated with these spacetime points are zero. In other words, those points have no matter or energy density.

The solution to the Poisson equation represents the metric \(g_{\mu\nu}^*\) in this static, weak gravitational field.

In the weak-field approximation, and assuming non-relativistic velocities ($v \lll c$), the geodesic equation reduces to the equation \cite{Einstein2}:

\begin{equation} \label{Eq 93}
\boxed{\frac{d^2 x}{dt^2} = -\frac{1}{2} \frac{\partial g_{44}^*}{\partial x},}
\end{equation}

\noindent which relates the acceleration of a test particle's position to the spatial gradient of the gravitational potential \(g_{44}^*\). Equation (\ref{Eq 93}) represents the equation of motion for a particle moving in a spacetime described by the metric perturbation $g_{44}^*$.

The left-hand side represents the acceleration of a test particle's position vector $x$ for time $t$. This describes how the position of a particle changes under the influence of gravity. The right-hand side involves the derivative of the gravitational potential \(g_{44}^*\) with respect to the spatial coordinate $x$. This derivative characterizes the spatial gradient of the gravitational potential.

However, in formulating equation \eqref{Eq 93} in 1913, Einstein took a different approach as he had not yet written the geodesic equation, which he would only pen a year later. The geodesic equation describes the motion of particles under the influence of gravity. While equation (\ref{Eq 93}) captures the motion of a particle influenced by gravity, it is a simplified form in the context of Hamiltonian formalism. 
Einstein applied the principle of stationary action, which states that the variation of the spacetime interval $\int ds$ is set to zero \cite{Einstein2}:

\begin{equation} \label{Eq 96}
\delta \int ds = 0.
\end{equation}

\noindent He then expressed (in the weak-field approximation) the spacetime interval $\int ds$ in terms of the metric tensor $g_{\mu \nu}$ and the coordinate differentials $dx_\mu$:  

\begin{equation} \label{Eq 97}
\boxed{ds^2 = \sum_{\mu\nu} g_{\mu\nu} dx_\mu dx_\nu.}
\end{equation}

\noindent He considered the variation of the Hamiltonian $H$:

\begin{equation}  \label{Eq 98}
H = -\frac{ds}{dt},
\end{equation}

\noindent for time $t$, as indicated by:  

\begin{equation} \label{Eq 316}
\delta \int dH dt = 0.
\end{equation}

\noindent This variation is related to the equations of motion of particles in the weak gravitational field.

Combining the principles of stationary action [equation (\ref{Eq 96})], spacetime interval [equation (\ref{Eq 97})], and the variation of the Hamiltonian [equation (\ref{Eq 98})], Einstein derived equation (\ref{Eq 93}), equations of motion for particles moving in the weak gravitational field \cite{Einstein2}. 

If Einstein had possessed the geodesic equation in 1913, he could have derived equation (\ref{Eq 93}) in the following manner. 
Starting with Equation (\ref{Eq 90}), we can consider a particle moving in this field in a weak gravitational field. We are interested in the motion along a single spatial dimension, say $x$. The geodesic equation for the $x$-component of motion can be written as:

\begin{equation} \label{Eq 99}
\frac{d^2 x}{dt^2} + \Gamma^x_{44} = 0.
\end{equation}

Where: $\frac{d^2 x}{dt^2}$ is the acceleration of the particle in the $x$ direction; $\Gamma^x_{44}$ is the Christoffel symbol associated with the $x$-coordinate and the time coordinate $t$, and $\frac{dx}{dt}$ is the velocity of the particle in the $x$-direction.

Now, considering that $g_{44}^*$ represents the gravitational potential (specifically, the time-time component of the metric perturbation), and we are interested in motion in the $x$-direction, we can relate $\Gamma^x_{44}$ to $g_{44}^*$:

\begin{equation}
\Gamma^x_{44} = -\frac{1}{2}\frac{\partial g_{44}^*}{\partial x}.
\end{equation}

This is a connection between the Christoffel symbol and the metric perturbation. Substituting this into the geodesic equation (\ref{Eq 99}), we get equation (\ref{Eq 93}). 
The acceleration $\frac{d^2 x}{dt^2}$ of the particle is influenced by the gravitational potential $g_{44}^*$ and the square of its velocity $\left(\frac{dx}{dt}\right)^2$. 
Einstein derived this equation in his general relativity in 1916. See equation (\ref{Eq 102}) in Section \ref{6}.

The derivation of equation (\ref{Eq 93}) indicates the development stage of Einstein's gravitational theory during that period. 
Einstein's journey toward the development of general relativity involved a series of steps and refinements in his understanding of gravity and the mathematical framework to describe it. In the early stages, as seen in his 1913 \emph{Entwurf} theory, he was still working within the framework of special relativity and classical mechanics. He was heavily influenced by the equations of motion from Newtonian gravity. As he continued to explore and refine his theory, he departed from the special relativitsic and classical mechanics framework. He moved toward the more elegant and comprehensive framework of differential geometry, which ultimately led him to the geodesic equation and the final formulation of general relativity \cite{Weinstein1}. 

Einstein studied the linearized \emph{Entwurf} field equations in a weak-field limit. He was looking for solutions describing how a small perturbation in the metric $g_{\mu\nu}^*$ affects the motion of particles and light. The perturbation $g_{44}^*$ was related to the gravitational potential, and he was interested in how it is influenced by mass and energy. Einstein's equations in the weak-field approximation involve the wave operator $\Box$ [equation (\ref{Eq 92})] and describe the behavior of metric perturbations $g_{\mu\nu}^*$ [equation (\ref{Eq 95})] in a weak gravitational field. Given these equations, one might ponder whether Einstein had previously considered gravitational waves within the framework of his \emph{Entwurf} theory. While these equations share some mathematical similarities with the equations of gravitational waves in Einstein's general relativity, the appearance of the wave operator $\Box$ in [equation (\ref{Eq 92})] is a mathematical feature of the weak-field approximation in the \emph{Entwurf} theory. It represents how perturbations in the metric evolve in response to matter sources. 
However, at this point, Einstein did not explicitly interpret these equations as describing gravitational waves as he did later in 1916. In 1913, during his work on the \emph{Entwurf} theory, Einstein was primarily focused on developing a theory of gravity consistent with the principles of special relativity and Newtonian gravity. While exploring the behavior of metric perturbations and their propagation, he had not yet fully recognized the concept of gravitational waves as distinct propagating disturbances in spacetime. 

\section{Einstein falls down the hole and climbs back} \label{4}

In 1914, Einstein refined the 1913 \emph{Entwurf} theory, demonstrating the validity of its field equations, particularly in adapted coordinate systems, as detailed in reference \cite{Einstein4}.
Einstein commenced this refinement by considering the energy-momentum balance for matter in a gravitational field, as presented in reference \cite{Einstein4}:

\begin{equation} \label{Eq 49}
\boxed{\sum_\nu \frac{\partial \mathcal{T}_{\sigma\nu}}{\partial x_\nu} = \frac{1}{2} \sum_{\mu\nu\tau} \frac{\partial g_{\mu\nu}}{\partial x_\sigma} \gamma_{\mu\tau} \mathcal{T}_{\tau\nu}.}
\end{equation}

\noindent $\mathcal{T}_{\sigma\nu}$ represents the energy-momentum tensor of matter in the gravitational field: 

\begin{equation} 
\mathcal{T}_{\sigma\nu} =  \sum_{\mu} \sqrt{-g}g_{\sigma\mu}\Theta_{\mu\nu}.
\end{equation}

\noindent Equation (\ref{Eq 49}) describes how this tensor relates to the derivatives of the metric $\gamma$ and the components of $\mathcal{T}$.

Einstein derived a less general form of what he referred to as "the 'Entwurf' equations" \cite{Einstein4}:

\begin{equation} \label{Eq 131}
\sum_{\alpha\beta\mu}\frac{\partial}{\partial x_\alpha} \left(\sqrt{-g}\, \gamma_{\alpha\beta} g_{\sigma\mu} \frac{\partial g_{\mu\nu}}{\partial x_\beta}\right) = -\kappa (\mathcal{T}_{\sigma\nu} + t_{\sigma\nu}).
\end{equation}

\noindent $t_{\sigma\nu}$, the stress-energy pseudo-tensor of the gravitational field is represented by the equation:

\begin{equation} \label{Eq 150}
t_{\sigma\nu} =  \sum_{\mu} \sqrt{-g} g_{\sigma\mu} \theta_{\mu\nu}.    
\end{equation}

This new form represents the curvature of spacetime as a function of the metric coefficients $\gamma_{\alpha \beta}$, the metric tensor $g_{\mu\nu}$, and the tensors $\mathcal{T}_{\sigma\nu}$ and $t_{\sigma\nu}$. 

Einstein wrote the equation, which relates the stress-energy pseudo-tensor of the gravitational field $t_{\sigma\nu}$ to the derivatives of the metric and $\gamma$:

\begin{equation} \label{Eq 141}
-2\kappa t_{\sigma\nu} = \sqrt{-g}\left(\sum_{\beta\rho\tau} \gamma_{\beta\nu} \frac{\partial g_{\rho\tau}}{\partial x_\sigma} \frac{\partial \gamma_{\rho\tau}}{\partial x_\beta} - \frac{1}{2} \sum_{\alpha\beta\rho\tau} \delta_{\sigma\nu} \gamma_{\alpha\beta} \frac{\partial g_{\rho\tau}}{\partial x_\alpha} \frac{\partial \gamma_{\rho\tau}}{\partial x_\beta}\right).
\end{equation}

\noindent Equation (\ref{Eq 141}) represents the stress-energy pseudo-tensor $t_{\sigma\nu}$ [equation (\ref{Eq 150})]. Einstein expresses $t_{\sigma\nu}$ in two different ways and uses either equation depending on the context of his calculation. Substituting $t_{\sigma\nu}$ from equation (\ref{Eq 150}) into equation (\ref{Eq 141}) you get equation (\ref{Eq 150}) back.

\noindent In equation (\ref{Eq 141}), when $\sigma$ and $\nu$ are equal, the Kronecker delta  $\delta_{\sigma\nu}$ is equal to $1$, and when they are not equal, $\delta_{\sigma\nu}$ equals $0$. $\delta_{\sigma\nu}$ is used in the second summation term. When $\sigma = \nu$, then $\delta_{\sigma\nu}=1$, and the term simplifies to: 

\begin{equation}
- \frac{1}{2} \sum_{\alpha\beta\rho\tau} 1  \cdot \gamma_{\alpha\beta} \frac{\partial g_{\rho\tau}}{\partial x_\alpha} \frac{\partial \gamma_{\rho\tau}}{\partial x_\beta}.    
\end{equation}

When $\sigma$ and $\nu$ are equal, $\delta_{\sigma\nu}$ becomes $0$ and the term disappears from the sum. So, the Kronecker delta $\delta_{\sigma\nu}$ acts as a switch that selects certain terms in the equation based on whether $\sigma$ and $\nu$ are equal or not. When they are equal, it selects the term; when they are not equal, it sets the term to zero. 

Recall that equation (\ref{Eq 145}) and equation (\ref{Eq 146}) represent expressions for $2\kappa \theta_{\mu\nu}$ and $2\kappa t_{\mu\nu}$ respectively, in terms of various derivatives of the metric coefficients $\gamma_{\alpha\beta}$ and the metric tensor $g_{\mu\nu}$.

In equation \ref{Eq 146}, the stress-energy tensor component is calculated using the metric tensor $g_{\mu \nu}$. In equation \eqref{Eq 141}, we have the Kronecker delta $\delta_{\sigma\nu}$ and the square root of the determinant of the metric tensor $\sqrt{-g}$ appearing instead of the metric tensor $g_{\mu \nu}$ that is present in the expression (\ref{Eq 146}). In equation \ref{Eq 141}, we replaced the metric tensor $g_{\mu \nu}$ with the Kronecker delta $\delta_{\sigma \nu}$.
The difference between equation \eqref{Eq 141} and equation \eqref{Eq 146} is related to the properties of the metric tensor and the determinant of the metric tensor.
These properties play a crucial role as Einstein defined them in 1914:\cite{Einstein11}:

\begin{equation} \label{Eq 23-1}
\sum_{\sigma} g_{\mu\sigma} g^{\nu\sigma} = \delta_{\mu}^{\nu},
\end{equation}

\noindent This equation expresses the orthogonality of the metric tensor and its inverse. It tells us that the contraction of the indices of the metric tensor with its inverse gives the Kronecker delta $\delta_{\mu}^{\nu}$, which is a diagonal matrix with elements equal to $1$ on the diagonal and $0$ off the diagonal:

\begin{equation}
\delta_{\mu}^{\nu} = \begin{cases} 
1 & \text{if } \mu = \nu, \\
0 & \text{if } \mu \neq \nu,
\end{cases}
\end{equation}

\noindent and, according to the multiplication theorem for determinants:

\begin{equation}
\left| \sum_{\sigma} g_{\mu\sigma} g^{\nu\sigma} \right| = |\delta_{\mu}^{\nu}| = 1.
\end{equation}

\noindent From this, it follows:

\begin{equation} \label{Eq 100}
|g_{\mu\nu}| \cdot |g^{\mu\nu}| = 1.
\end{equation}

\noindent This equation states that the product of the determinant of the metric tensor and the determinant of its inverse is equal to $1$.

The difference between equation \eqref{Eq 141} and equation \eqref{Eq 146} arises from the use of the Kronecker delta $\delta_{\sigma\nu}$ and the square root of the determinant of the metric tensor $\sqrt{-g}$ in equation \eqref{Eq 141}, as opposed to the metric tensor $g_{\mu \nu}$ used in equation \eqref{Eq 146}. This difference is related to equations \eqref{Eq 23-1} and \eqref{Eq 100}, 

Einstein states that the "'Entwurf' equations," referring to the \emph{Entwurf} equations for the stress-energy pseudo-tensor, namely equation (\ref{Eq 145}) $2\kappa\theta_{\mu\nu}$ and equation (\ref{Eq 146}) $2\kappa t_{\mu\nu}$, lead to the derivation of equation (\ref{Eq 141}) \cite{Einstein4}. His objective is to establish that his \emph{Entwurf} field equations maintain the conservation of energy-momentum in adapted coordinate systems.

Einstein proceeds by indicating that he conducted a differentiation with respect to $x_\nu$ (he took the partial derivative for $x_\nu$) on both sides of the equation (\ref{Eq 131}). Concurrently, he considered the following expression:

\begin{equation} \label{Eq 4}
\boxed{\sum_{\nu} \frac{\partial}{\partial x_\nu} (\mathcal{T}_{\sigma\nu} + t_{\sigma\nu}) = 0.}
\end{equation}

\noindent This equation represents the conservation of energy-momentum in the \emph{Entwurf} theory, which states that the covariant divergence of the stress-energy-momentum tensor is zero.
In simple terms, equation (\ref{Eq 4}) tells us how energy-momentum changes in response to the curvature of spacetime. 
In a generally covariant theory, the vanishing divergence of the energy-momentum tensor is a direct mathematical consequence of the field equations rather than an additional requirement. 
Yet, in his attempts to build a bridge between the conservation laws and his gravitational field equations, Einstein found apparent inconsistencies when his field equations were not generally covariant. This led him to look for compensating terms to ensure conservation, affecting the mathematical form of his proposed \emph{Entwurf} field equations.

Following the differentiation of equation (\ref{Eq 131}) with respect to $x_\nu$ and taking into consideration equation (\ref{Eq 4}), Einstein derived the following relations \cite{Einstein4}:

\begin{equation} \label{Eq 140}
\boxed{B_\sigma=\sum_{\alpha\beta\mu\nu}\frac{\partial^{2}}{\partial x_\nu x_\alpha} \left(\sqrt{-g}\, \gamma_{\alpha\beta} g_{\sigma\mu} \frac{\partial g_{\mu\nu}}{\partial x_\beta}\right) = 0.}
\end{equation}

\noindent $B_\sigma$ is a set of four differential conditions for the metric tensor $g_\mu \nu$. The $B_\sigma$ differential conditions are nothing but coordinate restrictions because they specify certain relationships that must hold between the second partial derivatives of the metric components for the spacetime coordinates $x_\alpha$ and $x_\nu$. These restrictions impose additional constraints on the metric tensor $g_{\mu \nu}$ and ensure that the \emph{Entwurf} theory's solutions conform to these restrictions. $B_\sigma$ means that equation (\ref{Eq 131}) is modified and adapted to satisfy the coordinate restrictions $B_\sigma$ \cite{Weinstein1}. 
\vspace{1mm} 

In 1913-1914, Einstein also presented several versions of the hole argument against general covariance, implying that generally covariant field equations were impermissible \cite{Einstein11}. 

Suppose $g_{\mu \nu}$ represents the metric tensor in the coordinate system $K$, and $g'_{\mu \nu}$ represents it in another coordinate system $K'$.
Einstein imagined a certain region of spacetime (a hole) where we do not know the exact values of the metric tensor. Outside this hole, $g_{\mu \nu}$ [the field $G(x)$] and $g'_{\mu \nu}$ [the field $G'(x')$] describe the same gravitational field, i.e., they are just different coordinate representations of the same physical reality. Inside the hole, however, they might differ.
Einstein's clever insight or maneuver trick was the following: he proposed swapping the coordinates for the values inside the hole. He took the expressions for the gravitational field in $K'$ (described by $g'_{\mu \nu}$) but evaluated them at the coordinate points of $K$. So, even though it is the field from $K'$, it is being described in the coordinates of $K$.
Mathematically, this involves replacing $x'$ (coordinates of $K'$) with $x$ (coordinates of $K$) for the values of $g'_{\mu \nu}$ inside the hole. This results in a new field, which we can denote as $G'(x)$. The maneuver implies that there can be two different solutions [$G(x)$ and $G'(x)$] for the gravitational field outside the hole, even though they are based on the same coordinate $x$ in the region of the hole.
This raised a concern: if general covariance were true, and the field equations were generally covariant, then there would not be a unique solution for the metric inside the hole. 

The "hole" restricts the usable coordinate systems to only adapted ones.
These adapted systems justify the restricted covariance of Einstein's \emph{Entwurf} equations and explain why fully generally covariant equations could be problematic: they allow too much freedom and ambiguity in their solutions.
By 1914, Einstein demonstrated the compatibility of his hole argument with the constraints of adapted coordinate systems. This framework justified the constrained covariance of the field equations and Einstein's \emph{Entwurf} gravitational tensor [equation (\ref{Eq 64})] and provided a rationale for generally rejecting covariant field equations \cite{Einstein11}, \cite{Weinstein1}, \cite{Weinstein2}.
The hole argument was a challenge that Einstein posed against the concept of general covariance in his developing \emph{Entwurf} theory. It demonstrated, to Einstein's thinking at the time, that if the field equations were generally covariant, then they could admit multiple, distinct solutions (different metric tensors) that represented the same physical situation inside of a given region (the hole), but different situations outside that region.  
One could posit that Einstein's continued defense of the \emph{Entwurf} theory had shades of dogmatism. But not quite. While Einstein was deeply committed to testing his theories, trying to salvage the \emph{Entwurf} theory, he leaned on the crutches of adapted coordinate systems and the hole argument.
\vspace{1mm} 

In 1914, the hole argument provided Einstein with a refuge against the constraints of general covariance. It allowed him to delve into the intricacies of the \emph{Entwurf} field equations. This pursuit continued until November 1915, when a sudden realization prompted him to reconsider his approach, acknowledging the limitations of his previous path. While ensnared by the clutches of the \emph{Entwurf} theory, Einstein authored a comprehensive review article dedicated to this theory. In this article, Einstein presented the action, which is a functional integral, as represented by \cite{Einstein11}:

\begin{equation} \label{12}
\boxed{J = \int L \sqrt{-g} \, d\tau.}     
\end{equation}

\noindent The integral includes the Lagrangian density of the gravitational field, $L$.

Einstein wrote equations for the stress-energy pseudo-tensor of the gravitational field, which necessitated a coordinate restriction $S_{\sigma \nu}=0$. He proposed a specific Lagrangian based on derivatives of the metric tensor:

\begin{equation} \label{Eq 10}
\boxed{L = -\sum_{\mu\rho\tau\nu} g^{\tau\nu} \Gamma_{\mu\tau}^{\rho} \Gamma_{\rho\nu}^{\mu}.}  
\end{equation}  

This Lagrangian is not represented in terms of the Christoffel symbols. 
$\Gamma_{\nu\sigma}^{\tau}$ in equation (\ref{Eq 10}) represents the components of the gravitational field, which are defined as the gradient of the metric tensor:

\begin{equation} \label{Eq 180}
\boxed{\Gamma_{\nu\sigma}^{\tau} = \frac{1}{2}\sum_\mu g^{\tau\mu}\frac{\partial {g _{\mu \nu}}}{\partial x_\sigma}.}
\end{equation} 

In the \emph{Entwurf} theory, Einstein initially expressed the components of the gravitational field in terms of the gradient of the metric tensor, as described in equation (\ref{Eq 180}). It is based on just one term of the three that comprise the full Christoffel symbols, indicating the simplification Einstein employed at that stage. In 1912, in the \emph{Zurich notebook}, Einstein had already written the Riemann-Christoffel tensor using the Christoffel symbols but gave it up in favor of the \emph{Entwurf} theory, see section \ref{1}. He needed to go through the \emph{Entwurf} theory to understand the full value of the Christoffel symbols in his theory.
In 1915, Einstein used the Christoffel symbols to represent the connection coefficients. The Christoffel symbols define the covariant derivative, geodesic equations, and the Riemann curvature tensor. This reflects how Einstein's understanding and the mathematical framework he used evolved. In the \emph{Entwurf} theory, Einstein formulated a theory of gravitation without resorting to the full machinery of Riemannian geometry, which is why there were differences in the connection used. However, as Einstein continued his work in November 1915, he moved towards using the full Christoffel symbols and the Riemann curvature tensor in his final general relativity formulation, recognizing Riemannian geometry's significance in describing curved spacetime (see discussion in section \ref{5}).

Given its specific form, Einstein believed that the Lagrangian equation (\ref{Eq 10}) would naturally lead to his \emph{Entwurf} field equations under adapted coordinate systems. The fact is that Einstein assumed that this Lagrangian was the only one leading to the \emph{Entwurf} field equations satisfying the coordinate restriction $S_{\sigma \nu}=0$. He thought that deriving the field equations through the variational approach reinforced the validity of his \emph{Entwurf} gravitational tensor. However, it was eventually discovered that this assumption was incorrect. 

In the 1914 \emph{Entwurf} theory, Einstein employed tensor densities. These mathematical objects behave similarly to tensors under coordinate transformations but include an additional scaling factor associated with the volume element, represented by the determinant of the metric tensor, denoted as $g$. This extra factor allows them to account for changes in volume. In essence, tensor densities do not transform in the same manner as tensors; rather, they incorporate an extra factor tied to the volume element of the manifold.

Einstein began with the Lagrangian, denoted as L, which depends on the contravariant fundamental metric tensor $g^{\mu\nu}$ and its first derivatives, $g_{\sigma}^{\mu\nu}$. The expression ${g_\sigma^{\mu\nu} \equiv \frac{\partial g^{\mu\nu}}{\partial x_\sigma}}$ represents the components of the metric tensor's derivative with respect to the coordinate $x_\sigma$.    
Einstein considered a scenario where the gravitational field associated with $g^{\mu\nu}$ is varied by an infinitesimal amount, resulting in the replacement of the metric tensor components $g^{\mu\nu}$ by $g^{\mu\nu} + \delta g^{\mu\nu}$. Importantly, the perturbation $\delta g^{\mu\nu}$ vanishes at the boundaries of an infinitesimal local region in spacetime. In this context, the Lagrangian $L$ transforms to $L + \delta L$, and the action $J$ transforms to $J + \delta J$.
The variational principle, as described in \cite{Einstein11}, is expressed as follows:

\begin{equation} \label{Eq 195}
\delta J = \int d\tau \sum_{\mu \nu} \delta g^{\mu \nu} \mathcal{G}_{\mu \nu},
\end{equation}

\noindent In this equation, $\delta J$ represents the variation of the action associated with the metric tensor $g^{\mu\nu}$.

Einstein recognized that the integrand, when divided by the square root of the determinant of the metric tensor, denoted as $\sqrt{-g}$, remains invariant. This insight led to the formulation of the invariant expression:

\begin{equation} \label{Eq 196}
\frac{1}{\sqrt{-g}} \sum_{\mu \nu} \delta g^{\mu \nu} \mathcal{G}_{\mu \nu}.
\end{equation}

This invariant form accounts for the transformation properties of the metric tensor and its associated variations.

\noindent Einstein derived the following equation:

\begin{equation} \label{Eq 197}
\delta J = \int d\tau \sum_{\mu\nu} \delta g^{\mu\nu} \left( \frac{\partial L\sqrt{-g}}{\partial g^{\mu\nu}} - \sum_{\sigma} \frac{\partial}{\partial x_{\sigma}} \left( \frac{\partial L\sqrt{-g}}{\partial g_{\sigma}^{\mu\nu}} \right) \right),    
\end{equation}

\noindent which represents the variation of the action for the metric tensor components $\delta g^{\mu \nu}$, corresponding to the Euler-Lagrange equations for the \emph{Entwurf} Lagrangian density $L$. These equations describe how the metric tensor components $g^{\nu \nu}$ must behave to extremize the action. 

Einstein defined the term in the brackets of equation (\ref{Eq 197}) as the \emph{Entwurf} gravitational tensor $\mathcal{G}_{\mu \nu}$ \cite{Einstein11}:

\begin{equation} \label{Eq 64}
\boxed{\mathcal{G_{\mu\nu}} = \frac{\partial L\sqrt{-g}}{\partial g^{\mu\nu}} - \sum_{\sigma} \frac{\partial}{\partial x_{\sigma}} \left( \frac{\partial L\sqrt{-g}}{\partial g_{\sigma}^{\mu\nu}} \right).}  
\end{equation}

\noindent This equation defines $\mathcal{G_{\mu\nu}}$, which characterizes in the \emph{Entwurf} theory the curvature of spacetime due to the distribution of mass and energy. $\mathcal{G_{\mu\nu}}$ is defined in terms of the variational derivatives of the action, equation (\ref{12}) concerning the metric tensor $g_{\mu \nu}$ and its first derivatives. $\sqrt{-g} L$ is the Lagrangian density and the square root of the determinant of the metric tensor. The derivative $\frac{\partial (\sqrt{-g} L)}{\partial g^{\mu\nu}}$ is the partial derivative of the Lagrangian density $\sqrt{-g} L$ with respect to the components of the metric tensor. The term $\frac{\partial (\sqrt{-g} L)}{\partial g_{\sigma}^{\mu\nu}}$ represents a partial derivative of $\sqrt{-g} L$ with respect to $g_{\sigma}^{\mu\nu}$, where recall that $g_{\sigma}^{\mu\nu}= \frac{\partial{g^{\mu \nu}}}{\partial{x^\sigma}}$ (the metric tensor's derivative with respect to the coordinate $x_\sigma$).

However, when we take variational derivatives with respect to $g_{\mu \nu}$ and its derivatives, the resulting expression can have the property of a standard tensor rather than a tensor density. Given $\mathcal{G_{\mu\nu}}$ is derived from $\sqrt{-g} L$, but due to the variational differentiation, it is a tensor rather than a tensor density. The factors of $\sqrt{-g}$ in the derivative terms are canceled out in forming the field equations, producing an object with tensorial transformation properties.

Einstein imposed the divergence-free condition to restrict the covariance of the \emph{Entwurf} field equations. He stressed that the \emph{Entwurf} gravitational tensor [equation (\ref{Eq 64})] is covariant only in adapted coordinate systems \cite{Einstein11} [see equation (\ref{Eq 140})]:
\vspace{1mm} 

\begin{equation} \label{Eq 115}
\boxed {B_\sigma = \sum_{\alpha \sigma \nu}\frac{\partial^2}{\partial x_{\sigma} \partial x_{\alpha}} \left( g^{\nu\alpha} \frac{\partial L \sqrt{-g}}{\partial g_{\sigma}^{\mu\nu}} \right) = 0.}    
\end{equation}
\vspace{1mm} 

\noindent Alternatively, Einstein wrote equation (\ref{Eq 115}) as follows:
\vspace{1mm} 

\begin{equation} \label{Eq 12}
\boxed{\sum_{\alpha \beta} \frac{\partial^2}{\partial x^\alpha \partial x^\alpha} \left( \sqrt{-g} g^{\alpha\beta} \Gamma^\nu_{\beta\sigma} \right) = 0.}    
\end{equation}
\vspace{1mm} 

Einstein's \emph{Entwurf} field equations prior to imposing the restriction $B_\sigma$ are listed below \cite{Einstein11}:  

\begin{equation} \label{Eq 186}
\boxed{\mathcal{G_{\mu\nu}} = -\kappa \mathcal{T}_{\sigma}^{\nu}.}  
\end{equation} 

\noindent Equations (\ref{Eq 186}) involve the correlation between Einstein's \emph{Entwurf} gravitational tensor density equation (\ref{Eq 64}) $\mathcal{G_{\mu\nu}}$ and the stress-energy tensor density $\mathcal{T}_{\sigma}^{\nu}$.

\noindent We can fully contract (i.e., summing over repeated indices) the \emph{Entwurf} field equations (\ref{Eq 186}): 

\begin{equation} \label{Eq 160}
\sum_{\tau \alpha} \left( -g^{\nu \tau} \frac{\partial L\sqrt{-g}}{\partial g^{\sigma \tau}} - g_{\alpha}^{\nu \tau} \frac{\partial L\sqrt{-g}}{\partial g_{\alpha}^{\sigma \tau}} \right) +\sum_{\alpha\tau} \frac{\partial}{\partial x_{\alpha}} \left( g^{\nu \tau} \frac{\partial L\sqrt{-g}}{\partial g_{\alpha}^{\sigma \tau}} \right) = -\kappa \mathcal{T}_{\sigma}^{\nu}.
\end{equation}

\noindent Equation (\ref{Eq 160}) is derived from the variation of the action with respect to the metric components $g_{\alpha}^{\sigma \tau}$ and the second term involves first derivatives with respect to the coordinates $x_\sigma$.

We can take the divergence of both sides of the equation (\ref{Eq 160}). To do this, we apply the divergence operator $\frac{\partial}{\partial x_{\alpha}}$ to each term on both sides of the equation. We, therefore, impose the condition $B_\sigma$ [equation (\ref{Eq 140})]. This condition simplifies the left-hand side of the equation, making it equal to zero. This simplification is represented as:

\begin{equation}
\sum_{\nu} \frac{\partial}{\partial x_{\alpha}} \left(g^{\nu \tau}\frac{\partial \sqrt{-g}}{\partial g^{\sigma \tau}_{\alpha}}\right) = 0.     
\end{equation}

\noindent The divergence of the term on the left-hand side of the equation (\ref{Eq 160}) is set to zero under the condition $B_\sigma$. Equation (\ref{Eq 160}) yields the divergence of this term:

\begin{equation} 
\sum_{\tau \alpha} \left( -g^{\nu \tau} \frac{\partial L\sqrt{-g}}{\partial g^{\sigma \tau}} - g_{\alpha}^{\nu \tau} \frac{\partial L\sqrt{-g}}{\partial g_{\alpha}^{\sigma \tau}} \right) + \kappa \mathcal{T}_{\sigma}^{\nu} = 0.
\end{equation}

\noindent We then obtain the equation:

\begin{equation} \label{Eq 118}
\boxed{\sum_{\nu} \frac{\partial}{\partial x_{\alpha}} \left( T_{\sigma}^{\nu} + \frac{1}{k} \sum_{\tau\alpha} \left( -g^{\nu\tau} \frac{\partial L\sqrt{-g}}{\partial g^{\sigma\tau}} - g_{\alpha}^{\nu\tau} \frac{\partial L\sqrt{-g}}{\partial g_{\alpha}^{\sigma\tau}} \right) \right) = 0.}
\end{equation}

We can rewrite equation (\ref{Eq 118}) by introducing $t^\nu_\sigma$:

\begin{equation} \label{Eq 117}
\boxed{-t_{\alpha}^{\tau} = \frac{1}{\kappa} \sum_{\tau \alpha} \left(g^{\nu \tau} \frac{\partial L\sqrt{-g}}{\partial g^{\sigma \tau}} + g_{\alpha}^{\nu \tau} \frac{\partial L\sqrt{-g}}{\partial g_{\alpha}^{\sigma \tau}} \right).}    
\end{equation}

Einstein replaced equations (\ref{Eq 160}) by the following equation \cite{Einstein11}: 
\vspace{1mm} 

\begin{equation} \label{Eq 11}
\boxed{\sum_{\alpha \beta} \frac{\partial}{\partial x^\alpha} \left( \sqrt{-g} g^{\alpha\beta} \Gamma^\nu_{\beta\sigma} \right) = -\kappa (T^\nu_\sigma + t^\nu_\sigma),}    
\end{equation}
\vspace{1mm} 

\noindent where, $T^\nu_\sigma$ is the stress-energy tensor density and $t^\nu_\sigma$ is the stress-energy pseudo-tensor density of the gravitational field \cite{Einstein11}:
\vspace{1mm} 

\begin{equation}  \label{Eq 15}
t_{\sigma}^{\nu} = \frac{\sqrt{-g}}{\kappa} \sum_{\mu\rho\tau\tau'} \left( g^{\nu\tau} \Gamma_{\mu\sigma}^{\rho} \Gamma_{\rho\tau}^{\mu} - \frac{1}{2} \delta_{\sigma}^{\nu} g^{\tau\tau'} \Gamma_{\mu\tau}^{\rho}, \Gamma_{\rho\tau'}^{\mu} \right).
\end{equation} 
\vspace{1mm} 

We can rewrite equation (\ref{Eq 160}) in the form of equation (\ref{Eq 11}). To rewrite equation (\ref{Eq 160}) in the form of equation (\ref{Eq 11}), we introduce $t^\nu_\sigma$, rearrange terms and combine them, and finally obtain equation (\ref{Eq 11}).

Einstein wrote the covariant formulation of the energy-momentum balance for matter in a gravitational field \cite{Einstein11}: 

\begin{equation} \label{Eq 191}
\boxed{\sum_{\nu} \frac{\partial \mathcal{T}_{\tau}^{\nu}}{\partial x_{\nu}} = \frac{1}{2} \sum_{\mu\tau\nu} g^{\tau\mu} \frac{\partial g_{\mu\nu}}{\partial x_{\sigma}} \mathcal{T}_{\tau}^{\nu}.}
\end{equation}

\noindent The stress-energy tensor of continuously distributed masses neglecting the effects of surface forces $\mathcal{T}_{\tau}^{\nu}$ is represented by:

\begin{equation} \label{Eq 211}
\mathcal{T}_{\tau}^{\nu} = \rho_0 \sqrt{-g} \frac{dx_{\nu}}{ds} \sum_{\mu} g_{\tau\mu} \frac{dx_{\mu}}{ds}.
\end{equation}

When surface forces are ignored, $\mathcal{T}_{\tau}^{\nu}$ describes the bulk properties of the continuous mass distribution, including its energy density, momentum density, and stress. This simplification is used to focus on $\mathcal{T}_{\tau}^{\nu}$ without considering the details of how it interacts with external boundaries and surfaces.

Equation (\ref{Eq 191}) states that the divergence of the stress-energy tensor $\mathcal{T}^{\nu}_{\tau}$ with respect to the spacetime coordinates $x_\nu$ vanishes. On the right-hand side, we have a term involving the metric tensor $g^{\tau\mu}$ and its partial derivatives for spacetime coordinates $x_\sigma$, as well as the stress-energy tensor $\mathcal{T}^{\nu}_{\tau}$.
Compare to equation (\ref{Eq 49}).
Without a gravitational field ($g^{\mu\nu}$ = const.), the term on the right-hand side of the equation (\ref{Eq 191}) simplifies because the metric tensor and its derivatives do not vary for spacetime coordinates. This simplification results in the equation resembling the energy-momentum conservation law in special relativity. 

The choice made by Einstein to include the gradient of the metric tensor $\frac{\partial g_{\mu\nu}}{\partial x_{\sigma}}$ on the right-hand side of equation (\ref{Eq 191}) was a key insight in the development of Einstein's theory of gravitation. It led him to choose the metric tensor gradient [$\Gamma_{\nu\sigma}^{\tau}$ equation (\ref{Eq 180})] to represent the gravitational field's components in his \emph{Entwurf} theory.
Einstein had already formulated equations (\ref{Eq 191}) and (\ref{Eq 211}) on page 5R of his \emph{Zurich Notebook} \cite{Einstein7}.
Einstein stated that the \emph{Entwurf} gravitational tensor $\mathcal{G}_{\mu \nu}$ [equation (\ref{Eq 64})] follows the same equation form as equation (\ref{Eq 191}) \cite{Einstein11}:

\begin{equation}\label{Eq 183}
\sum_{\nu \tau} \frac{\partial g^{\tau \nu}}{\partial x_{\nu}} \mathcal{G}_{\sigma \tau} + \frac{1}{2} \sum_{\mu \nu} \frac{\partial g^{\mu \nu}}{\partial x_{\sigma}} \mathcal{G}_{\mu \nu} = 0,
\end{equation}

\noindent The second term on the left-hand side of the equation (\ref{Eq 183}) involves the components of the gravitational field $\Gamma^\tau_{\nu\sigma}$, which are defined as the gradient of the metric tensor according to equation (\ref{Eq 180}). This term involves the derivatives of the metric tensor $g^{\mu \nu}$ with respect to spacetime coordinates and the \emph{Entwurf} gravitational tensor $\mathcal{G}$.
\vspace{1mm} 

With the condition (\ref{Eq 115}), Einstein rewrote equation (\ref{Eq 64}) as a new coordinate condition $C_\sigma=0$ intended to restrict the choice of coordinates to those that satisfy the condition \cite{Einstein11}:

\begin{equation} \label{Eq 120}
C_\sigma = B_\sigma - \sum_{\nu} \frac{\partial S_{\sigma}^{\nu}}{\partial x_ \nu} = 0.    
\end{equation}

\noindent $C_\sigma=0$ is a coordinate restriction rather than a consequence of the \emph{Entwurf} field equations. Einstein believed that when this condition was satisfied, the form of the theory was simplified and made valid for adapted coordinate systems \cite{Einstein11}. Using equations (\ref{12}), (\ref{Eq 195}), and (\ref{Eq 64}) Einstein derived $S_{\sigma}^{\nu}$: 

\begin{equation} \label{Eq 322}
S_{\sigma}^{\nu} = \sum_{\mu\tau } \left(g^{\nu \tau} \frac{\partial L\sqrt{-g}}{\partial g^{\sigma \tau}} + g_{\mu}^{\nu \tau} \frac{\partial L\sqrt{-g}}{\partial g_{\mu}^{\sigma \tau}} +\frac{1}{2}\delta_{\sigma}^{\nu} L\sqrt{-g} - \frac{1}{2}g_{\sigma}^{\mu \tau} \frac{\partial L\sqrt{-g}}{\partial g_{\nu}^{\mu \tau}}\right).    
\end{equation}

\noindent Now, we can define:

\begin{equation} \label{Eq 360}
S_{\sigma}^{\nu} = \sum_{\mu\tau} \left(\boxed{g^{\nu \tau} \frac{\partial L\sqrt{-g}}{\partial g^{\sigma \tau}} + g_{\mu}^{\nu \tau} \frac{\partial L\sqrt{-g}}{\partial g_{\mu}^{\sigma \tau}}} +\boxed{\frac{1}{2}\delta_{\sigma}^{\nu} L\sqrt{-g} - \frac{1}{2}g_{\sigma}^{\mu \tau} \frac{\partial L\sqrt{-g}}{\partial g_{\nu}^{\mu \tau}}}\right).    
\end{equation}

Einstein emphasized that satisfying his \emph{Entwurf} field equations \eqref{Eq 186} necessitates fulfilling \emph{two} conditions (or restrictions) $B_\sigma = 0$ and:

\begin{equation}  \label{Eq 23}
S_{\sigma}^{\nu} \equiv 0.    
\end{equation}

\noindent If $S_{\sigma}^{\nu}=0$, then the equation (\ref{Eq 120}) simplifies to:

\begin{equation} 
B_\sigma - \sum_{\nu} \frac{\partial 0}{\partial x_ \nu} = 0.    
\end{equation}

\noindent Since the derivative of zero with respect to any variable is zero, the equation becomes:

\begin{equation} 
B_\sigma - \sum_{\nu} 0 = B_\sigma - 0 = B_\sigma  =0.    
\end{equation}

\noindent So, if $S_{\sigma}^{\nu}=0$, it implies that $B_\sigma$ must also be zero for the equation to be satisfied.
\vspace{1mm} 
 
Einstein divided equation \eqref{Eq 360} into two parts:
 
1) The first and second terms on the right-hand side of the equation (\ref{Eq 322}) represent a stress-energy pseudo tensor $t_{\alpha}^{\tau}$, \eqref{Eq 117}.

2) The third and fourth terms on the right-hand side of the equation (\ref{Eq 322}) represent a stress-energy pseudo tensor $t_{\sigma}^{\nu}$:

\begin{equation}\label{Eq 184}
\boxed{-t_{\sigma}^{\nu} = \frac{1}{2\kappa} \left( \delta_{\sigma}^{\nu} L\sqrt{-g} - \sum_{\mu \tau} g_{\sigma}^{\mu \nu} \frac{\partial L\sqrt{-g}}{\partial g_{\nu}^{\mu \tau}} \right).}
\end{equation}

\noindent The $\delta_{\sigma}^{\nu}$ is the Kronecker delta, which equals $1$ when $\sigma=\nu$ and $0$ otherwise.

With the newly introduced pseudo tensor equation (\ref{Eq 117}), we can bring equation (\ref{Eq 118}) into the form:

\begin{equation} \label{Eq 321}
\boxed{\sum_{\nu} \frac{\partial}{\partial x_{\alpha}} \left( T_{\sigma}^{\nu} + t_{\sigma}^{\nu}\right) = 0.}
\end{equation}

\noindent This equation represents the result of taking the divergence of both sides of the equation (\ref{Eq 160}) and imposing the two conditions $B_\sigma=0$ and $S_{\sigma}^{\nu}=0$ on the left-hand side. The first term on the left-hand side $\sum_{\nu}\partial x_{\alpha}T_{\sigma}^{\nu}$ represents the divergence of the stress-energy tensor. The second term on the left-hand side $\sum_{\nu}\partial x_{\alpha}t_{\sigma}^{\nu}$ represents the stress-energy pseudo tensor. 
Under this limitation, this equation represents energy-momentum conservation in the \emph{Entwurf} Theory. 
Einstein believed that the conditions $B_\sigma =0$ and $S_{\sigma \nu}=0$, combined with his \emph{Entwurf} field equations, would guarantee the energy-momentum conservation law for both matter and the gravitational field.

With the equation for the Lagrangian (equation (\ref{Eq 10})), Einstein expressed equation (\ref{Eq 184}) in the form of equation (\ref{Eq 15}). 
By expressing the left-hand side of equation (\ref{Eq 15}) in this way, we are rewriting the terms involving $\Gamma_{\nu\sigma}^{\tau}$ [the components of the gravitational field [defined as the gradient of the metric tensor in equation (\ref{Eq 180})] in terms of the metric tensor and its derivatives. 

Einstein believed (incorrectly, as it later turned out) that enforcing certain mathematical conditions, like the divergence-free condition equations (\ref{Eq 12}), would lead to a restricted form of the field equations that would be covariant only under certain transformations, i.e., the adapted ones. 

Einstein eventually recognized in 1915 that he had selected an unsuitable Lagrangian function; see equation (\ref{Eq 10}) $+$ equation (\ref{Eq 180}). The \emph{Entwurf} Lagrangian function did not consistently yield \emph{Entwurf} field equations invariant for adapted coordinate systems. Einstein found that the variational formalism applied to the \emph{Entwurf} Lagrangian did not have a unique solution as he initially believed. Taking the variational derivative of the Lagrangian equation (\ref{Eq 10}) concerning the metric might lead to multiple possible field equations. Among those are the covariant Einstein field equations of general relativity. It is important to note that the Lagrangian in the 1915 general relativity looks structurally similar to equation (\ref{Eq 10}). However, the two underlying meanings and implications are significantly different due to the definitions of the connection terms. In the 1915 Lagrangian, $\Gamma$ is the Christoffel symbol derived from the metric tensor. 
This meant to Einstein in the autumn of 1915 that covariance concerning adapted coordinate systems was misguided, prompting Einstein to require general covariance. \cite{Weinstein1}, \cite{Weinstein2}.

Between 1914 and 1915, the \emph{Entwurf} theory came under intense scrutiny from Einstein's peers. They argued that his inability to produce generally covariant field equations stemmed from inherent flaws in his gravitational theory. Yet Einstein felt his approach was justifiable, chiefly due to possessing the clever hole argument. However, in the autumn of 1915, Einstein realized that the issue also affected other calculations and derivations related to his \emph{Entwurf} theory. He surmised there must be a miscalculation in his work on rotation. He believed that the same error that impacted rotation also caused Mercury's perihelion to be affected, as he conveyed in letters to Erwin Freundlich and Otto Naumann \cite{CPAE8}, Doc. 123, \cite{CPAE8}, Doc 124. Einstein had already linked these two issues in the \emph{Einstein-Besso manuscript}.

In the fall of 1915, Einstein revisited the calculations from pages $41$-$42$ of the \emph{Einstein-Besso manuscript} \cite{CPAE4}, Doc. 14. 
Upon revisiting his previous calculations, Einstein realized that the rotating system's description was incompatible with his \emph{Entwurf} vacuum field equation (\ref{Eq 5}) \cite{CPAE8}, Doc. 123.
Einstein found that: 

\begin{equation} \label{Eq 6}
\kappa t_{44} = -\omega^2.    
\end{equation}

In 1913, Einstein thought that $\kappa t_{44}$ had the following relation to $\omega^2$: $\kappa t_{44}=(-)4\omega^2$ [equation (\ref{Eq 2}), section \ref{3}].

Einstein aimed to guarantee that equation (\ref{Eq 5}) held for a rotating system. He plugged in equation (\ref{Eq 6}) to get a relation for the $\Delta g_{44}$:

\begin{equation}
\Delta g_{44} - 2\omega^2 - \omega^2.       
\end{equation}

In 1913, Einstein held the belief that $\Delta g_{44}=-4\omega^2$ (as indicated in equation (\ref{Eq 311}), section \ref{3}). However, in 1915, he realized that his initial assumption regarding $\Delta g_{44}$ was incorrect, and it differed from his subsequent findings:

\begin{equation} \label{Eq 7}
\Delta g_{44}=-3\omega^2.    
\end{equation}

Einstein proposed a solution for $g_{44}$ in the form of a term proportional to the squared distance from the axis of rotation, equation (\ref{Eq 8}) [section \ref{3}]. By equating equation (\ref{Eq 313}) and equation (\ref{Eq 7}), we can determine the value of the constant $C$:

\begin{equation}
\Delta g_{44}=4C\omega^2 = -3\omega^2.    
\end{equation}

\noindent Consequently, Einstein found that the constant $C$ has a value of $-\frac{3}{4}$. In other words, $C$ is not equal to 1. This marks the culmination of Einstein's calculation, where he provides a solution for the metric component $g_{44}$ for a rotating system:

\begin{equation} 
g_{44} = 1- \frac{3}{4} \omega^2(x^2 + y^2).    
\end{equation}

Einstein's above-derived solution for the $g_{44}$ component from the \emph{Entwurf} field equations differed from the one achieved by directly transforming the Minkowski metric into the rotating coordinate system, i.e., equation (\ref{Eq 9}). 

\section{Einstein returns to the November tensor} \label{6}

In November 1915, after much reflection, Einstein distanced himself from the \emph{Entwurf} field equations. He omitted any reference to the hole argument in his November 1915 theory, realizing its objective was not achieved. This argument neither protected the \emph{Entwurf} theory from criticism nor from being debunked through observation and experiment. 

In his review article of 1914, Einstein further developed his hole argument. He showed that two different solutions, $G(x)$ and $G'(x)$, can coexist outside the "hole" (in spacetime), but inside the hole, they coincide. This suggests that generally covariant field equations alone cannot uniquely determine what happens inside the hole, as the choice of coordinate systems matters. By November 1915, upon returning to general covariance, Einstein grasped that two distinct gravitational fields in the same space-time region were implausible. 

After dissociating himself from the field equations of the \emph{Entwurf} theory, Einstein recognized that there are no preferred structures or features, such as the "hole" in spacetime.
He understood that until a specific metric tensor field is defined, there is no well-defined spacetime. In other words, spacetime has no inherent structure; its geometry is determined by the distribution of matter and energy described by the metric tensor field.
This realization led Einstein to the conclusion that you cannot specify a structure on the spacetime manifold a priori. In essence, there is no pre-existing "hole" or other features in spacetime that are independent of the metric tensor field.
Einstein's key insight is that no spacetime framework exists until the gravitational field equations (which determine the metric tensor field) are solved. The metric tensor field defines the geometry of spacetime, and different solutions to these equations can describe different spacetime geometries.

The lengthy and intricate journey from 1912 culminated with Einstein revisiting his original concept—the 1912 November tensor. 
Instead of discarding his 1914 \emph{Entwurf} theory, Einstein refined it in the first relativity paper he submitted on November 4, 1915, to the \emph{Proceedings of the Royal Prussian Academy of Sciences, Berlin}. His 1916 review article, \emph{The Foundation of the General Theory of Relativity} further elaborated this refinement \cite{Einstein5}. Analyzing the \emph{Entwurf} theory, one can find that it already encompasses the foundational mathematical structure of general relativity, except for its field equations. Notably, Einstein's November submissions heavily drew from the 1914 \emph{Entwurf} theory's formalism.

So, what refinements did Einstein introduce? Einstein proposed that his gravitational tensor be invariant under unimodular transformations. 
Einstein recognized that the square root of the determinant of the metric tensor, $\sqrt{-g}$, equals $1$. By postulating $\sqrt{-g}=1$, Einstein effectively restricted his attention to transformations with a determinant of $1$. These are unimodular transformations. By considering only unimodular transformations, Einstein sought a middle ground between full general covariance and the simpler special relativistic invariance. Under unimodular transformations, the determinant of the metric tensor remains constant and, in Einstein’s first November paper, it is equal to $-1$. 
In the \emph{Entwurf} theory, he introduced the concept of tensor densities, which are like tensors but have an extra factor of $\sqrt{-g}$. Considering only unimodular transformations, this factor disappears from many fundamental equations, simplifying calculations \cite{Einstein1}. 

In his 1914 \emph{Entwurf} theory, Einstein described the Riemann tensor with the help of the Kronecker delta (which can take values $+1$ or $-1$ based on permutations of its indices). While the Kronecker delta is mathematically straightforward, its use as a representation for the Riemann tensor is non-standard and is bound to introduce complications when performing tensor operations. There were, therefore, inconsistencies in Einstein’s representations of the Rieman tensor \cite{Einstein11}. By 1915, when Einstein postulated $\sqrt{-g}=1$, these problems were resolved.
He could ignore the effects of volume scaling on his equations by focusing on unimodular transformations. The equations retained their form under unimodular transformations, simplifying the mathematics and eliminating inconsistencies.

The main issue with the \emph{Entwurf} theory was that its field equations were not generally covariant, i.e., that the field equations did not take the same form in all coordinate systems. General covariance is a property that Einstein believed a proper theory of gravitation should possess. But, at the time, he thought that achieving such covariance would lead to equations that did not conserve energy and momentum or lead to the Newtonian limit. Thus, he initially settled for the restricted covariance of the \emph{Entwurf} theory.
The revolutionary aspect of the November 4, 2015 paper and the main difference of the new paper compared to the previous \emph{Entwurf} theory lies in the field equations. 

In the first November 1915 paper, Einstein focused on the Ricci tensor, derived from the Riemann-Christoffel tensor [see equation (\ref{Eq 38})]:

\begin{equation} \label{Eq 40}
\boxed{R^{\rho}_{\ \sigma \mu \nu} = \partial_{\mu} \Gamma^{\rho}_{\nu \sigma} - \partial_{\nu} \Gamma^{\rho}_{\mu \sigma} + \sum_{\lambda} \left( \Gamma^{\rho}_{\mu \lambda} \Gamma^{\lambda}_{\nu \sigma} - \Gamma^{\rho}_{\nu \lambda} \Gamma^{\lambda}_{\mu \sigma} \right).}
\end{equation}

\noindent This fourth-rank tensor is expressed in terms of Christoffel symbols, which are functions of the metric tensor and its derivatives. Einstein wrote a formula that provided the components of this tensor. By contracting the Riemann-Christoffel tensor, we get the Ricci tensor, a second-rank tensor:

\begin{equation} \label{Eq 270}
\{\rho \mu, \lambda \nu\} = \sum_{\rho} g^{\mu\rho} \left( \boxed{\partial_{\rho} \Gamma^{\rho}_{\nu\sigma} - \Gamma^{\rho}_{\nu\lambda} \Gamma^{\lambda}_{\rho\sigma}} + \boxed{\Gamma^{\rho}_{\rho\lambda} \Gamma^{\lambda}_{\nu\sigma} - \partial_{\nu} \Gamma^{\rho}_{\rho\sigma}} \right).
\end{equation}

When we multiply the contravariant metric tensor components $g^{\mu\rho}$ with the covariant metric tensor components $g_{\mu\rho}$, we obtain the Kronecker delta $\delta^{\mu}_{\nu}$:

\begin{equation}
g^{\mu\rho} g_{\mu\rho} = \delta^{\mu}_{\nu}.   
\end{equation}

Following this relationship, Einstein wrote equation (\ref{Eq 270}) in the following form \cite{Einstein1}:

\begin{equation} \label{Eq 323}
\boxed{\{\rho \mu, \lambda \nu\}=G_{\mu \nu}=R_{\mu \nu}+S_{\mu \nu},}
\end{equation}

\noindent which shows this contraction resulting in the tensor $G_{\mu \nu}$. In this equation, $R_{\mu \nu}$ and $S_{\mu \nu}$ represent components of this tensor. Specifically, $R_{\mu \nu}$ is determined by the first and second terms of the equation (\ref{Eq 296}) [see section \ref{1}], while $S_{\mu \nu}$ is defined by this equation's third and fourth terms. Einstein grasped that the two parts of $G_{\mu \nu}$ had distinct physical meanings \cite{Einstein1}.
Indeed, Einstein's \emph{Zurich Notebook} from 1912 indicates that he was already working on the decomposition of $G_{\mu \nu}$ into the Ricci tensor $R_{\mu \nu}$ and the other part, $S_{\mu \nu}$. Furthermore, equation (\ref{Eq 40}) is the 1912 equation (\ref{Eq 38}. Remember that Einstein had already worked with transformations that kept the determinant of $g$ equal to $1$ in 1912 \cite{Einstein7} (see section \ref{1}). 
Under such transformations, all the quantities $G_{\mu \nu}$, $R_{\mu \nu}$ and $S_{\mu \nu}$ behaved as tensors. Einstein recognized that the two parts of $S_{\mu \nu}$ were tied to the derivative of the logarithm of $\sqrt{-g}$. When he postulated $\sqrt{g} = 1$, this term went to zero, leaving the Ricci tensor $R_{\mu \nu}$ as the primary curvature measure. Once again, he had performed this derivation three years earlier \cite{Einstein7}, see section \ref{1} equations (\ref{Eq 296}) and (\ref{Eq 37}). If $S_{\mu \nu}$ = 0, the equation (\ref{Eq 323}) simplifies to the following form:

\begin{equation}
G_{\mu \nu}=R_{\mu \nu}=\partial_{\rho} \Gamma^{\rho}_{\nu\sigma} - \Gamma^{\rho}_{\nu\lambda} \Gamma^{\lambda}_{\rho\sigma}.    
\end{equation}

Einstein postulated that this Ricci tensor $R_{\mu \nu}$ was his new gravitational tensor and was directly related to the distribution of matter and energy in the universe, represented by the energy-momentum tensor $T_{\mu \nu}$. His new field equations give this relationship:

\begin{equation} \label{Eq 289}
\boxed{R_{\mu \nu} = - \kappa T_{\mu \nu}.}      
\end{equation}

The constraint $\sqrt{g} =1$ meant that the field equations were not generally covariant. Recall that general covariance is the principle that the laws of physics should be the same in all coordinate systems. Einstein realized this was a limitation and actively worked to generalize his theory further \cite{Einstein1}, \cite{Weinstein1}, \cite{Weinstein2}. 

Einstein derived his field equations using a variational approach. He started with an action (an integral of a Lagrangian density $\mathcal{L}$ over spacetime). The variation of the action for $g_{\mu \nu}$ is zero for the physical trajectory. The Euler-Lagrange equations are then applied to this action to obtain the equations of motion, leading to Einstein's field equations. 
Einstein formulated his field equations in a Lagrangian form and then established that they respect energy-momentum conservation. 

In November 1915, when Einstein formulated his general relativity, he used a Lagrangian similar to the one from the 1914 \emph{Entwurf} paper, equation (\ref{Eq 10}). However, there were crucial differences in how he treated certain terms and applied the variational principle. These changes allowed Einstein to derive the now-famous field equations. In 1915, Einstein's Lagrangian involved the Christoffel symbols in modern notation:

\begin{equation}
\Gamma^{\lambda}_{\mu \nu} = \frac{1}{2} g^{\lambda \sigma} \left( \partial_{\mu} g_{\nu \sigma} + \partial_{\nu} g_{\mu \sigma} - \partial_{\sigma} g_{\mu \nu} \right).    
\end{equation}

In Einstein's notation, which uses different conventions for index placement and summation, the Christoffel symbols are [see equation (\ref{Eq 278}), section \ref{1}]: \cite{Einstein1}:

\begin{equation} \label{Eq 16}
\boxed{\Gamma^\sigma_{\mu\nu} = -\frac{1}{2} \sum_{\alpha} g^{\sigma\alpha} \left( \frac{\partial g_{\mu\alpha}}{\partial x_\nu} + \frac{\partial g_{\nu\alpha}}{\partial x_\mu} - \frac{\partial g_{\mu\nu}}{\partial x_\alpha} \right).}
\end{equation}

Einstein incorporated into equation (\ref{Eq 49}) the Christoffel symbols, equation (\ref{Eq 16}):

\begin{equation} \label{eq 212}
\sum_{\alpha}\frac{\partial T_\sigma^\alpha}{\partial x_\alpha} = -\sum_{\alpha\beta}\Gamma^\alpha_{\sigma\beta} T_\alpha^\beta.
\end{equation}

In Einstein's 1915 general relativity formulation, the Lagrangian he used for the gravitational field is derived from the Riemann curvature tensor and the Christoffel symbols. 
$\sqrt{-g}$ guarantees that the Lagrangian is a scalar density, which is necessary for the action to be invariant under general coordinate transformations (diffeomorphisms). In 1915, Einstein assumed that $\sqrt{-g}=1$, which simplified the form of equation (\ref{12}).  

Einstein derived from the Euler-Lagrange equation with the inclusion of the stress-energy tensor on the right side the following equation \cite{Einstein1}:

\begin{equation} \label{Eq 13}
\sum_\alpha \frac{\partial}{\partial x_\alpha} \left( \frac{\partial \mathcal{L}}{\partial g^{\mu\nu}_\alpha} \right) - \frac{\partial \mathcal{L}}{\partial g^{\mu\nu}} = -\kappa T_{\mu\nu}.    
\end{equation}

\noindent Einstein then delved into the structure of the energy-momentum content of the gravitational field. He contracted equation (\ref{Eq 13}) and obtained \cite{Einstein1}:

\begin{equation} \label{Eq 331}
\boxed{\sum_{\alpha \mu \nu} \frac{\partial}{\partial x_\alpha} \left( g^{\mu\nu}_\sigma \frac{\partial \mathcal{L}}{\partial g^{\mu\nu}_\alpha} \right) - \frac{\partial \mathcal{L}}{\partial x_\sigma}} = -\kappa \sum_{\mu\nu}T_{\mu\nu} g^{\mu\nu}_\sigma.    
\end{equation}

\noindent He defined the stress-energy pseudo-tensor of the gravitational field \cite{Einstein1}:

\begin{equation} \label{Eq 190}
\boxed{-2\kappa t^\alpha_\sigma = \delta^\alpha_\sigma \mathcal{L} - \sum_{\mu\nu} g_{\sigma}^{\mu\nu} \frac{\partial \mathcal{L}}{\partial g^{\mu\nu}_\alpha}.}
\end{equation}

The pseudo-tensor is not unique; its exact form can differ depending on the definition. Still, its divergence should always yield a conservation law for the combined energy-momentum of matter and the gravitational field. This is crucial for the conservation of energy-momentum in general relativity. The term involving $\delta^\alpha_\sigma \mathcal{L}$ can be seen as a trace term involving the value of $\mathcal{L}$ itself. The term involving the derivatives of $\mathcal{L}$ for the components of $g_{\mu \nu}$ captures how the Lagrangian responds to small variations in the metric. These terms come from applying the Euler-Lagrange equation to the action based on $\mathcal{L}$ and $g_{\mu \nu}$.

\noindent If we take the divergence of both sides of the equation (\ref{Eq 190}) with respect to the index $\alpha$, we get: 

\begin{equation} \label{Eq 334}
2\kappa \sum_\alpha \frac{\partial}{\partial x_\alpha}(t^\alpha_\sigma) = \boxed{\sum_{\alpha \mu\nu} \frac{\partial}{\partial x_\alpha}\left(g_{\sigma}^{\mu\nu}\frac{\partial \mathcal{L}}{\partial g^{\mu\nu}_\alpha}\right)-\frac{\partial \mathcal{L}}{\partial x_\sigma}.}
\end{equation} 

Plugging equation (\ref{Eq 334} into equation (\ref{Eq 331}) brings it to the following form: 

\begin{equation} \label{Eq 335}
\sum_\alpha \frac{\partial}{\partial x_\alpha}(t^\alpha_\sigma) =\boxed{\frac{1}{2}\sum_{\mu\nu} \frac{\partial g^{\mu\nu}}{\partial x_\sigma}T_{\mu\nu}.}    
\end{equation}

\noindent where remember that, $\frac{\partial g^{\mu\nu}}{\partial x_\sigma}\equiv g_{\sigma}^{\mu\nu}$.

Now, Einstein reformulated his \emph{Entwurf} [equation (\ref{Eq 49})] energy-momentum balance for matter in a gravitational field as follows \cite{Einstein11}:

\begin{equation} \label{Eq 330}
\boxed{\sum_\lambda \frac{\partial T^{\lambda}_{\sigma}}{\partial x_\lambda} = -\frac{1}{2} \sum_{\mu\nu} \frac{\partial g^{\mu\nu}}{\partial x_\sigma} T_{\mu\nu}.}
\end{equation}

Inserting equation (\ref{Eq 330}) into (\ref{Eq 335}) leads to the conservation of total energy-momentum in a gravitational field \cite{Einstein1}: 

\begin{equation} \label{Eq 14}
\boxed{\sum_{\nu} \frac{\partial}{\partial x_{\nu}} (T_{\sigma}^{\nu} + t_{\sigma}^{\nu}) = 0.}
\end{equation}

This expression includes a summation over all spacetime indices $\nu$, indicating that the energy-momentum conservation holds in all spacetime directions. Einstein emphasizes here the idea that energy-momentum conservation applies at each point in spacetime.  

In his papers of 1915 on general relativity, Einstein took inspiration from his earlier \emph{Entwurf} theory and equations \cite{Einstein11} to formulate the field equations using the Christoffel symbols and pseudo-tensors. This suggests a certain continuity in his thinking and the development of his gravitational theory. Equation (\ref{Eq 117}) from the \emph{Entwurf} theory represents a pseudo-tensor and includes derivatives of the Lagrangian density for the metric components. In his paper of 1915 \cite{Einstein1}, Einstein derived an equation resembling (\ref{Eq 117}) but expressed it with the Christoffel symbols equation (\ref{Eq 16}): 

\begin{equation} \label{Eq 200}
\boxed{t_{\sigma}^{\nu} = \frac{1}{2} \delta_{\sigma}^{\nu} \sum_{\mu\nu\alpha\beta} g^{\mu\nu} \Gamma_{\mu\beta}^{\alpha} \Gamma_{\nu\alpha}^{\beta} - \sum_{\mu\nu\alpha} g^{\mu\nu} \Gamma_{\mu\sigma}^{\alpha} \Gamma_{\nu\alpha}^{\nu} .}
\end{equation}

Equation (\ref{Eq 184}) from the \emph{Entwurf} theory also represents a pseudo-tensor and describes the relationship between the metric, the Lagrangian density, and the gravitational field. In his paper of 1915 \cite{Einstein1}, Einstein derived an equation resembling equation (\ref{Eq 184}): equation (\ref{Eq 190}). 

Let us delve into the contrast between Einstein's general theory of relativity, submitted on November 4, 1915 \cite{Einstein1}, and his earlier \emph{Entwurf} theory \cite{Einstein11} regarding energy-momentum conservation.
In Einstein's earlier \emph{Entwurf} theory, energy-momentum conservation, equation (\ref{Eq 14}), was guaranteed only in adapted coordinates, a restrictive condition that makes the theory limited. If we can only express a physical principle in specific coordinate systems, it is arguably not a truly universal law of physics \cite{Einstein11}. 
Now, recall that equation (\ref{Eq 11}) can be considered the field equations that account for both matter and the gravitational field. Equation (\ref{Eq 15}) provides a specific expression for the pseudo-tensor of the gravitational field, and equation (\ref{Eq 12}) is a condition for the conservation of energy-momentum in the \emph{Entwurf} theory. Still, it is only valid in adapted coordinates. 
In Einstein's November 4, 1915, general theory of relativity, energy-momentum conservation is derived without the adapted coordinates constraints on the coordinate system. However, Einstein still restricted the covariance of his field equations to the subclass unimodular transformations. Unlike the adapted coordinates, the restriction to unimodular coordinates was Einstein's attempt to simplify the formulation of his gravitational theory, avoiding certain complexities and achieving a conservation law for energy and momentum. 

After demonstrating his field equations satisfy the principle of energy momentum, Einstein dedicated himself to deriving the weak-field limit. He wrote his field equations in the following form:

\begin{equation} \label{Eq 202}
\sum_{\alpha} \frac{\partial \Gamma_{\mu\nu}^{\alpha}}{\partial x_{\alpha}} + \sum_{\alpha\beta} \Gamma_{\mu\beta}^{\alpha} \Gamma_{\nu\alpha}^{\beta} = -\kappa T_{\mu\nu}.    
\end{equation}

First, Einstein contracted the field equations (\ref{Eq 202}): he multiplied the field equations (\ref{Eq 202}) by the inverse metric $g^{\mu\nu}$ and then summed over both indices $\mu$ and $\nu$ to obtain a scalar equation: 

\begin{equation} \label{Eq 17}
\sum_{\alpha\beta} \frac{\partial^2 g^{\alpha\beta}}{\partial x_\alpha \partial x_\beta} - \sum_{\sigma\tau\alpha\beta} g^{\sigma\tau} \Gamma_{\sigma\beta}^\alpha \Gamma_{\tau\alpha}^\beta + \sum_{\alpha\beta} \frac{\partial}{\partial x_\alpha} \left( g^{\alpha\beta} \frac{\partial \log\sqrt{-g}}{\partial x_\beta} \right) = -\kappa T.    
\end{equation}

\noindent This scalar equation is the contracted field equation. It represents a scalar equation for the metric components $g^{\alpha\beta}$ and involves several terms:
the Laplacian of the metric components $\frac{\partial^2 g^{\alpha\beta}}{\partial x_\alpha \partial x_\beta}$.
Terms involving products of the metric components $g^{\sigma\tau}$ and the Christoffel symbols $\Gamma_{\sigma\beta}^\alpha$ and $\Gamma_{\tau\alpha}^\beta$.
A term related to the derivative of the determinant of the metric tensor $\frac{\partial}{\partial x_\alpha} \left( g^{\alpha\beta} \frac{\partial \log\sqrt{-g}}{\partial x_\beta} \right)$, which arises from the determinant of the metric tensor and its derivative.
The right-hand side is a scalar multiple of the energy-momentum tensor $T$.

Second, Einstein multiplied the original field equations (\ref{Eq 202}) by the inverse metric $g^{\mu\nu}$ and summed only over index $\nu$ to obtain another form of the contracted field equations:

\begin{equation} \label{Eq 215}
\sum_{\alpha\nu}\frac{\partial}{\partial x_\alpha}\left(g^{\nu\lambda}\Gamma_{\mu\nu}^\alpha\right) - \boxed{\sum_{\alpha\beta\nu}g^{\nu\beta}\Gamma_{\nu\mu}^\alpha\Gamma_{\beta\alpha}^\lambda} = -\kappa T_\mu^\tau.    
\end{equation}

This equation involves the divergence of a quantity $\frac{\partial}{\partial x_\alpha}\left(g^{\nu\lambda}\Gamma_{\mu\nu}^\alpha\right)$
and the connection terms $\sum_{\alpha\beta\nu}g^{\nu\beta}\Gamma_{\nu\mu}^\alpha\Gamma_{\beta\alpha}^\lambda$.
The right-hand side represents the components of the energy-momentum tensor $T_\mu^\tau$.

Einstein then plugged the left-hand side of the stress-energy pseudo-tensor [equation (\ref{Eq 200})]:

\begin{equation} \label{Eq 326}
t_{\sigma}^{\nu} - \frac{1}{2} \delta_{\sigma}^{\nu} \sum_{\mu\nu\alpha\beta} g^{\mu\nu} \Gamma_{\mu\beta}^{\alpha} \Gamma_{\nu\alpha}^{\beta} = \boxed{-\sum_{\mu\nu\alpha} g^{\mu\nu} \Gamma_{\mu\sigma}^{\alpha} \Gamma_{\nu\alpha}^{\nu}.} 
\end{equation}

\noindent into equation (\ref{Eq 215}):

\begin{equation} \label{Eq 234}
\sum_{\alpha\nu}\frac{\partial}{\partial x_\alpha}\left(g^{\nu\lambda}\Gamma_{\mu\nu}^\alpha\right) - \frac{1}{2} \delta_{\mu}^{\lambda} \sum_{\mu\nu\alpha\beta} g^{\mu\nu} \Gamma_{\mu\beta}^{\alpha} \Gamma_{\nu\alpha}^{\beta} = -\kappa (T_\mu^\lambda + t_\mu^\lambda).    
\end{equation}

Now, he took the divergence of equation (\ref{Eq 234}) with respect to the spacetime coordinate $x_\mu$:

\begin{equation} 
\frac{\partial}{\partial x_\mu}\left(\sum_{\alpha\nu}\frac{\partial}{\partial x_\alpha}\left(g^{\nu\lambda}\Gamma_{\mu\nu}^\alpha\right) - \frac{1}{2} \delta_{\mu}^{\lambda} \sum_{\mu\nu\alpha\beta} g^{\mu\nu} \Gamma_{\mu\beta}^{\alpha} \Gamma_{\nu\alpha}^{\beta}\right) = -\kappa \frac{\partial}{\partial x_\mu}(T_\mu^\lambda + t_\mu^\lambda).
\end{equation}

We take the derivative inside the parentheses with respect to $x_\mu$ by applying the product rule and summing over indices $\mu$, $\nu$, $\alpha$, and $\beta$:

\begin{equation} \label{Eq 327}
\frac{\partial}{\partial x_\mu}
\left(\sum_{\alpha\beta}\frac{\partial^2 g^{\alpha\beta}}{\partial x_\alpha \partial x_\beta} - \sum_{\nu \lambda \alpha \beta} g^{\nu\lambda}\Gamma_{\mu\nu}^\alpha\Gamma_{\alpha\lambda}^\beta\right) - \frac{1}{2} \boxed{\delta_{\mu}^{\lambda}} \sum_{\mu\nu\alpha\beta} \frac{\partial}{\partial x_\mu}\left(g^{\mu\nu} \Gamma_{\mu\beta}^{\alpha} \Gamma_{\nu\alpha}^{\beta}\right) = 0.
\end{equation}

\noindent I would like to offer two comments regarding this equation:

1) According to the conservation of momentum and energy, equation (\ref{Eq 14}), the divergence of the stress-energy tensor should be zero. This means that the source terms on the right-hand side do not contribute to the divergence on the left-hand side, and the equation simplifies accordingly.

2) The divergence of the Kronecker delta takes the following form:

\begin{equation}
\frac{\partial}{\partial x_\mu}\delta_{\mu}^{\lambda} = \frac{\partial x^\lambda}{\partial x_\mu}.
\end{equation}

\noindent However, we do not need to take the divergence of the Kronecker delta because the Kronecker delta term on the left-hand side of the equation (\ref{Eq 234}) is already in its final form and does not involve differentiation with respect to coordinates. Hence, the Kronecker delta $\delta_{\mu}^{\lambda}$ only equals $1$ when $\mu = \lambda$ and equals zero otherwise (i.e., when $\mu \neq \lambda$). 

Thus, the equation (\ref{Eq 327}) simplifies to the following equation involving the second derivatives of the metric and derivatives of the Christoffel symbols \cite{Einstein1}:

\begin{equation} \label{Eq 325}
\frac{\partial}{\partial x_\mu}\left(\sum_{\alpha\beta}\frac{\partial^2 g^{\alpha\beta}}{\partial x_\alpha \partial x_\beta} - \sum_{\sigma \tau \alpha \beta}g^{\sigma\tau}\Gamma_{\beta\sigma}^\alpha\Gamma_{\alpha\tau}^\beta\right) = 0.
\end{equation}

The above equation implies that the entire expression enclosed within the parentheses should equal zero:  

\begin{equation} \label{Eq 217}
\sum_{\alpha\beta}\frac{\partial^2 g^{\alpha\beta}}{\partial x_\alpha \partial x_\beta} - \sum_{\sigma \tau \alpha \beta}g^{\sigma\tau}\Gamma_{\beta\sigma}^\alpha\Gamma_{\alpha\tau}^\beta = 0,
\end{equation}

Einstein wrote the first term of the equation (\ref{Eq 217}), while disregarding the second term \cite{Einstein1}:

\begin{equation} \label{Eq 225}
\sum_{\alpha\beta}\frac{\partial^2 g^{\alpha\beta}}{\partial x_\alpha \partial x_\beta} = 0.    
\end{equation}

\noindent This equation relates to the second mixed partial derivatives of the metric tensor components $g^{\alpha\beta}$ and represents a condition on the curvature of spacetime itself.
It is a simplified form that only includes the second-order derivatives of the metric tensor $g^{\alpha\beta}$ for the coordinates $x_\sigma$. It does not involve the Christoffel symbols or their derivatives. 

Einstein contends that equation (\ref{Eq 225}) serves as a first-order approximation of equation \eqref{Eq 217}. He asserts that, when considering the first approximation of (\ref{Eq 225}), we possess the liberty to impose the following condition \cite{Einstein1}:

\begin{equation} \label{227}
\boxed{\sum_\beta\left(\frac{\partial g^{\alpha\beta}}{\partial x_\alpha}\right) = 0.}    
\end{equation}

This is the familiar Hertz condition, denoted as equation (\ref{Eq 201}), which Einstein resurrected from his 1912 \emph{Zurich notebook}, rescuing it from obscurity. It is a condition on the first partial derivatives of the metric tensor components for the coordinates. If we take the derivative of this equation for $x_\nu$ we obtain equation (\ref{Eq 225}). 

Einstein then derived the equation:

\begin{equation} \label{Eq 228} 
\frac{1}{2} \sum_{\alpha} \frac{\partial^2 g_{\mu\nu}}{\partial x_{\alpha}^2} = \kappa T_{\mu\nu},    
\end{equation}

\noindent from his field equations \eqref{Eq 202}.

Let us derive this equation. Starting with the following equation:

\begin{equation} \label{Eq 233}
\sum_{\alpha} \frac{\partial \Gamma_{\mu\nu}^{\alpha}}{\partial x_{\alpha}} = -\kappa T_{\mu\nu},    
\end{equation}

\noindent can we derive equation (\ref{Eq 228})?

\noindent First, we will take the derivatives of the Christoffel symbols $\Gamma_{\mu\nu}^{\alpha}$ in equation (\ref{Eq 16}) and use the Hertz condition equation (\ref{227}). We start with equation (\ref{Eq 233}) and take the derivative of the Christoffel symbols with respect to $x_\alpha$:

\begin{equation} \label{Eq 341}
\frac{\partial \Gamma_{\mu\nu}^{\alpha}}{\partial x_{\alpha}} = \frac{1}{2} \frac{\partial}{\partial x_\alpha} \left(g^{\alpha\beta} \left(\frac{\partial g_{\beta\mu}}{\partial x_{\nu}} + \frac{\partial g_{\beta\nu}}{\partial x_{\mu}} - \frac{\partial g_{\mu\nu}}{\partial x_{\beta}}\right)\right).
\end{equation}

\noindent Now, we use the product rule for derivatives. The result we obtain is the following. 
\noindent Here we have the first term on the right-hand side:

\begin{equation}
\frac{1}{2} \boxed{\left(\frac{\partial g^{\alpha\beta}}{\partial x_\alpha}\right)} \left(\frac{\partial g_{\beta\mu}}{\partial x_{\nu}} + \frac{\partial g_{\beta\nu}}{\partial x_{\mu}} - \frac{\partial g_{\mu\nu}}{\partial x_{\beta}}\right).    
\end{equation}

\noindent Here we have the second term on the right-hand side:

\begin{equation}
+ \frac{1}{2} g^{\alpha\beta} \left(\frac{\partial}{\partial x_\alpha}\left(\frac{\partial g_{\beta\mu}}{\partial x_{\nu}}\right) + \frac{\partial}{\partial x_\alpha}\left(\frac{\partial g_{\beta\nu}}{\partial x_{\mu}}\right) - \frac{\partial}{\partial x_\alpha}\left(\frac{\partial g_{\mu\nu}}{\partial x_{\beta}}\right)\right).    
\end{equation}

\noindent Now, we can simplify the first term on the right-hand side using the Hertz condition from equation \eqref{227}:

\begin{equation}
\sum_\beta\left(\frac{\partial g^{\alpha\beta}}{\partial x_\alpha}\right) = 0,
\end{equation}

\noindent which means that the first term drops out, and we are left with the second term:

\begin{equation}
\frac{1}{2} g^{\alpha\beta} \left(\frac{\partial}{\partial x_\alpha}\left(\frac{\partial g_{\beta\mu}}{\partial x_{\nu}}\right) + \frac{\partial}{\partial x_\alpha}\left(\frac{\partial g_{\beta\nu}}{\partial x_{\mu}}\right) - \frac{\partial}{\partial x_\alpha}\left(\frac{\partial g_{\mu\nu}}{\partial x_{\beta}}\right)\right).
\end{equation}

\noindent Now, we can rewrite this in a more compact form:

\begin{equation} \label{Eq 351}
\Gamma_{\mu\nu}^{\alpha}= \frac{1}{2} g^{\alpha\beta} \frac{\partial^2 g_{\beta\mu}}{\partial x_\alpha \partial x_\nu} + \frac{1}{2} g^{\alpha\beta} \frac{\partial^2 g_{\beta\nu}}{\partial x_\alpha \partial x_\mu} - \frac{1}{2} g^{\alpha\beta} \frac{\partial^2 g_{\mu\nu}}{\partial x_\alpha \partial x_\beta}.
\end{equation}

\noindent This expression was not the one Einstein used now. These are the second derivatives of the metric tensor, and this expression is the derivative of the Christoffel symbols $\Gamma_{\mu\nu}^{\alpha}$ with respect to the coordinates $x_\alpha$. Einstein would opt for this path in a few months \cite{Einstein5}.

So, let us pursue an alternative path. Let us expand the Christoffel symbols $\Gamma_{\mu\nu}^{\alpha}$ in terms of their components. Starting with equation (\ref{Eq 16}), we will first work on the first term involving: $\frac{\partial g_{\beta\mu}}{\partial x_{\nu}}$. We expand the first term using the product rule for partial derivatives:

\begin{equation}
\frac{1}{2} g^{\alpha\beta} \frac{\partial g_{\beta\mu}}{\partial x_{\nu}} = g_{\beta\mu},\nu.    
\end{equation}

\noindent Now, let us work on the second term involving $\frac{\partial g_{\beta\nu}}{\partial x_{\mu}}$:

\noindent We expand the second term using the product rule for partial derivatives:

\begin{equation}
\frac{1}{2} g^{\alpha\beta} \frac{\partial g_{\beta\nu}}{\partial x_{\mu}} =  g_{\beta\nu}, \mu.   
\end{equation}

\noindent Now, let us rewrite the Christoffel symbols with these expanded terms:

\begin{equation}
\Gamma_{\mu\nu}^{\alpha} = \frac{1}{2} g_{\beta\mu},\nu + \frac{1}{2} g_{\beta\nu}, \mu - \boxed{\frac{1}{2} \frac{\partial g_{\mu\nu}}{\partial x_{\beta}}.}
\end{equation}

\noindent To express the third term as the second partial derivative of the metric tensor components with respect to $x_\alpha$, we take the derivative of the third term:   

\begin{equation}
\frac{1}{2} \sum_{\alpha} \frac{\partial^2 g_{\mu\nu}}{\partial x_{\alpha}^2}.
\end{equation}

Despite Einstein's efforts, he could not yet deduce the Newtonian limit from his field equations (\ref{Eq 202}). However, unlike in 1912, he did not dismiss the November tensor. On the contrary, he had confidence in it and sought to enhance the covariance of his field equations and expand them. 

\section{Einstein expands the Covariance of his theory}

Let us revisit equation (\ref{Eq 17}. If the divergence of the left-hand side of the equation (\ref{Eq 217}) is zero, it implies that the sum of the terms on the left-hand side must be equal to zero for all values of indices $\alpha$ and $\beta$. In other words, each term in the sum must be zero. Therefore, the second term involving the Christoffel symbols must also be zero for each combination of indices $\sigma$, $\tau$, $\alpha$, and $\beta$.
Mathematically, this is expressed as equation (\ref{Eq 217}). 
If these conditions are met, it will result in Einstein's derivation encountering difficulties because equation (\ref{Eq 17} reduces to the following form:

\begin{equation} \label{Eq 216}
\boxed{\sum_{\alpha\beta} \frac{\partial}{\partial x_\alpha} \left( g^{\alpha\beta} \frac{\partial \log\sqrt{-g}}{\partial x_\beta} \right) = -\kappa T.}    
\end{equation}

In an Addendum to the November 4, 1915 paper, submitted on November 11, 1915, Einstein showed that the equation (\ref{Eq 216}) cannot be satisfied for all coordinate systems because if $\sqrt{-g}=1$ in a coordinate system, then $\log 1 = 0$, and the right-hand side of the equation becomes $\kappa T = 0$. This would imply that the trace of the stress-energy tensor $\sum_\sigma T^{\sigma}_{\sigma}$ must vanish in that coordinate system. Hence, equation (\ref{Eq 216}) is a kind of coordinate condition that cannot be satisfied in all coordinate systems. 
Einstein then wrote the full Einstein field equation (in the presence of matter and energy). He showed that when the field equations are expressed in terms of the determinant of the metric tensor $\sqrt{-g}$, they become coordinate-independent, i.e., they have the same form in all coordinate systems \cite{Einstein6}:

\begin{equation} \label{Eq 39}
\boxed{G_{\mu\nu} = \kappa T_{\mu\nu},}
\end{equation}

In a vacuum (where $T_{\mu\nu} = 0$), this reduces to: 

\begin{equation} \label{Eq 39-1}
\boxed{G_{\mu\nu} = 0.}
\end{equation} 

\noindent which represents spacetime curvature due to gravitational fields in the absence of matter. In the final paper (\cite{Einstein14}), Einstein mentioned that he had previously considered equations (\ref{Eq 39}) with Grossmann three years earlier in the \emph{Zurich Notebook}\cite{Einstein7}. Grossmann provided Einstein with the necessary mathematical tools, specifically the Riemann–Christoffel tensor and the Ricci tensor [\ref{Eq 38}]. In turn, Einstein utilized these tools to derive the November tensor (\ref{Eq 37}) and other derivations, which are documented in his \emph{Zurich notebook}. These details are elaborated upon in section \ref{1}.

As his work progressed, Einstein realized the November 11, 1915 restrictions were unnecessary. He discovered that the full, unrestricted theory naturally ensured the conservation of energy-momentum, provided he included the trace of the energy-momentum tensor $T$ in his field equations. This realization led him to the generally covariant field equations under the restriction of unimodular coordinates \cite{Einstein14}:

\begin{equation} \label{Eq 20}
\boxed{G_{\mu\nu} = -\kappa \left( T_{\mu\nu} - \frac{1}{2} g_{\mu\nu} T \right), \sqrt{-g}=1.}    
\end{equation}

\noindent $G_{\mu\nu}$ is the Einstein tensor, $T_{\mu\nu}$ is the energy-momentum tensor (describing the distribution of matter and energy), and $g_{\mu\nu}$ is the metric tensor. 
The trace of the stress-energy tensor $T_{\mu\nu}$ is represented by:

\begin{equation}
\boxed{T = \sum_{\rho\sigma} g^{\rho\sigma} T_{\rho\sigma} = \sum_{\sigma} T^\sigma_\sigma.}    
\end{equation}

\noindent The left-hand side $\sum_{\rho\sigma}g^{\rho\sigma} T_{\rho\sigma}$ uses the indices $\rho$ and $\sigma$ to sum over all components of the stress-energy tensor $T_{\mu\nu}$ with the metric tensor $g^{\rho\sigma}$ contracting the indices. This notation reflects Einstein's use of the summation symbol $\sum$ at the time. It is worth noting that Einstein had not yet introduced the Einstein summation convention, which he would invent less than a year later. With the summation convention, repeated indices imply summation. The second part $\sum_{\sigma} T^\sigma_\sigma$ is a simplified way to express the same trace by summing over the diagonal components of $T_{\mu\nu}$ where $\sigma$ appears as both indices. It emphasizes that we are summing only over the components where $\mu$ and $\nu$ are equal, which correspond to the diagonal elements of the tensor. 

Einstein simplified the determinant of the metric tensor $\sqrt{-g}$ and imposed an unimodular condition $\sqrt{-g}=1$. He chose coordinates in which $\sqrt{-g}$ is set to a constant value $1$. Choosing unimodular coordinates ensures that the volume element in these coordinates remains constant and does not change as we move through spacetime. It can be viewed as a choice of gauge in general relativity. Gauge choices do not affect physical observables, but they can simplify the form of equations.

As mentioned in section \ref{1}, Einstein had previously formulated an equation \eqref{Eq 276} in his \emph{Zurich notebook} of 1912 that bears a striking resemblance in form to his field equations \eqref{Eq 20} of 1915. Whether Einstein retained this knowledge from 1912 or had completely forgotten about it, one cannot deny that his act of formulating such a field equation in 1912 indicates his thought process.

On November 20, 1915, David Hilbert presented a paper titled "The Foundations of Physics" to the Göttingen Academy of Sciences, in which he included his version of the gravitational field equations of general relativity. Just five days later, on November 25, 1915, Einstein presented his version of the gravitational field equations \eqref{Eq 20-1}. 
There have been claims that Hilbert published the correct form of the field equations just days before Einstein did. There has been much debate surrounding whether Hilbert outpaced Einstein in the race to derive the correct final form of the field equations. People have argued that what is noteworthy is that six days earlier, on November 20, 1915, Hilbert derived the same field equations that Einstein had been searching for over an extended period. Some have asserted that the brilliant and highly skilled mathematician elegantly arrived at the correct form of these equations, unlike Einstein's path, which involved various restrictions, conditions, approximations and challenges.

Scholars in the history of science have dedicated significant effort to scrutinizing Hilbert's manuscripts and publications. They meticulously examined the equations in his handwritten documents, compared them with those in his published works, determined publication dates precisely, and analyzed the consistency between manuscript equations and their counterparts in his published papers. To streamline the discussion, I will provide a concise summary of the overarching findings resulting from this extensive research. 
It is important to note that Hilbert's presentation on November 20 did not initially contain a generally covariant theory. His paper was only officially published on March 31, 1916 \cite{Hilbert}. Notably, in his published paper, Hilbert made corrections based on Einstein's final field equations from November 25, 1915, which were published in his paper on December 2, 1915 \cite{Einstein14}.
Hilbert revised his November 20, 1915, paper sometime between December 1915 and March 1916, incorporating the gravitational field equations in terms of the Einstein tensor [equation \eqref{Eq 383}] after reading Einstein's November 25, 1915 paper. In the published version of his paper from March 1916, Hilbert added a reference to Einstein's November 25, 1915 paper, acknowledging that the resulting field equations of gravitation appeared to align with the theory of general relativity established by Einstein in his recent papers \cite{Stachel}; \cite{Corry}.

In his November 25, 1915 paper, Einstein wrote equations (\ref{Eq 20}) as follows \cite{Einstein14}:

\begin{equation} \label{Eq 20-1}
\boxed{G_{im} = -\kappa \left( T_{im} - \frac{1}{2} g_{im} T \right), \sqrt{-g}=1.}    
\end{equation}

\noindent Let us stick to Einstein's original notation. Einstein rewrote his field equations as follows \cite{Einstein14}: 

\begin{equation} \label{Eq 342}
\sum_l \frac{\partial \Gamma_{im}^l}{\partial x_l} + \sum_{\rho l} \Gamma_{i\rho}^l \Gamma_{ml}^\rho = -\kappa(T_{im} - \frac{1}{2} g_{im} T), \quad \sqrt{-g} = 1.
\end{equation}

He contracted his field equations (\ref{Eq 20-1}). He contracted the indices $i$ and $m$, summing over them. He rearranged the terms to separate the Christoffel symbols and the metric tensor terms. He then obtained the contracted form of equations (\ref{Eq 20-1}) in the particular coordinate system where $\sqrt{-g}=1$ :

\begin{equation} \label{Eq 352}
\sum_{\alpha\beta} \frac{\partial^2 g^{\alpha\beta}}{\partial x_\alpha \partial x_\beta} - \sum_{\sigma\tau\alpha\beta} g^{\sigma\tau} \Gamma_{\sigma\beta}^\alpha \Gamma_{\tau\alpha}^\beta + \sum_{\alpha\beta} \frac{\partial}{\partial x_\alpha} \left(g^{\alpha\beta} \frac{\partial \log(\sqrt{-g})}{\partial x_\beta}\right) = \kappa T.
\end{equation}

Remember that Einstein wrote in the \emph{Zurich notebook} the condition equation (\ref{Eq 295}). If $\sqrt{-g} = 1$ the condition in equation (\ref{Eq 295}) becomes zero. Hence, the third term drops out, leaving us with the following equation:

\begin{equation} \label{Eq 343}
\sum_{\alpha\beta} \frac{\partial^2 g^{\alpha\beta}}{\partial x_\alpha \partial x_\beta} - \sum_{\sigma\tau\alpha\beta} g^{\sigma\tau} \Gamma_{\sigma\beta}^\alpha \Gamma_{\tau\alpha}^\beta = \kappa T.
\end{equation}

The second term in this equation equals the derivative of the stress-energy pseudo tensor [equation (\ref{Eq 200}) [see explanation in section \ref{6}]. Equation (\ref{Eq 343}) is brought to the following form: 

\begin{equation}
\sum_{\alpha\beta} \frac{\partial^2 g^{\alpha\beta}}{\partial x_\alpha \partial x_\beta} - \kappa(T + t)= 0.
\end{equation}

\noindent Now, let us proceed in accordance with Einstein's guidance \cite{Einstein14}:

\noindent 1) We multiply equation \eqref{Eq 342} by $\frac{\partial g^{im}}{\partial x_\sigma}$ and sum over $i$ and $m$:

\begin{equation} \label{Eq 353}
\sum_l \left(\sum_{im}\frac{\partial g^{im}}{\partial x_\sigma} \frac{\partial \Gamma_{im}^l}{\partial x_l}\right) + \sum_{\rho l} \left(\sum_{im}\frac{\partial g^{im}}{\partial x_\sigma} \Gamma_{i\rho}^l \Gamma_{ml}^\rho\right) = -\kappa\sum_{im}\frac{\partial g^{im}}{\partial x_\sigma}\left(T_{im} - \frac{1}{2} g_{im} T\right).
\end{equation}

\noindent 2) Equation \eqref{Eq 330} reads: \cite{Einstein14}:
 
\begin{equation}
\sum_{\lambda} \frac{\partial T_{\sigma}^{\lambda}}{\partial x_{\lambda}} =-\frac{1}{2} \sum_{\mu \nu} \frac{\partial g^{\mu \nu}}{\partial x_\sigma} T_{\mu \nu}.    
\end{equation}

\noindent 3) Einstein plugged the right-hand side of equation \eqref{Eq 330} into the right-hand side of \eqref{Eq 353}. He then took the derivative of equation (\ref{Eq 200} and plugged it into the left-hand side of equation \eqref{Eq 353}. This manipulation, which is not straightforward, led him to the energy-momentum conservation equation (\ref{Eq 14}). 

A few months later, Einstein no longer required the above manipulations as he was able to achieve conservation of energy-momentum straightforwardly \cite{Einstein5}. 

\noindent Einstein wrote his field equations in the form of equations (\ref{Eq 342}):

\begin{equation} \label{Eq 346}
\frac{\partial \Gamma_{\mu\nu}^{\alpha}}{\partial x_{\alpha}} + \Gamma_{\mu\beta}^{\alpha} \Gamma_{\nu\alpha}^{\beta} = -\kappa\left(T_{\mu\nu} - \frac{1}{2} g_{\mu\nu} T\right) \quad \sqrt{-g} = 1.
\end{equation}

In the 1916 paper, Einstein introduced the summation convention \cite{Einstein5}. Equation (\ref{Eq 346}) uses the summation convention where necessary, particularly in the repeated indices within the Christoffel symbols and their derivatives. The indices $\mu$ and $\nu$ on the left-hand side are free indices and do not imply summation: the first term on the left-hand side $\frac{\partial \Gamma^\alpha_{\mu\nu}}{\partial x^\alpha}$ contains the index $\alpha$, which is repeated as a subscript and a superscript. This indicates summation over $\alpha$. So, the summation convention applies to the first term. In the second term $\Gamma^\alpha_{\mu\beta} \Gamma^\beta_{\nu\alpha}$ the indices $\alpha$ and $\beta$ are repeated as both subscripts and superscripts, implying summation over these indices as well. The right-hand side of the equation contains expressions with indices $\mu$ and $\nu$ ($T_{\mu \nu}$ and $g_{\mu \nu}$). These indices are free, not repeated, and not subject to summation.

\noindent In his first November 4 paper, Einstein contracted his field equations leading to equation (\ref{Eq 215}) \cite{Einstein1}. 
In his 1916 paper, "Foundations of General Relativity," he contracted equations (\ref{Eq 346}) and then substituted the equation (\ref{Eq 200}) (the pseudo stress-energy tensor) into the obtained result. 
The final result is \cite{Einstein5}:

\begin{equation} \label{Eq 347}
\frac{\partial}{\partial x_{\alpha}} \left(g^{\sigma\beta} \Gamma_{\mu\beta}^{\alpha} - \frac{1}{2}\delta_{\mu}^{\sigma}g^{\lambda\beta} \Gamma_{\lambda\beta}^{\alpha}\right)= -\kappa\left(T_{\mu}^{\sigma} + t_{\mu}^{\sigma}\right).
\end{equation}

\noindent Einstein took the derivative $\frac{\partial}{\partial x_\sigma}$ of equation \eqref{Eq 347}:

\begin{equation} \label{Eq 348}
\frac{\partial}{\partial x_\sigma}\left(\frac{\partial}{\partial x_{\alpha}} \left(g^{\sigma\beta} \Gamma_{\mu\beta}^{\alpha} - \frac{1}{2}\delta_{\mu}^{\sigma}g^{\lambda\beta} \Gamma_{\lambda\beta}^{\alpha}\right)\right) = \frac{\partial}{\partial x_\sigma}\left(-\kappa\left(T_{\mu}^{\sigma} + t_{\mu}^{\sigma}\right)\right),
\end{equation}

\noindent and demonstrated that the left-hand side of equation (\ref{Eq 348}) is an identity:

\begin{equation} \label{Eq 349}
\frac{\partial^2}{\partial x_\alpha\partial x_\sigma}\left(g^{\sigma\beta} \Gamma_{\mu\beta}^{\alpha} - \frac{1}{2}\delta_{\mu}^{\sigma}g^{\lambda\beta} \Gamma_{\lambda\beta}^{\alpha}\right) \equiv 0.    
\end{equation}

\noindent Hence, if the identity in equation (\ref{Eq 349}) vanishes, the field equations (\ref{Eq 346}) satisfy equation (\ref{Eq 14}), i.e., the conservation of energy-momentum \cite{Einstein5}. 

Equation \eqref{Eq 349} involves the second partial derivatives of the terms $g^{\sigma\beta} \Gamma_{\mu\beta}^{\alpha}$ and 
$\frac{1}{2}\delta_{\mu}^{\sigma}g^{\lambda\beta} \Gamma_{\lambda\beta}^{\alpha}$ with respect to $x_\alpha$ and $x_\sigma$. There are valid reasons to question whether equation \eqref{Eq 349} is the contracted Bianchi identity. Equation \eqref{Eq 349} resembles the contracted Bianchi identity. In the contracted Bianchi identity, we have a contraction of the Riemann tensor and its derivatives. Equation \eqref{Eq 349} involves derivatives of the metric and Christoffel symbols. 
However, equation \eqref{Eq 349} represents the contracted Bianchi identity in unimodular coordinates $\sqrt{-g}=1$. 

In his paper of March 1916, "The Foundations of Physics", Hilbert provided the following field equation \cite{Hilbert}:

\begin{equation} \label{Eq 383}
-\frac{1}{\sqrt{g}}\frac{\partial \sqrt{g}L}{\partial g^{\mu \nu}} = K_{\mu \nu} - \frac{1}{2}g_{\mu \nu} K.
\end{equation}

\noindent If we set the left-hand side (what is now known as the Einstein tensor) of the equation \eqref{Eq 383} to zero:

\begin{equation} \label{Eq 383-1}
K_{\mu \nu} - \frac{1}{2}g_{\mu \nu} K=0.
\end{equation}

\noindent and then take the derivative of this equation, we would obtain the contracted Bianchi identities, valid for unimodular coordinates.  

As I show in section \ref{7}, in a subsequent short paper \cite{Einstein13} to the March 1916 one \cite{Einstein5}, Einstein provided further evidence to support his previous findings and showed that the contracted Bianchi identity arises as a consequence of the variational principle associated with the Einstein-Hilbert action [equation \eqref{Eq 357}]. Einstein demonstrated that there is no need for the unimodular restriction when deriving the contracted Bianchi identity from equation \eqref{Eq 357}. The contracted Bianchi identity equation \eqref{Eq 110} holds without the requirement of using unimodular coordinates or any specific coordinate system restrictions.

In the same review paper \cite{Einstein5}, Einstein finally achieved his long-awaited goal. He demonstrated that his new general theory of relativity connects to Newtonian physics in the weak field and slow-motion limits. In this limit, the theory reduces to the familiar Newtonian gravitational equations. Still, it provides a more comprehensive framework for understanding gravity near massive objects or at high speeds. Einstein demonstrated that when you consider the weak-field approximation and slow-moving particles (i.e., velocities much less than the speed of light), the geodesic equation leads to Newton's equations of motion as a first-order approximation.

The geodesic equation determines the path a free-falling particle would trace as it moves within this gravitational field \cite{Einstein1}, \cite{Einstein5}:

\begin{equation} \label{Eq 33}
\boxed{\frac{d^2 x_\tau}{ds^2} = \Gamma_{\mu\nu}^\tau \frac{dx_\mu}{ds} \frac{dx_\nu}{ds}.}
\end{equation}

\noindent where $\frac{d^2x_\tau}{ds^2}$ represents the acceleration of a particle along a geodesic, $\Gamma^\tau_{\mu \nu}$ represents the Christoffel symbols, and $\left(\frac{dx_\mu}{ds}\right)\left(\frac{dx_\nu}{ds}\right)$ represents the $4$-velocity of the particle.
Equation (\ref{Eq 33}) relates the curvature of spacetime (described by the Christoffel symbols $\Gamma$) to the motion of particles.
The Christoffel symbols in the geodesic equation encapsulate the information about the curvature of spacetime due to the presence of mass energy.  

Comparing the predicted orbits from these geodesic equations with the classical Newtonian mechanics' predictions, Einstein could highlight the corrections introduced by general relativity. Specifically, he focused on the perihelion precession of Mercury, which had a known discrepancy when using Newtonian mechanics (see discussion in section \ref{10}). 

When $\mu = \nu = 4$ (representing the time component of spacetime), we get the following three equations \cite{Einstein5}:

1) The geodesic equation \eqref{Eq 33} reduces to the geodesic equation for a particle's acceleration along a geodesic, specifically dealing with the $44$ component of the Christoffel symbols:

\begin{equation} \label{Eq 41-1}
\frac{d^2 x_\tau}{ds^2} = \Gamma^\tau_{44}.    
\end{equation}

\noindent $x_\tau$ takes the values from $0$ to $3$, corresponding to the different spacetime coordinates. $\frac{d^2 x_\tau}{ds^2}$ represents the second derivative of the coordinate $x_\tau$ with respect to the proper time $s$ along the geodesic. This describes how the coordinate $x_\tau$ changes as the particle moves on a geodesic through spacetime. $\Gamma^\tau_{44}$ indicates the time component of the Christoffel symbol, i.e, it represents how the curvature of spacetime in the time direction ($x^4$) affects the acceleration of a test particle along the $\tau^\text{th}$ coordinate $(x^\tau)$.  

Setting $s = t$ in equation \eqref{Eq 41-1} brings it to the form: 

\begin{equation} \label{Eq 41}
\frac{d^2 x_\tau}{dt^2} = \Gamma^\tau_{44}.    
\end{equation}

\noindent This is the first equation.

2) The second and third equations are the same as equation (\ref{Eq 41}) but modified to the weak-field approximation (and non-relativistic velocities) by incorporating gravitational effects: 

\begin{equation} \label{Eq 42}
\frac{d^2 x_\tau}{dt^2} = -\begin{bmatrix}
44 \\
\tau
\end{bmatrix} = -\frac{1}{2} \left(\frac{\partial g_{4\tau}}{\partial x_4} - \frac{\partial g_{4\tau}}{\partial x_4} + \frac{\partial g_{44}}{\partial x_\tau}\right), \quad \text{for},  \tau=1,2,3.
\end{equation}  \quad \text{and:}

\begin{equation}
\frac{d^2 x_4}{dt^2} = -\begin{bmatrix}
44 \\
4    
\end{bmatrix} \quad \text{for}, \tau = 4.
\end{equation}

\noindent These three equations correspond to the Newtonian equations of motion for a particle under the influence of gravity.  

The first and second terms in equation (\ref{Eq 42}) cancel each other out, and the equation (\ref{Eq 42}) simplifies to \cite{Einstein5}:

\begin{equation} \label{Eq 102}
\boxed{\frac{d^2 x_\tau}{dt^2} = -\frac{1}{2} \frac{\partial g_{44}}{\partial x_\tau},}
\end{equation}

In this specific case, equation (\ref{Eq 42}) results in a simplification where the terms involving $\frac{\partial g_{4\tau}}{\partial x_4}$ cancel out. 

\noindent Equation \eqref{Eq 102} represents the equations of motion for a material point in the Newtonian theory of gravity. It is derived from the geodesic equation in the weak-field limit, where velocities are much smaller than the speed of light. $\frac{d^2 x_\tau}{dt^2}$ represents the acceleration of a particle with respect to time. The particle is accelerated in the gravitational potential ($\frac{g_{44}}{2}$).

Next, Einstein writes the relationship between the $44$ component of the stress-energy tensor ($T_{44}$) and the density of matter ($\rho$) in the weak-field approximation \cite{Einstein5}:

\begin{equation} \label{Eq 355}
T_{44} = \left(\frac{dx_4}{dt}\right)^2 \rho = \rho = T.    
\end{equation}

\noindent In this equation $\frac{dx_4}{dt}$ represents the square of the component of the $4$-velocity with respect to time and $T$ represents the trace of the stress-energy tensor. 
Equation \eqref{Eq 355} is essentially the same as the equation for pressureless dust equation \eqref{Eq 356} that Einstein had formulated in 1912 (see section \ref{1}). So, the content of both equations is quite the same. There is a component ($T_{44}$) of the stress-energy tensor where $x_4$ corresponds to time, and $\rho$ is the energy density of the dust, which is assumed to be constant in the specific case of the weak field approximation. 

Einstein derives Poisson's equation from the field equations \eqref{Eq 346} in the Newtonian approximation. The terms with $\Gamma$ are Christoffel symbols and their derivatives. The right-hand side represents the distribution of energy and momentum in spacetime with a modification by subtracting half the trace $T$ of the energy-momentum tensor. 
Einstein derives Poisson's equation by focusing on the specific component $\mu = \nu = 4$ and under the conditions of the weak-field approximation. 
In the weak-field approximation, Einstein simplifies the left-hand side of the equation (\ref{Eq 346}) for the $44$ component. Simplifying and calculating the left-hand side of the equation (\ref{Eq 346} involves taking partial derivatives of the Christoffel symbols $\Gamma^\mu_{\nu\alpha}$ for spacetime coordinates. 

Einstein first provides a general representation for the left-hand side of the equation (\ref{Eq 346}), where $\tau$ can take values $1$, $2$, $3$, or $4$ \cite{Einstein5}:

\begin{equation} \label{Eq 44}
\frac{\partial}{\partial x_\tau} \Gamma^\tau_{\mu\nu} = \frac{\partial}{\partial x_1} \left[{\begin{matrix} \mu\nu \\ 1 \end{matrix}}\right] + \frac{\partial}{\partial x_2} \left[{\begin{matrix} \mu\nu \\ 2 \end{matrix}}\right] + \frac{\partial}{\partial x_3} \left[{\begin{matrix} \mu\nu \\ 3 \end{matrix}}\right] - \frac{\partial}{\partial x_4} \left[{\begin{matrix} \mu\nu \\ 4 \end{matrix}}\right].
\end{equation}

\noindent for $\tau=1,2,3,4$.    

\noindent He focuses on the specific component $\mu = \nu = 4$ of equation (\ref{Eq 44}), representing the calculation of $\frac{\partial^2 g_{44}}{\partial x_1^2}$ and similar terms:

\begin{equation} 
\frac{\partial}{\partial x_\tau} \Gamma^\tau_{44} = \boxed{\frac{\partial}{\partial x_1} \left[{\begin{matrix} 44 \\ 1 \end{matrix}}\right] + \frac{\partial}{\partial x_2} \left[{\begin{matrix} 44 \\ 2 \end{matrix}}\right] + \frac{\partial}{\partial x_3} \left[{\begin{matrix} 44 \\ 3 \end{matrix}}\right]} - \frac{\partial}{\partial x_4} \left[{\begin{matrix} 44 \\ 4 \end{matrix}}\right].
\end{equation}

\noindent He obtains an equation that only involves spatial derivatives and removes the time components entirely. This equation involves derivatives of $g_{44}$ for spatial coordinates \cite{Einstein5}:

\begin{equation} \label{Eq 57}
\frac{1}{2} \left(\frac{\partial^2 g_{44}}{\partial x_1^2} + \frac{\partial^2 g_{44}}{\partial x_2^2} - \frac{\partial^2 g_{44}}{\partial x_3^2}\right) = -\frac{1}{2} \nabla^2 g_{44}.    
\end{equation}

\noindent It represents the final result for the left-hand side of equation (\ref{Eq 346}), which simplifies to $\frac{1}{2}\left(\frac{\partial^2 g_{44}}{\partial x_1^2} + \frac{\partial^2 g_{44}}{\partial x_2^2} - \frac{\partial^2 g_{44}}{\partial x_3^2}\right)$, corresponding to the 44 component of the metric tensor.

After the above intermediate steps are used to calculate the left-hand side of the field equations (\ref{Eq 346}), for the specific component $\mu = \nu = 4$, we calculate the right-hand side. Plugging equation \eqref{Eq 355} into the right-hand side of the field equations (\ref{Eq 346}) gives:

\begin{equation} \label{Eq 304}
-\kappa\left(\rho-\frac{1}{2}\rho\right).    
\end{equation}

\noindent So, considering the Newtonian limit, combining \eqref{Eq 57} and \eqref{Eq 304} leads to the Poisson equation \cite{Einstein5}: 

\begin{equation} \label{Eq 45}
\boxed{\nabla^2 g_{44} = \kappa \rho.}    
\end{equation}

This equation is equivalent to equation (\ref{Eq 51}) (the speed of light is set to $c = 1$). Equation (\ref{Eq 45}) is the final result, where the Laplacian $\nabla^2g_{44}$ of the gravitational potential $g_{44}$, and $\kappa\rho$ is related to the mass density, consistent with the Newtonian theory of gravity.

\section{Einstein generalizes his theory} \label{7}

Einstein's journey to his general relativity was one of multiple detours and changes in approach. His equations during 1915-1916 show various stages in the mathematical formulation of the gravitational theory, especially as it transitioned from the older \emph{Entwurf} theory to the 1916 general relativity \cite{Einstein13}. 

Einstein's development of the general theory of relativity involved iterating through several of his previous mathematical formulations. It is interesting to note the parallels between the 1914 \emph{Entwurf} theory and the 1915 and 1916 forms of Einstein's general relativity. In some sense, Einstein's \emph{Entwurf} theory was a stepping stone. It was an intermediate theory, which, while not being the final correct theory of gravity, helped Einstein refine his thoughts and ideas. As demonstrated in the previous sections and as will be further illustrated here, there is a remarkable resemblance between Einstein's equations in the \emph{Entwurf} theory and equations in his 1915 and 1916 general relativity. The resemblance between the two suggests that Einstein's early ideas were not completely abandoned but evolved and refined. The transition from using tensor densities in the \emph{Entwurf} theory to using tensors (components change, but the geometric entity they represent remains invariant) in the 1915 theory was crucial. This shift allowed for more general coordinate systems, abandoning the need for adapted coordinates for unimodular ones. 

The action integral in the \emph{Entwurf} theory was in the form of an equation (\ref{12}). $L$ is the Lagrangian density and $d \tau$ is the spacetime volume element. This formulation made it so that the action would be invariant under coordinate transformations, a crucial feature. However, the equations were limited to specific coordinate systems when Einstein imposed $\sqrt{-g}=1$ in 1915. 
Indeed, in his general theory of relativity of 1915-1916, Einstein derived the field equations through a metric variational principle. However, his treatment was unsatisfactory because it relied on the coordinate condition $\sqrt{-g}=1$. In his review paper, "The Foundation of the General Theory of Relativity" \cite{Einstein5}, Einstein apologized for developing covariant field equations with respect to a coordinate system where $\sqrt{-g}=1$.

In the manuscript of the 1916 review paper titled "The Foundation of the General Theory of Relativity," this apology can be found in a note on page $40$a, which is only partially written, with the bottom half remaining empty. In it, Einstein mentioned that the most general laws governing the gravitational field and matter had been derived, but only for a coordinate system where $\sqrt{-g}=1$. This approach led to significantly simplifying formulas and calculations without sacrificing the requirement of general covariance. This was achieved by specializing the coordinate system from generally covariant equations.
Einstein still pondered whether the field equations could be formulated without assuming $\sqrt{-g}=1$ to achieve conservation of energy and momentum. He eventually found that both conservation principles held, but he chose not to include these comprehensive considerations in the 1916 review paper because they did not contribute anything objectively new.
The detailed considerations that Einstein decided not to communicate in the 1916 review paper \cite{Einstein5} were contained in an unpublished five-page appendix to part D of \cite{Einstein5} titled "Appendix: Formulation of the Theory on the Basis of a Variational Principle" \cite{CPAE6}, Doc. 31. This five-page Appendix forms the core of Einstein's short paper of 1916 titled "Hamilton's Principle and the General Theory of Relativity" \cite{Einstein13}. 

In March 1916, Hilbert published his paper on the foundations of physics. Hilbert's approach extensively employed a metric variational principle using a Lagrangian density. Hilbert expressed the action integral as \cite{Hilbert}:

\begin{equation} \label{Eq 380}
\int L \sqrt{-g} d\omega, \quad i = 1, 2, 3, 4.
\end{equation}

\noindent Where $L \sqrt{-g}$ represents the Lagrangian density. 

\noindent Hilbert illustrated that the Ricci tensor $K_{\mu \nu}$ allows the following invariant \cite{Hilbert}:

\begin{equation} \label{Eq 384}
K = g^{\mu \nu} K_{\mu \nu}.  
\end{equation}

\noindent Where $K_{\mu \nu}$ is the Ricci tensor, and $K$ is the Ricci scalar.

Subsequently, Einstein revisited the problem in his short paper titled "Hamilton's Principle and the General Theory of Relativity" of 1916 \cite{Einstein13}. In this paper, he presented the "comprehensive considerations" as promised in his 1916 review paper, "The Foundation of the General Theory of Relativity" \cite{Einstein5}. Through this work, Einstein successfully relinquished the restrictive unimodular coordinate condition $\sqrt{-g}=1$, allowing him to treat the variational principle without this restriction \cite{Einstein13}; \cite{Weinstein3}.

In the brief paper published in 1916, Einstein derived the field equations of general relativity employing a variational principle. Einstein formulated the equations without the restriction $\sqrt{-g}=1$, instead expressing the Lagrangian as related to the Ricci scalar $R$ \cite{Einstein13}:

\begin{equation} 
\mathcal{L} = \sqrt{-g} R.     
\end{equation}

\noindent Here, $R$ is the Ricci scalar (a contraction of the Ricci tensor):

\begin{equation} \label{Eq 385}
R = g^{\mu \nu} R_{\mu \nu}.    
\end{equation}

\noindent This is equation \eqref{Eq 384}. 

In 1915, Einstein assumed that $\sqrt{-g}=1$, which simplifies the form of the Lagrangian density to just: $\mathcal{L} = R$.
The equations were limited to specific coordinate systems.
Inspired by Hilbert's approach published in his paper of 1916 \cite{Hilbert} [equation \eqref{Eq 357} equals Hilbert's equation \eqref{Eq 380}], Einstein reconsidered his original equations. Instead of his \emph{Entwurf} action equation (\ref{12}), he defined a new action involving only the Ricci scalar, now known as the Einstein-Hilbert action \cite{Einstein13}: 

\begin{equation} \label{Eq 357}
S =\int \mathcal{L} \, d\tau, \quad \text{where} \quad L = \frac{\mathcal{L}}{\sqrt{-g}}.
\end{equation}

\noindent This action is invariant under all coordinate transformations, not just those for which $\sqrt{-g}=1$. Upon varying this action for the metric tensor, Einstein derived the Einstein field equations in their now-familiar form without the unimodular restriction.

As Hilbert demonstrated and Einstein realized the Ricci scalar, 
$K$ (represented by $R$ in modern notation) is a natural candidate for the Lagrangian when it comes to gravitational action, especially since it is an invariant under coordinate transformations \cite{Hilbert}.

In his paper \cite{Einstein13}, Einstein possibly was inspired by the formalism of the 1914 \emph{Entwurf} theory \cite{Einstein11} by circumventing the need for adapted coordinate systems. It can be seen as Einstein taking the core ideas from the \emph{Entwurf} theory of 1914 \cite{Einstein11}, correcting its shortcomings, and combining them with newer insights from the November 1915 theory \cite{Einstein1} and\cite{Einstein14}, to arrive at the final theory of the 1916 general relativity \cite{Einstein13}.
Specifically, the \emph{Entwurf} gravitational tensor [equation (\ref{Eq 64})] is covariant only in adapted coordinate systems \cite{Einstein11} [see equation (\ref{Eq 140})] under the restriction (\ref{Eq 115}).
Einstein wrote the variational principle equation (\ref{Eq 195}), and he recognized that the integrand divided by $\sqrt{-g}$ would remain invariant, leading to the invariant form equation (\ref{Eq 196}).  
Recall that Einstein's \emph{Entwurf} field equations were equations (\ref{Eq 186}). 
In these equations $\mathcal{G}_{\mu \nu}$ is the gravitational tensor density defined by equation (\ref{Eq 64}), while $T^{\tau}_{\nu}$ is the stress-energy tensor density. And equations (\ref{Eq 117}) and (\ref{Eq 184}) both represent the stress-energy pseudo-tensor of the gravitational field in the \emph{Entwurf} theory. They are derived from the Lagrangian equation (\ref{Eq 10}) and its derivatives for the metric tensor.
Einstein imposed two conditions or restrictions equations \eqref{Eq 115} and (\ref{Eq 23}) which led him to equations (\ref{Eq 184}) and (\ref{Eq 117}) \cite{Einstein11}. 

In 1916, Einstein introduced a modified version of the gravitational tensor \cite{Einstein13}:

\begin{equation} \label{Eq 24}
G = \frac{\mathcal{G}}{\sqrt{-g}},    
\end{equation}

\noindent which he argued to be invariant under arbitrary transformations.

\noindent So, Einstein wrote the action \eqref{Eq 357} in the form \cite{Einstein13}:

\begin{equation} \label {Eq 105}
S  =\int \mathcal{G} \, d\tau = \int \mathcal{G}^* \, d\tau + F,
\end{equation}

\noindent and showed that the variations of these two actions are equivalent, i.e., that the action remains invariant under coordinate transformations.

Einstein then derived an expression for $\mathcal{G}^*$. His starting point was the infinitesimal coordinate transformation $x_\nu$, which he had already written in his \emph{Entwurf} review paper of 1914 \cite{Einstein11}, \cite{Einstein13}:

\begin{equation} \label{Eq 368}
x_\nu' = x_\nu + \Delta x_\nu.
\end{equation}

\noindent Equation \eqref{Eq 368} represents how the coordinates $x_\nu$ transform under an infinitesimal change $\Delta x_\nu$. These transformations are localized, i.e., they only affect a limited region of spacetime. $\Delta x_\nu$ vanishes outside some arbitrarily chosen region of spacetime.

Einstein derived the transformation laws for the components of the metric tensor $g^{\mu \nu}$, and their derivatives $g^{\mu \nu}_{\sigma}$ under the coordinate transformations \cite{Einstein11}, \cite{Einstein13}:
 
\begin{equation} \label{Eq 359}
\Delta g^{\mu\nu} = g^{\mu\alpha} \frac{\partial \Delta x_\nu}{\partial x_\alpha} + g^{\nu\alpha} \frac{\partial \Delta x_\mu}{\partial x_\alpha},
\end{equation}

\begin{equation} \label{Eq 362}
\Delta g_{\sigma}^{\mu\nu} = \frac{\partial \Delta g^{\mu\nu}}{\partial x_\sigma} - g_{\alpha}^{\mu\nu} \frac{\partial \Delta x_\alpha}{\partial x_\sigma}.
\end{equation}
\vspace{1mm} 

\noindent Equations \eqref{Eq 359} and \eqref{Eq 362} represent the variations in the metric tensor components due to the coordinate transformations given by $\Delta g^{\mu\nu}$ and $\Delta g_{\sigma}^{\mu\nu}$.

Einstein then wrote the expression: 

\begin{equation}
\sqrt{-g}\Delta\left(\frac{\mathcal G}{\sqrt{-g}}\right),    
\end{equation}

\noindent which represents the variation of $\frac{\mathcal{G}}{\sqrt{-g}}$ [equation \eqref{Eq 24}] with respect to the metric tensor components $g^{\mu \nu}$. To understand how $\frac{\mathcal{G}}{\sqrt{-g}}$  varies with changes in the metric tensor components, Einstein introduced the variation operator $\Delta$. The result of applying $\Delta$ to $\frac{\mathcal{G}}{\sqrt{-g}}$  involves terms that account for changes in both the metric components and $\mathcal{G}$. 

The result of the variation of $\frac{\mathcal{G}}{\sqrt{-g}}$ with respect to the metric components is the equation below, equation \eqref{Eq 106}. This equation includes terms that account for changes in the metric components $\Delta g^{\mu\nu}$ and $\Delta g_{\sigma}^{\mu\nu}$ as well as the variation of the coordinate changes $\Delta x_\sigma$ with respect to the coordinates $x_\nu$ \cite{Einstein13}:

\begin{equation} \label{Eq 106}
S_\sigma^\nu \frac{\partial \Delta x_\sigma}{\partial x_\nu} + 2 \frac{\partial \mathcal{G}^*}{\partial g_{\alpha}^{\mu\sigma}} g^{\mu\nu} \frac{\partial^2 \Delta x_\sigma}{\partial x_\nu \partial x_\alpha}.   
\end{equation}

Einstein derives $S_\sigma^\nu$, which is defined as follows \cite{Einstein13}:

\begin{equation} \label{Eq 107}
S_\sigma^\nu = 2 \left(\frac{\partial \mathcal{G}^*}{\partial g^{\mu\sigma}} g^{\mu\nu} + \frac{\partial \mathcal{G}^*}{\partial g_{\alpha}^{\mu\sigma}} g_{\alpha}^{\mu\sigma}\right) + \delta_\sigma^\nu \mathcal{G}^* - \frac{\partial \mathcal{G}^*}{\partial g_{\nu}^{\mu\alpha}} g_{\sigma}^{\mu\alpha}.    
\end{equation}

Recall that Einstein considered transformations \eqref{Eq 368} that only affected a limited region of spacetime and vanished in an infinitesimal neighborhood of the boundary. He argued that the value of the integral of the action [equation \eqref{Eq 105}] over the boundary does not change during the transformation ($\Delta(F)=0$). Einstein concluded that the invariance of the action under coordinate transformations implies that:

\begin{equation} \label{Eq 358}
\Delta\left(\int \mathcal G \, d\tau\right) = \Delta\left(\int \mathcal{G}^* \, d\tau\right). 
\end{equation}

\noindent In the equation \eqref{Eq 358}, Einstein is considering the variation of two integrals with respect to the coordinate transformations, the left-hand side: $\Delta\left(\int \mathcal G \, d\tau\right)$ and the right-hand side: $\Delta\left(\int \mathcal{G}^* \, d\tau\right)$. Einstein's goal is to show that the left-hand side must vanish, i.e., that the variation of the integral on the left is zero under coordinate transformations. The reasoning behind Einstein's statement is the following. Since both $\frac{\mathcal{G}}{\sqrt{-g}}$ and $\sqrt{-g}\tau$ (the infinitesimal proper time interval) are invariants, any variations in these quantities due to coordinate transformations should cancel out when we take the variation on both sides of the equation. If the left-hand side of equation \eqref{Eq 358} vanishes ($\Delta\left(\int \mathcal G \, d\tau\right)$), the action remains invariant under coordinate transformations. 
In short, equation \eqref{Eq 358} expresses the invariance of the action under coordinate transformations. It states that the change in the action on the left-hand side equals the change in the action on the right-hand side when we change coordinates. 

The variation of the action involves how $\mathcal{G}^*$ changes due to coordinate transformations, which is described by equation \eqref{Eq 106}. This is where the terms related to $\Delta \mathcal{G}^*$ (the change in \(\mathcal{G}^*\) due to coordinate transformations) and derivatives of $\mathcal{G}^*$ come into play. 
$S_\sigma^\nu$ from equation \eqref{Eq 107} plays a role in understanding how $\mathcal{G}^*$ and the metric tensor components interact during these transformations. The fact that $\frac{\partial \Delta x_\sigma}{\partial x_\nu}$  can be represented as the Kronecker delta $\delta_{\sigma}^{\nu}$ simplifies some of the calculations and demonstrates that the variations are localized to specific coordinates.

Einstein defined the gravitational pseudo-tensor $t_{\sigma}^{\nu}$ in terms of $\mathcal{G}^*$ and its derivatives for the metric tensor. He derived two equivalent forms of the stress-energy pseudo-tensor of the gravitational field from equation (\ref{Eq 107}). He first obtained \(S_{\sigma}^{\nu} = 0\) from equation (\ref{Eq 106}). Einstein showed that the right-hand side of the equation (\ref{Eq 106}) must vanish when all $\frac{\partial^2 \Delta x_\sigma}{\partial x_\nu \partial x_\alpha}$ are allowed to vary freely. In other words, for any arbitrary variations of the coordinates $\Delta x_\sigma$, the equation should hold. This means that both terms on the right-hand side must independently vanish.
This leads to the conclusion that the first term of equation \eqref{Eq 106} $S_\sigma^\nu \frac{\partial \Delta x_\sigma}{\partial x_\nu}$ must vanish for any arbitrary variations. Therefore, it implies that $S_\sigma^\nu$ itself must be identically zero. This results in the identity condition: 

\begin{equation} \label{Eq 333}
S_\sigma^\nu \equiv 0.    
\end{equation}

In this context, equation \eqref{Eq 107} takes the following form:

\begin{equation} \label{Eq 108}
-\left( g^{\mu\nu} \frac{\partial \mathcal{G}^*}{\partial g^{\mu\sigma}} + g_{\alpha}^{\mu\nu} \frac{\partial \mathcal{G}^*}{\partial g_{\alpha}^{\mu\sigma}} \right) = \frac{1}{2} \left( \delta_{\sigma}^{\nu} G^* - g_{\sigma}^{\mu\alpha} \frac{\partial \mathcal{G}^*}{\partial g_{\nu}^{\mu\alpha}}\right).    
\end{equation}

\noindent The term on the right-hand side of equation (\ref{Eq 108}) leads to the following form of the stress-energy pseudo-tensor equation \cite{Einstein13}:

\begin{equation} \label{Eq 26}
t_{\sigma}^{\nu} = \frac{1}{2} \left( \delta_{\sigma}^{\nu} G^* - g_{\sigma}^{\mu\alpha} \frac{\partial \mathcal{G}^*}{\partial g_{\nu}^{\mu\alpha}} \right),    
\end{equation}

\noindent The term on the left-hand side of equation (\ref{Eq 108}) is also $t_{\sigma}^{\nu}$, and represents the second form of the stress-energy pseudo-tensor $t_{\sigma}^{\nu}$ \cite{Einstein13}: 

\begin{equation} \label{Eq 27}
t_{\sigma}^{\nu} = -\left( g^{\mu\nu} \frac{\partial \mathcal{G}^*}{\partial g^{\mu\sigma}} + g_{\alpha}^{\mu\nu} \frac{\partial \mathcal{G}^*}{\partial g_{\alpha}^{\mu\sigma}} \right).    
\end{equation}

The $S_{\sigma}^{\nu}=0$ terms [equations (\ref{Eq 333})] encapsulate constraints and conditions that the theory must satisfy. These terms are derived from the variation of the $G^*$ [equation (\ref{Eq 358})], a scalar invariant. They ensure that the theory meets the principles of invariance under coordinate transformations,
$S_{\sigma}^{\nu}=0$ encapsulates aspects of how the action integral of $\mathcal{G}^*$ varies under infinitesimal coordinate transformations. It quantifies how changes in the metric tensor affect $\mathcal{G}^*$ and its derivatives.
$S_{\sigma}^{\nu}$ characterizes how $\mathcal{G}^*$ varies when the metric is changed. These equations provide insights into the relationship between metric variations and the corresponding changes in the gravitational field represented by $\mathcal{G}^*$.
Equation (\ref{Eq 333}) implies that the change in $\mathcal{G}^*$ is zero: 

\begin{equation} \label{Eq 328}
\Delta \mathcal{G}^* = 0.    
\end{equation}

This is a significant outcome because it demonstrates that under infinitesimal coordinate transformations, $\mathcal{G}^*$ remains invariant. The equation (\ref{Eq 333}) shows that under infinitesimal coordinate transformations, the quantity $\mathcal{G}^*$ remains invariant. In other words, when $S_{\sigma}^{\nu}=0$, it means that the terms and conditions that arise from the variation of the action, as expressed by $S_{\sigma}^{\nu}$, are satisfied and result in identically zero equations. This, in turn, supports the assertion that both sides of equation \eqref{Eq 358} are equal to each other, implying the invariance of $\mathcal{G}^*$ under coordinate transformations. So, equation \eqref{Eq 333} is consistent with and supportive of the conclusion that equation \eqref{Eq 358} demonstrates the invariance of $\mathcal{G}^*$ under infinitesimal coordinate transformations \eqref{Eq 368}.

$S_{\sigma}^{\nu}$ are instrumental in deriving the Bianchi identity. Using $S_{\sigma}^{\nu} \equiv 0$ [equation \eqref{Eq 333}], Einstein derives an equation for the second derivatives of the metric changes. This is an essential step towards deriving a form of the contracted Bianchi identity  \cite{Einstein13}: 

\begin{equation} \label{Eq 110}
\boxed{\frac{\partial^2}{\partial x_{\nu} \partial x_{\alpha}} \left( g^{\mu\nu} \frac{\partial \mathcal{G}^*}{\partial g_{\alpha}^{\mu\sigma}} \right) = 0.}    
\end{equation}

Let us perform a brief derivation. We start by using equations (\ref{Eq 106}) and \eqref{Eq 328}.
Since \(\Delta \mathcal{G}^* = 0\), we have:

\begin{equation}
0 = \frac{\partial \Delta \mathcal{G}^*}{\partial \Delta x_{\sigma}} + 2 \frac{\partial \mathcal{G}^*}{\partial g_{\alpha}^{\mu\sigma}} g^{\mu\nu} \frac{\partial^2 \Delta x_{\sigma}}{\partial x_{\nu} \partial x_{\alpha}}.    
\end{equation}

\noindent Now, let us isolate the \(S_{\sigma}^{\nu}\) term:

\begin{equation}
S_{\sigma}^{\nu} = -2 \frac{\partial\mathcal{G}^*}{\partial g_{\alpha}^{\mu\sigma}} g^{\mu\nu} \frac{\partial^2 \Delta x_{\sigma}}{\partial x_{\nu} \partial x_{\alpha}} \cdot \left( \frac{1}{\frac{\partial \Delta x_{\sigma}}{\partial x_{\nu}}} \right).    
\end{equation}

\noindent However, the term in parentheses on the right-hand side can be simplified further. \(\frac{\partial \Delta x_{\sigma}}{\partial x_{\nu}}\) represents how the new coordinates \(x_{\nu}\) change concerning the infinitesimal coordinate changes \(\Delta x_{\sigma}\), and this can be approximated as a Kronecker delta:

\begin{equation}
\frac{\partial \Delta x_{\sigma}}{\partial x_{\nu}} \approx \delta_{\sigma}^{\nu}.    
\end{equation}

\noindent So, we can replace \(\frac{1}{\frac{\partial \Delta x_{\sigma}}{\partial x_{\nu}}}\) with \(\delta_{\sigma}^{\nu}\). Now, the equation becomes:

\begin{equation}
S_{\sigma}^{\nu} = -2 \frac{\partial \mathcal{G}^*}{\partial g_{\alpha}^{\mu\sigma}} g^{\mu\nu} \frac{\partial^2 \Delta x_{\sigma}}{\partial x_{\nu} \partial x_{\alpha}} \cdot \delta_{\sigma}^{\nu}.    
\end{equation}

\noindent Since the Kronecker delta \(\delta_{\sigma}^{\nu}\) equals $1$ when \(\sigma = \nu\) and $0$ otherwise, this equation simplifies to:

\begin{equation}
S_{\sigma}^{\nu} = -2 \frac{\partial \mathcal{G}^*}{\partial g_{\alpha}^{\mu\sigma}} g^{\mu\nu} \frac{\partial^2 \Delta x_{\sigma}}{\partial x_{\nu} \partial x_{\alpha}} \cdot 1.    
\end{equation}

\noindent And finally, we have:

\begin{equation}
\boxed{S_{\sigma}^{\nu} = -2 \frac{\partial \mathcal{G}^*}{\partial g_{\alpha}^{\mu\sigma}} g^{\mu\nu} \frac{\partial^2 \Delta x_{\sigma}}{\partial x_{\nu} \partial x_{\alpha}}.}    
\end{equation}

\noindent This is the expression for \(S_{\sigma}^{\nu}\) when \(\Delta \mathcal{G}^* = 0\) is satisfied.

\noindent But since $S_{\sigma}^{\nu}=0$, then the last equation simplifies as follows:

\begin{equation}
0 = -2 \frac{\partial \mathcal{G}^*}{\partial g_{\alpha}^{\mu\sigma}} g^{\mu\nu} \frac{\partial^2 \Delta x_{\sigma}}{\partial x_{\nu} \partial x_{\alpha}}.    
\end{equation}

\noindent and since $S_{\sigma}^{\nu}=0$, the left-hand side is zero, and this implies:

\begin{equation} 
0 = \frac{\partial \mathcal{G}^*}{\partial g_{\alpha}^{\mu\sigma}} g^{\mu\nu} \frac{\partial^2 \Delta x_{\sigma}}{\partial x_{\nu} \partial x_{\alpha}},    
\end{equation}

\noindent According to the invariant $\int \mathcal{G^*} d \tau$ this is rewritten as:

\begin{equation} \label{Eq 400}
0 = \int \frac{\partial \mathcal{G}^*}{\partial g_{\alpha}^{\mu\sigma}} g^{\mu\nu} \frac{\partial^2 \Delta x_{\sigma}}{\partial x_{\nu} \partial x_{\alpha}} d\tau.    
\end{equation}

\noindent In other words, Einstein performed two successive partial integrations with respect to $\tau$ and then defined: 

\begin{equation} \label{Eq 401}
\Delta x_\sigma = \frac{\partial \mathcal{G}^*}{\partial g_{\alpha}^{\mu\sigma}} g^{\mu\nu}.   
\end{equation}

\noindent So, with this definition, Einstein obtained the Bianchi identity equation \eqref{Eq 110} from equation \eqref{Eq 400}.

I want to offer a comment on this matter:
The contracted Bianchi identities that relate the Riemann curvature tensor to the Ricci tensor ensure the theory's internal consistency and energy-momentum conservation. 
Through the contracted Bianchi identities, Einstein showed that the $S_{\sigma}^{\nu}=0$ terms directly relate to the gravitational field's energy and momentum. 
\vspace{1mm} 

Equation \eqref{Eq 107} bears a resemblance to equation \eqref{Eq 322} from Einstein's 1914 \emph{Entwurf} paper \cite{Einstein11}. However, it is important to note that the derivation of both equations differs significantly despite the presence of some common transformations and equations, as extensively discussed in this section. Therefore, I will refrain from reiterating the points discussed in this section and focus on the identities. 

The contracted Bianchi identity (\ref{Eq 110}) and the 1914 \emph{Entwurf} $B_\sigma=0$ equation (\ref{Eq 115}) have a similar mathematical form. They both involve second derivatives of the metric and some functional of the metric for the coordinates. However, they serve a different purpose. $B_\sigma=0$ equation (\ref{Eq 115}) was introduced as a coordinate restriction in the \emph{Entwurf} theory. When equation (\ref{Eq 115}) is satisfied, it ensures that the integral equation (\ref{Eq 12}) is invariant under coordinate transformations that leave $B_\sigma=0$ equation (\ref{Eq 115}) unchanged. 
In 1914, Einstein introduced a coordinate condition $C_\sigma=0$, equation (\ref{Eq 12}) intended to restrict the choice of coordinates to those that satisfy the condition \cite{Einstein11}, see equation (\ref{Eq 120}).
On the other hand, in Einstein's 1916 general theory of relativity \cite{Einstein13}, the equation $S_{\sigma}^{\nu}=0$ [equation \eqref{Eq 333}] arises in the context of the derivation of equations \eqref{Eq 106}, and the contracted Bianchi identity [equation (\ref{Eq 110})]. Equation (\ref{Eq 110}) is an expression that arises as a consequence of the variational principle. In contrast to his 1916 work on general relativity, Einstein's derivations in the 1914 \emph{Entwurf} paper revolved around reconciling the two constraints: $B_\sigma=0$ [equation (\ref{Eq 115})] and $S_{\sigma}^{\nu}=0$ [equations (\ref{Eq 23})], while preserving energy-momentum conservation. In his 1916 paper, he transformed these constraints into the contracted Bianchi identities and the condition $S_{\sigma}^{\nu}=0$.

In the \emph{Entwurf} theory, equations (\ref{Eq 23}) and (\ref{Eq 115}) can be seen as a precursor to the conservation equations (\ref{Eq 116}). In general relativity, one obtains conservation laws from the contracted Bianchi identities and the Einstein field equations. The Bianchi identities ensure the Einstein field equations' left-hand side automatically satisfies the energy-momentum conservation equation (\ref{Eq 14}). 

Indeed, at first glance, the equations of the \emph{Entwurf} theory \cite{Einstein11} and Einstein's 1916 theory \cite{Einstein13} may seem very similar, especially when comparing the pseudo-tensors for gravitational field energy from both the \emph{Entwurf} theory and the 1916 equations. However, there are notable differences in interpretation. The 1916 formulation ties the curvature of spacetime directly to the presence of energy and momentum, making it a foundational element. 

In the 1916 paper \cite{Einstein13}, Einstein writes the generally covariant equations representing the energy-momentum balance for matter in a gravitational field:

\begin{equation} \label{Eq 116}
\boxed{\sum_{\nu} \frac{\partial \mathcal{T}_{\tau}^{\nu}}{\partial x_{\nu}} = \frac{1}{2} \sum_{\mu\nu\tau} g^{\mu\nu} \frac{\partial g_{\mu\nu}}{\partial x_{\sigma}} \mathcal{T}_{\tau}^{\nu}.} 
\end{equation}

In equation (\ref{Eq 116}), $\mathcal{T}_{\tau}^{\nu}$ represents the components of the energy-momentum tensor, and $g^{\mu\nu}$ represents the components of the metric tensor in general relativity. The equation describes how the components of the energy-momentum tensor relate to the derivatives of the metric $g$ in Einstein's generally covariant field equations.
Compare equation (\ref{Eq 116}) to equation (\ref{Eq 49}) from section \ref{4}. Equation (\ref{Eq 49}) is written in a covariant form, where the indices are in lower positions. Equation (\ref{Eq 116}) is written in a contravariant form, where the indices are in upper positions. When Einstein transitioned from the \emph{Entwurf} theory to general relativity, the primary change was in the field equations. The key difference between the two theories lies in the field equations used to describe how gravity is generated and how it affects spacetime geometry. So, equation (\ref{Eq 49}) from the \emph{Entwurf} theory and equation (\ref{Eq 116}) from general relativity are equivalent and describe the same physical processes in a covariant manner \cite{Weinstein1}, \cite{Weinstein1}.

\section{1915: The Perihelion of Mercury} \label{10}

In 1915, Einstein correctly explained the anomalous perihelion precession of Mercury \cite{Einstein15}. He started with his vacuum field equations in the absence of matter and energy [equations \eqref{Eq 39-1}]:

\begin{equation} \label{Eq 30}
\boxed{\sum_{\alpha} \frac{\partial \Gamma_{\mu\nu}^{\sigma}}{\partial x_{\alpha}} + \sum_{\alpha\beta} \Gamma_{\mu\beta}^{\sigma} \Gamma_{\nu\alpha}^{\beta} = 0.}
\end{equation}

The special theory of relativity uses the flat spacetime metric, known as the Minkowski metric. A simple coordinate system can represent this by the diagonal matrix $(-1, -1, -1, 1)$. Einstein's approach was to look at perturbations around this flat metric. So, rather than solving the complex equations for an exact metric, he solved it for a metric close to the flat Minkowski metric, especially where the gravitational field is weak. These perturbations can be considered small changes to the Minkowski metric. 

Einstein put forward conditions for the Sun's gravitational field:
It is static, i.e., it doesn't change with time. It is spherically symmetric: To a good approximation, the Sun is a perfect sphere, and its gravitational field should mirror this symmetry. It is asymptotically flat: far away from the Sun (or at infinity), spacetime should look flat, i.e., like Minkowski spacetime. 
Under these conditions and to the first approximation, Einstein found the metric \cite{Einstein15}:

\begin{equation} \label{Eq 29}
g_{\rho\sigma} = -\delta_{\rho\sigma} - \alpha\frac{ x_{\rho} x_{\sigma}}{r^3}, \quad g_{44} = 1 - \frac{\alpha}{r}.
\end{equation}

\noindent $r$ is the radial distance from the Sun, and $\alpha$ is a parameter defined by equation (\ref{Eq 73}).

Einstein examined the Christoffel symbols as the gravitational field's components. With the first-order metric, equation (\ref{Eq 29}), these symbols indicate how vectors change as they move around in curved space-time. 

\noindent Einstein derived the given formula \cite{Einstein15}:

\begin{equation} \label{Eq 82}
\Gamma_{\rho\sigma}^\tau = -\alpha \left( \delta_{\rho\sigma} \frac{x_\tau}{r^2} - \frac{3}{2} \frac{x_\rho x_\sigma x_\tau}{r^5} \right),
\end{equation}

\noindent In the first-order approximation of the Christoffel symbols, which represent the components of the gravitational field, Einstein considered indices $\rho, \sigma, \tau \in {1, 2, 3}$.
Specifically, for the gravitational field's time components, denoted by $\Gamma_{44}^\sigma$, the following equation applies: 

\begin{equation} \label{Eq 31}
\Gamma_{44}^\sigma = -\frac{\alpha}{2} \frac{x_\sigma}{r^3},
\end{equation}

\noindent where $\sigma$ takes on values from the set ${1, 2, 3}$. Equation \eqref{Eq 31} describes the behavior of the time components of the gravitational field, characterized by $\Gamma_{44}^\sigma$, and reveals their relationship to the variables $\alpha$, $x_\sigma$, and $r$.

Einstein then plugged the known components of the gravitational field into the vacuum field equation \eqref{Eq 30} to derive further insights about the gravitational field in the vicinity of the Sun. By only considering terms where $\mu = \nu = 4$ on the left-hand side of equation (\ref{Eq 30}), and inserting the results from equation (\ref{Eq 31}), he derived \cite{Einstein15}:

\begin{equation}
\sum_\sigma \frac{\partial \Gamma_{44}^\sigma}{\partial x_\sigma} = \frac{\alpha^2}{2r^4}.
\end{equation}

\noindent This led to the second-order approximation:

\begin{equation} \label{Eq 32}
\Gamma_{44}^\sigma = -\frac{\alpha}{2} \frac{x_\sigma}{r^3} \left(1-\frac{\alpha}{r}\right).
\end{equation}

After obtaining the necessary gravitational field components, Einstein utilized the geodesic equation (\ref{Eq 33}) to describe the motion of a point mass (i.e., a planet) in the curved space-time around the Sun \cite{Einstein15}. 

Einstein assumed that the planet's velocity was much less than the speed of light. Hence, \( dx_1, dx_2, dx_3 \) are considered smaller than \( dx_4 \). \( dx_4 \) refers to the time component, and this assumption is equivalent to saying that the spatial velocities are much smaller than the speed of light.

Using the previously derived gravitational field components \( \Gamma_{44}^\nu \), equation (\ref{Eq 31}) Einstein arrives at the equations of motion \cite{Einstein15}:

\begin{equation} \label{Eq 34}
\boxed{\frac{d^2 x_\nu}{ds^2} = \Gamma_{44}^\nu = -\frac{\alpha}{2} \frac{x_\nu}{r^3}.}
\end{equation}
for $\nu = 1,2,3$, and
\begin{equation}
\frac{d^2 x_4}{ds^2} = 0.
\end{equation}

\noindent These equations represent how a test particle (or planet) would move under the influence of the gravitational field near the Sun. The equation \( \frac{d^2 x_4}{ds^2} = 0 \) signifies that there is not any acceleration in the time component.
Einstein derives a relativistic version of Newton's second law for gravitational force, in which the potential \( \Phi \) is given by \( \Phi = -\frac{\alpha}{2r} \). If we set $s = x_4$ to first-order approximation, then this simplifies equations \eqref{Eq 34} to a form resembling classical Newtonian gravity:

\begin{equation}
\boxed{\frac{d^2 x_\nu}{dt^2} =\frac{\Phi x_\nu}{r^2}.}    
\end{equation}
\vspace{1mm} 

\noindent $x_\nu$ represents the position of a particle, $r$ represents the distance between the center of mass and the particle, and the equation describes the motion of a particle under the influence of a spherically symmetric gravitational field.
\vspace{1mm} 

Einstein then expressed three key concepts \cite{Einstein15}:
\vspace{1mm} 

1) \emph{The "area law"}  in polar coordinates (i.e., Kepler's second law) \( B \): This states that the area swept out by the planet as it orbits the Sun remains constant:

\begin{equation} \label{Eq 66}
\boxed{B = r^2 \frac{d\phi}{ds}.}
\end{equation}

\noindent$r$ is the radial distance between the planet and the Sun, and $ \frac{d\phi}{ds}$ represents the rate of change of the angular coordinate $\phi$ for the path length $s$ of the planet. $B$ is the area law constant.

Consider Besso's "area law" \eqref{Eq 300}. To obtain Einstein's "area law" \eqref{Eq 66}, we would need some additional assumptions. Besso's equation alone does not directly lead to Einstein's equation \eqref{Eq 66}. We would need more context related to $B, r, \phi$, and $s$ to make that connection.  

2) \emph{The energy \( E \)}: A combination of kinetic and potential energies:

\begin{equation} \label{Eq 71}
E = \frac{1}{2} u^2 + \Phi.
\end{equation}

\noindent This equation represents the total energy ($E$) in terms of the kinetic energy $\frac{1}{2} u^2$ and the gravitational potential energy $\Phi$.
\vspace{1mm} 

3) \emph{The effective velocity of the planet}: 

\begin{equation} \label{Eq 376}
u^2 = \frac{dr^2 + r^2 d\phi^2}{ds^2},
\end{equation}

\noindent which describes the effective velocity of the planet in the curved spacetime due to the Sun's gravitational field.

After considering the first-order gravitational field components, Einstein incorporated the second-order components \( \Gamma_{44}^\sigma \) as derived previously by equation (\ref{Eq 32}). He then used the geodesic equation (\ref{Eq 33})
to compute the planet's motion with these second-order effects.

Eventually, Einstein simplified the equation of motion for \( \nu = 4 \), reaching the equation \cite{Einstein15}:

\begin{equation}
\frac{d^2 x_4}{ds^2} = -\left(\frac{dx_4}{ds}\right)^2.
\end{equation}

He derived equations indicating how the Sun's gravitational field influences the time component of the planet's motion (connected to its energy). 

Using equation (\ref{Eq 31}), Einstein derives the equation of motion in the first-order approximation. He then progresses to incorporate second-order effects captured by the subsequent equations. 
The motion equations are written in terms of Christoffel symbols. They are involved in the geodesic equation that describes the motion of a test particle in a curved spacetime.
When Einstein derives the motion equations in the first-order approximation, he linearizes the problem. The dominant term he focuses on is $\Gamma_{44}^\sigma$ since the component $x^4$ corresponds to the time coordinate in a four-dimensional spacetime, making it crucial for understanding gravitational effects. As Einstein moves to the second-order approximation, he is accounting for higher-order terms that capture more subtle effects of the curved spacetime on the planet's motion. This includes the linear and second-order contributions of $\Gamma_{44}^\sigma$ and other components of the Christoffel symbols. 

Under these approximations, the resulting equations yield Kepler's laws of planetary motion. This is significant because it ties Einstein's general relativity to classical mechanics, showing that under certain approximations, Einstein's theory reduces to the previously accepted laws of gravity. Einstein's derivation also shows that the "area law" holds even in the second-order approximation. 

We compare the relativistic motion equation:

\begin{equation} \label{Eq 75}
\boxed{\frac{d^2 x_\nu}{ds^2} = -\frac{\alpha}{2} \frac{x_\nu}{r^3} \left(1 + 3\left(\frac{B}{r}\right)^2\right),}    
\end{equation}

\noindent and the Newtonian motion equation (\ref{Eq 34}). The main distinction is the additional factors in the relativistic versions, which account for the effects of relativistic gravity. 

Equation (\ref{Eq 75}) is a component of the geodesic equation. The left-hand side represents the acceleration of the particle's position, and the right-hand side represents the gravitational forces acting on it.
Specifically, equation (\ref{Eq 75}) describes how the position $x_\nu$ (where $\nu$ represents a spacetime coordinate) of a test particle (in this case, Mercury) changes as it moves along its orbit under the influence of the Sun's gravitational field. It uses the components of the four-velocity $\frac{d x_\nu}{ds}$ to represent how the velocity changes along the geodesic path. 

Einstein then derives another orbit equation \cite{Einstein15}:

\begin{equation} \label{Eq 67}
\frac{{dr^2 + r^2 d\phi^2}}{{ds^2}} = 2E + \frac{\alpha}{r} + \frac{\alpha B^2}{r^3}.
\end{equation}

\noindent This equation incorporates on the left-hand side the effective velocity of the planet in the Sun's gravitational field $u^2$ [equation \eqref{Eq 376}], and on the right-hand side $E$ [equation (\ref{Eq 71})], $B$ [equation (\ref{Eq 66})], and $\alpha$ [equation (\ref{Eq 73})].

Using the area law, we get from equation \eqref{Eq 67} a relativistic version of the equation that describes planetary orbits in Newtonian gravity:

\begin{equation} \label{Eq 68}
\boxed{\left(\frac{d}{d\phi} \left(\frac{1}{r}\right)\right)^2 = \frac{2E}{B^2} + \frac{\alpha}{rB^2} - \frac{1}{r^2} + \frac{\alpha}{r^3}.}  
\end{equation}

\noindent The right-hand side of this equation represents the square of the derivative of $\frac{1}{r}$ for $\phi$. 

\noindent We then integrate equation (\ref{Eq 68}) for $\phi$:

\begin{equation}
\int\left(\frac{d}{d\phi}\left(\frac{1}{r}\right)\right)^2 d\phi = \int\left(2\frac{E}{B^2} + rB^2\alpha - \frac{1}{r^2} + r^3\alpha\right) d\phi.
\end{equation}

\noindent This equation represents the integral of the square of the derivative of $\frac{1}{r}$ with respect to $\phi$ on the left-hand side, equated to the integral on the right-hand side involving terms related to energy ($E$) and the area law constant ($B$).

This integration yields the expression for $\phi$:

\begin{equation} \label{Eq 70}
\phi = \pi\left(1 + \frac{3}{2}\frac{\alpha}{a(1-e^2)}\right).
\end{equation}

This equation shows that $\pi$ represents the angle swept out by the radius vector from perihelion to aphelion during one orbit. Therefore, the precession of the perihelion angle per revolution is given by:

\begin{equation} \label{Eq 81}
\boxed{\Delta \phi = 2\pi - \phi = 2\pi - \pi\left(1 + \frac{3}{2}\frac{\alpha}{a(1-e^2)}\right) = \pi\frac{3\alpha}{a(1-e^2)}.}
\end{equation}

Finally, a central achievement of this derivation is the equation (\ref{Eq 70}), which calculates the precession of the perihelion of Mercury. This precession was a long-standing problem in astronomy, and Einstein's theory of general relativity was able to explain this discrepancy. The deviation of $\phi$ and $\pi$ gives the amount by which Mercury's orbit precesses every time it orbits the sun, a value that couldn't be entirely accounted for with Newtonian physics \cite{Einstein15}. 

On November 18, 1915, just a week after November 11, 1915, Einstein presented his groundbreaking work on the perihelion motion of Mercury. It was a significant achievement.
The day afterward, on November 19, 1915, Hilbert, the competitor of Einstein, sent a letter to Einstein congratulating him for his success in explaining the perihelion motion. In the letter, Hilbert humorously and somewhat cynically remarked, "If I could calculate as rapidly as you do" \cite{CPAE8} Doc. 149.
In June 1913, Besso visited Einstein in Zurich, and together, they embarked on solving the \emph{Entwurf} gravitation equations, as detailed in section \ref{2}. Their collaborative effort was dedicated to finding solutions to the perplexing issue of advancing Mercury's perihelion within the gravitational field of a static sun. During this endeavor, Einstein guided Besso through the intricate calculations necessary for this pursuit.
The \emph{Entwurf} theory, at the time, predicted a perihelion advance of approximately $18{"}$, a notable discrepancy from the observed $45{"}$. By the close of 1915, Einstein decided to abandon the \emph{Entwurf} gravitational theory. Instead, he transitioned to his new 1915 generally covariant field equations, applying the fundamental framework of calculations derived from the \emph{Einstein-Besso manuscript}.

With his newly formulated general relativity theory, Einstein achieved the correct precession of Mercury's perihelion, demonstrating remarkable progress based on the methods developed during his collaboration with Besso two years earlier. However, it is worth noting that Einstein did not formally acknowledge Besso's earlier contributions in his 1915 paper addressing the anomalous precession of Mercury, omitting any mention of Besso's name \cite{Weinstein1}.

Besso and Einstein used the "area law" as a crucial step in calculating the precession of Mercury's perihelion. The "area law" was crucial in explaining Mercury's precession in the 1913 and 1915 derivations. While there are similarities between the two derivations, it is crucial to understand the context and significance of each. To obtain Einstein's "area law" \eqref{Eq 66} from Besso's "area law" \eqref{Eq 300}, we would need some additional assumptions. Besso's equation alone does not directly lead to Einstein's equation \eqref{Eq 66}. We would need more context related to $B, r, \phi$, and $s$ to make that connection. Now, $s$ is related to Einstein's geodesic equation, as it represents the parameterization of the path followed by a particle moving under the influence of gravity, such as a planet orbiting the Sun. In the \emph{Entwurf} theory, Einstein did not explicitly write the geodesic equation. In Besso's equation \eqref{Eq 300}, the parameter $s$ is related to the action and the equations of motion in the \emph{Entwurf} theory. 

Einstein derived the "area law" equation (\ref{Eq 66}) as part of his derivation, similar to Besso's calculations. However, Einstein used the geodesic equation [equation (\ref{Eq 75})] and the Christoffel symbols [equation (\ref{Eq 82})] to calculate the motion of Mercury in a curved spacetime caused by the Sun's gravitational field.
Besso derived the "area law" [equation (\ref{Eq 80}) or equation \eqref{Eq 300}] in the \emph{Entwurf} theory, which was not generally covariant.
Besso and Einstein, in 1913, integrated this law over the entire orbit of Mercury and obtained an expression for the precession of the perihelion: $\Delta \varphi$, equation (\ref{Eq 61}) in terms of constants and parameters of the theory, ultimately yielding a value of $18^{''}$ per revolution. 
Einstein in 1915 developed his general theory of relativity, which is generally covariant and provides a more accurate description of gravitation. 
After incorporating second-order effects, he arrived at an equation for the precession of the perihelion angle ($\delta \phi$) [equation (\ref{Eq 81})], which was also in terms of constants and parameters. 
Einstein's theory ultimately explained the anomalous precession of Mercury's orbit, and he obtained a value for ($\delta \phi$) that matched the observed value.

Einstein's achievement in the Mercury perihelion paper was built upon earlier groundwork, the \emph{Einstein-Besso manuscript}; see section \ref{2}. However, the subtle differences between the derivation in the two works necessitated careful extrapolation and adjustment to reach the equations and conclusions presented in Einstein's famous Mercury perihelion paper \cite{Einstein15}. 

\section{Deflection of light}

In his 1916 review article, Einstein computed the deflection of light rays passing through the gravitational field of the Sun, as predicted by his new theory of general relativity \cite{Einstein5}.

Einstein begins with the geodesic equation [see equation (\ref{Eq 33})]:

\begin{equation} 
\boxed{g_{\mu\nu} \frac{dx^\mu}{ds} \frac{dx^\nu}{ds} = 0.}
\end{equation}

\noindent In this equation, $g_{\mu\nu}$ represents the components of the metric tensor, and $\frac{dx^\mu}{ds}$ represents the four-velocity of the particle. The equation represents the geodesic equation for null (massless) particles, such as photons. It states that the metric tensor's contraction with the particle's four-velocity is zero along its path. 

Einstein then sets the line element $ds^2$ equal to zero for null geodesics, i.e., along the world lines of light rays [see equation (\ref{Eq 97})]:

\begin{equation} \label{Eq 378}
\boxed{ds^2 = \sum_{\mu\nu} g_{\mu\nu} dx^\mu dx^\nu = 0.}
\end{equation}

\noindent This reflects that the proper time along a light ray's path is zero, as light travels at the speed of light ($c$) in a vacuum.

In his review paper \cite{Einstein5}, Einstein defined the magnitude of the velocity of light $\gamma$ as experienced by an observer on Earth:

\begin{equation} \label{Eq 46}
\gamma = \sqrt{\left(\frac{dx^1}{dx^4}\right)^2 + \left(\frac{dx^2}{dx^4}\right)^2 + \left(\frac{dx^3}{dx^4}\right)^2}.
\end{equation}

\noindent $\gamma$ is calculated in terms of the derivatives of spacetime coordinates with respect to the coordinate representing time ($dx_4$). In this equation, $dx^\mu$ represents the infinitesimal change in the spacetime coordinates ($x^1, x^2, x^3, x^4$) along the path of a light ray, where $\mu$ ranges from $1$ to $3$. The expression $\left(\frac{dx^1}{dx^4}\right)^2 + \left(\frac{dx^2}{dx^4}\right)^2 + \left(\frac{dx^3}{dx^4}\right)^2$ represents the squared components of the velocity vector of light in a vacuum along the spatial coordinates, which are normalized by dividing by $dx^4$ to obtain a dimensionless quantity. 
This equation quantifies how the velocity of light changes in a curved spacetime. 
$\gamma$ depends on how the spacetime metric $g_{\mu \nu}$ varies along the path of the light ray. 

Einstein predicts that light should be deflected when passing near massive objects and writes equation (\ref{Eq 73}) that relates the angle of deflection ($\alpha$) of a light ray passing near a massive object (the Sun) to the mass of the object ($M$) and the fundamental constants Newton's gravitational constant ($G$) and the speed of light ($c$).
He calculates the light ray's path's total deflection ($B$) as it passes by the massive Sun. It is an integral of the derivative of $\gamma$ [equation (\ref{Eq 46})] with respect to one of the spatial coordinates ($x_1$), showing how much the light ray's trajectory is bent due to the curvature of spacetime \cite{Einstein5}:

\begin{equation} \label{Eq 47}
\boxed{B = \int_{-\infty}^{+\infty} \frac{\partial\gamma}{\partial x^1} dx^2,} 
\end{equation}

\noindent where $dx^2$ is the second coordinate component and is a spacetime coordinate $dx^\mu$. I am using Einstein's original notation and trying to use consistent notation. 

Einstein then employs the approximation for the components of the metric tensor $g_{\rho\sigma}$, as given in equation (\ref{Eq 29}), in the vicinity of a massive object. These components describe the curvature of spacetime caused by the mass of the Sun. The specific form of the metric components $g_{\rho\sigma}$ and $g_{44}$ in equation (\ref{Eq 29}) is derived from the solutions of the generally covariant vacuum field equations (\ref{Eq 30}) for the gravitational field of the Sun.  
Equation (\ref{Eq 29}) incorporates the Schwarzschild metric, which describes the spacetime around a spherically symmetric mass like the Sun. However, in his 1916 paper, Einstein deliberately decided not to employ the Schwartschild metric, which contained a singularity he strongly disliked \cite{Weinstein1}.

Einstein then calculates $\gamma$ [equation (\ref{Eq 46})] based on the metric components from equation (\ref{Eq 29}):

\begin{equation} \label{Eq 48}
\gamma = \frac{dx^2}{dx^4} = \sqrt{-\frac{g_{44}}{g_{22}}} = 1 - \frac{\alpha}{2r}\left(1 + \frac{x^2_2}{r^2}\right). 
\end{equation}

He then calculates the total deflection ($B$) [equation (\ref{Eq 47})] of a light ray passing near a massive object (the Sun) using the expression for $\gamma$ from equation (\ref{Eq 48}):

\begin{equation} 
B = \frac{\kappa M}{2\pi\Delta} = \frac{4GM}{c^2\Delta} = \frac{2\alpha}{\Delta}. 
\end{equation}

\noindent This equation relates the deflection to the mass of the Sun ($M$) and the distance of the light ray's closest approach ($\Delta$) to the Sun.

Finally, Einstein obtained the specific value for the deflection ($B$) in terms of $\alpha$ [equation (\ref{Eq 73})] and $\Delta$, providing a concrete prediction for the angle of deflection of light rays passing by the Sun \cite{Einstein5}:

\begin{equation} \label{Eq 50}
\boxed{B = \frac{2\alpha}{\Delta} = \frac{1.7 \text{ seconds of arc}}{\Delta}.} 
\end{equation}

Equation (\ref{Eq 50}) for light deflection depends on the metric components [equation (\ref{Eq 29})] and the metric variations determined by the vacuum field equations (\ref{Eq 30}). The metric variations refer to how the metric components change as one moves through spacetime. In the case of the deflection of light, these variations in the metric components result from the gravitational field's influence on the path of the light ray. Light follows the curvature of spacetime, and as it passes through regions of varying gravitational potential, as determined by equation (\ref{Eq 29}), its path is bent. So, in equation (\ref{Eq 50}), we have $B$, the deflection of light, which depends not only on the specific form of the metric components $g_{\rho\sigma}$ given in equation (\ref{Eq 29}) but also on how these metric components change along the path of the light ray. The variations in the metric components capture the effects of gravity on the trajectory of light, leading to the observed deflection. These variations are an essential part of the calculation and are a consequence of the solutions to the generally covariant vacuum field equations. 

In the \emph{Entwurf} theory, the calculated deflection angle ($B$) for the gravitational field of the Sun is found to be half the value compared to what is predicted by Einstein's generally covariant 1915-1916 general relativity \cite{Weinstein1}:

\begin{equation}
B = \int_{-\infty}^{+\infty} \frac{\partial\gamma}{\partial x^1} dx^2 = \frac{2GM}{c^2 \Delta} = \boxed{\frac{\alpha}{\Delta} = \frac{0.85 \text{ seconds of arc}}{\Delta}.}
\end{equation}

In his November 18, 1915, paper on the perihelion of Mercury, Einstein discovered that his newly formulated field equations could account for the deflection of a ray of light as it passed by the Sun. Employing Huygens' principle and an approximate calculation method previously used to determine Mercury's perihelion advance, Einstein reported a significant finding. By applying Huygens' principle to equations \eqref{Eq 378} and \eqref{Eq 29}, and through straightforward calculations (see the calculations in this section), he determined that a light ray passing near the Sun at a distance $\Delta$ would experience an angular deflection of magnitude $\frac{2 \alpha}{\Delta}$ equivalent to $1.7$ arc seconds. This finding contrasted his earlier calculations from 1907 and 1911, which had relied on the equivalence principle, and his 1913 \emph{Entwurf} theory calculations. These earlier calculations had consistently yielded a deflection of half the magnitude, $\frac{\alpha}{\Delta}$ (equivalent to $0.83$ arc seconds) Therefore, Einstein concluded that a ray of light passing close to the Sun should undergo a deflection of $1.7$ seconds of arc, twice the previously predicted $0.85$ seconds of arc.

It is noteworthy that this particular topic was not one that Einstein had worked on with his close friend Besso. In writing the 1915 perihelion of Mercury paper \cite{Einstein15} and acknowledging Besso's assistance, Einstein faced the challenge of striking a subtle balance. He needed to explain his substantial contributions while acknowledging Besso's valuable support. Given the competitive context with David Hilbert to establish the correct form of the field equations, which Einstein ultimately achieved before Hilbert in \cite{Einstein14}, Einstein's time constraints may have limited his ability to provide a more elaborate acknowledgment to his dear friend.

Nevertheless, Einstein likely harbored a sense of remorse for not mentioning Besso in his paper on the perihelion of Mercury. In a similar context, before co-authoring a work with Leopold Infeld, Einstein received a suggestion from the latter. Infeld proposed that they should review the existing scientific literature and acknowledge the previous work of scientists who had explored similar subjects. In response to this suggestion, Einstein laughed and exclaimed, "Oh, yes, do it by all means. Already I have transgressed in this regard on numerous occasions" \cite{Infeld}.

\section{Discussion}

Einstein's theory of general relativity is widely regarded as one of the most significant breakthroughs in the history of physics. It challenged established notions and expanded the boundaries of our understanding, unveiling a new vision of spacetime and gravity.

Many intriguing questions surround Einstein's groundbreaking achievements. Was the theory of general relativity solely the creation of Einstein, the solitary figure who would seclude himself in an office with his violin, pipe, and a stack of papers? Or was it the culmination of Einstein's multifaceted collaborations and interactions with other scientists? To what extent did Einstein's close friend Besso and his schoolmate Grossmann contribute to the mathematics underpinning the general theory of relativity? Additionally, we can explore broader questions about how scientists engage in debates and refine their ideas. Was the theory of general relativity a conceptual and physical innovation emerging from a synthesis of Einstein's interactions with friends and colleagues? Or was it the result of debates and conflicts between Einstein and his contemporaries? 

In my book "General Relativity Conflict and Rivalries,"  which extends from my Ph.D. thesis under the late Mara Beller's supervision, I presented an approach that zooms in on the work of the individual scientist Einstein and his interactions with friends and colleagues. 
Beller's research emphasizes that the scientists with whom Einstein interacted, either implicitly or explicitly, formed a complex web of collaboration \cite{Beller}. In my book, I unearth latent undercurrents that could not have been exposed by merely tracing Einstein's intellectual journey toward the general theory of relativity. The interconnectedness and nuanced meanings unveiled in my book shed light on the pivotal figures who influenced Einstein during this transformative period.
According to this perspective, the ongoing dialogues and exchanges between Einstein and his associates significantly contributed to the development of general relativity \cite{Weinstein3}. 

Conventional historical accounts often downplay the contributions of physicists like Max Abraham, Gunnar Nordström, Gustav Mie, and David Hilbert, who offered different perspectives and engaged in discussions about the theory of gravitation. Works that did not align with Einstein's overarching conceptual framework, particularly the heuristic equivalence principle and Mach's principle, were often overlooked in favor of celebrating Einstein's prodigious scientific accomplishments. In my book, I highlight the limitations of a simplistic hero-worship narrative and the challenges of tracing Einstein's intellectual path to the general theory of relativity. I argue that we should not dismiss Einstein's responses to the works of Max Abraham, Gunnar Nordström, Gustav Mie, Tullio Levi-Civita, David Hilbert, and others. Instead, Einstein's engagement with these works, whether explicit or implicit, represented a dynamic interaction that significantly aided him in developing the general theory of relativity \cite{Weinstein3}.

\emph{In this paper, I want to emphasize that Einstein's genius was not isolated but flourished within dynamic scientific discourse. His work was undoubtedly enhanced and complemented by contributions and discussions with friends and colleagues, especially Michele Besso and Marcel Grossmann, demonstrating the collaborative nature of scientific progress. Nevertheless, it is of paramount importance to note that, while these individuals made significant contributions, they did not hold an equivalent role in shaping the overarching framework of general relativity. This framework remained primarily Einstein's intellectual creation. This balanced perspective enhances our understanding of the collaborative dynamics crucial in developing groundbreaking theories}.

\section*{Acknowledgement}

This work is supported by ERC advanced grant number 834735.
I thank Prof. John Stachel for sitting with me for many hours discussing Einstein's general relativity and the history of general relativity.

\end{document}